\RequirePackage[l2tabu,orthodox]{nag}

\documentclass[headsepline,footsepline,footinclude=false,oneside,fontsize=11pt,paper=a4,listof=totoc,bibliography=totoc]{scrbook} % one-sided

\PassOptionsToPackage{table,svgnames,dvipsnames}{xcolor}

\usepackage[utf8]{inputenc}
\usepackage[T1]{fontenc}
\usepackage[sc]{mathpazo}
\usepackage[ngerman,american]{babel}
\usepackage[autostyle]{csquotes}
\usepackage[%
  backend=biber,
  sorting=none,
  url=true,
  style=numeric,
  maxnames=4,
  minnames=3,
  maxbibnames=99,
  giveninits,
  uniquename=init]{biblatex} % TODO: adapt citation style
\usepackage{graphicx}
\graphicspath{ {./images/} }
\usepackage{scrhack} % necessary for listings package
\usepackage{listings}
\usepackage{lstautogobble}
\usepackage{tikz}
\usepackage{pgfplots}
\usepackage{pgfplotstable}
\usepackage{booktabs}
\usepackage[final]{microtype}
\usepackage{caption}
\usepackage[hidelinks]{hyperref} % hidelinks removes colored boxes around references and links

\usepackage{amsmath}
\usepackage{algorithm}
\usepackage[noend]{algpseudocode}
\usepackage{lscape}
\usepackage{longtable}
\usepackage{makecell}

\usepackage{booktabs}
\usepackage{graphicx}

\usepackage{comment}

\bibliography{bibliography}

\setkomafont{disposition}{\normalfont\bfseries} % use serif font for headings
\BeforeTOCHead[toc]{{\cleardoublepage\pdfbookmark[0]{\contentsname}{toc}}}

\definecolor{TUMBlue}{HTML}{0065BD}
\definecolor{TUMSecondaryBlue}{HTML}{005293}
\definecolor{TUMSecondaryBlue2}{HTML}{003359}
\definecolor{TUMBlack}{HTML}{000000}
\definecolor{TUMWhite}{HTML}{FFFFFF}
\definecolor{TUMDarkGray}{HTML}{333333}
\definecolor{TUMGray}{HTML}{808080}
\definecolor{TUMLightGray}{HTML}{CCCCC6}
\definecolor{TUMAccentGray}{HTML}{DAD7CB}
\definecolor{TUMAccentOrange}{HTML}{E37222}
\definecolor{TUMAccentGreen}{HTML}{A2AD00}
\definecolor{TUMAccentLightBlue}{HTML}{98C6EA}
\definecolor{TUMAccentBlue}{HTML}{64A0C8}

\pgfplotsset{compat=newest}
\pgfplotsset{
  cycle list={TUMBlue\\TUMAccentOrange\\TUMAccentGreen\\TUMSecondaryBlue2\\TUMDarkGray\\},
}

\newcommand*{\getUniversity}{Technische Universität München}
\newcommand*{\getFaculty}{Department of Informatics}
\newcommand*{\getTitle}{Automatic Identification and Classification of Share Buybacks and their Effect on Short-, Mid- and Long-Term Returns}
\newcommand*{\getTitleGer}{Automatische Identifizierung und Klassifizierung von Aktienrückkaufprogrammen und ihre Auswirkungen auf kurz-, mittel- und langfristige Renditen}
\newcommand*{\getAuthor}{Thilo Reintjes}
\newcommand*{\getDoctype}{Bachelor's Thesis in Information Systems}
\newcommand*{\getSupervisor}{Prof. Dr. rer. nat. Ruben Mayer}
\newcommand*{\getAdvisor}{M.Sc. Herbert Woisetschläger}
\newcommand*{\getSubmissionDate}{15.09.2022}
\newcommand*{\getSubmissionLocation}{Munich}

\begin{document}

% Set page numbering to avoid "destination with the same identifier has been already used" warning for cover page.
% (see https://en.wikibooks.org/wiki/LaTeX/Hyperlinks#Problems_with_Links_and_Pages).
\pagenumbering{alph}
\begin{titlepage}
  % HACK for two-sided documents: ignore binding correction for cover page.
  % Adapted from Markus Kohm's KOMA-Script titlepage=firstiscover handling.
  % See http://mirrors.ctan.org/macros/latex/contrib/koma-script/scrkernel-title.dtx,
  % \maketitle macro.
  \oddsidemargin=\evensidemargin\relax
  \textwidth=\dimexpr\paperwidth-2\evensidemargin-2in\relax
  \hsize=\textwidth\relax

  \centering

  \IfFileExists{logos/tum.pdf}{%
    \includegraphics[height=20mm]{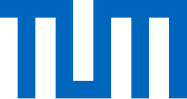}
  }{%
    \vspace*{20mm}
  }

  \vspace{5mm}
  {\huge\MakeUppercase{\getFaculty{}}}\\

  \vspace{5mm}
  {\large\MakeUppercase{\getUniversity{}}}\\

  \vspace{20mm}
  {\Large \getDoctype{}}

  \vspace{15mm}
  {\huge\bfseries \getTitle{}}

  \vspace{15mm}
  {\LARGE \getAuthor{}}

  \IfFileExists{logos/faculty.png}{%
    \vfill{}
    \includegraphics[height=20mm]{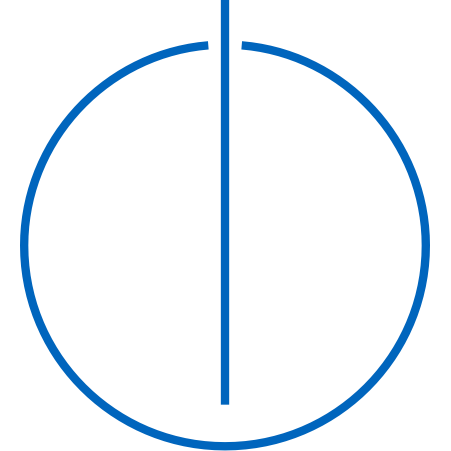}
  }{}
\end{titlepage}

\frontmatter{}

\begin{titlepage}
  \centering

  \IfFileExists{logos/tum.pdf}{%
    \includegraphics[height=20mm]{logos/tum.pdf}
  }{%
    \vspace*{20mm}
  }

  \vspace{4mm}
  {\huge\MakeUppercase{\getFaculty{}}}\\

  \vspace{4mm}
  {\large\MakeUppercase{\getUniversity{}}}\\

  \vspace{16mm}
  {\Large \getDoctype{}}

  \vspace{13mm}
  {\huge\bfseries \getTitle{}}

  \vspace{8mm}
  {\huge\bfseries \foreignlanguage{ngerman}{\getTitleGer{}}}

  \vspace{13mm}
  \begin{tabular}{l l}
    Author:          & \getAuthor{} \\
    Supervisor:      & \getSupervisor{} \\
    Advisor:         & \getAdvisor{} \\
    Submission Date: & \getSubmissionDate{} \\
  \end{tabular}

  \IfFileExists{logos/faculty.png}{%
    \vfill{}
    \includegraphics[height=20mm]{logos/faculty.png}
  }{}
\end{titlepage}

\thispagestyle{empty}
\vspace*{0.8\textheight}
\noindent
I confirm that this \MakeLowercase{\getDoctype{}} is my own work and I have documented all sources and material used.

\vspace{15mm}
\noindent
\getSubmissionLocation{}, \getSubmissionDate{} \hspace{50mm} \getAuthor{}

\cleardoublepage{}

\addcontentsline{toc}{chapter}{Acknowledgments}
\thispagestyle{empty}

\vspace*{20mm}

\begin{center}
{\usekomafont{section} Acknowledgments}
\end{center}

\vspace{10mm}

%TODO: Acknowledgments
I would like to express my sincere gratitude to Tim Jeck, Managing Director at Entrepreneurial Investment Partnership (EIP), for his continuous support and advice throughout my thesis. I would also like to express my gratefulness to my advisor Herbert Woisetschläger for his guidance, as well as to the Chair of Application and Middleware Systems for the computational resources made available to me.

\cleardoublepage{}

\chapter{\abstractname}

%TODO: Abstract

This thesis investigates share buybacks, specifically share buyback announcements. It addresses how to recognize such announcements, the excess return of share buybacks, and the prediction of returns after a share buyback announcement.
To achieve this, this thesis discusses ways to filter out share buyback announcements from an incoming news flow, analyze them and train a variety of machine learning models to predict future returns.
For this purpose, we illustrate two natural language processing (NLP) approaches for the automated detection of share buyback announcements. Even with very small amounts of training data, we can achieve an accuracy of up to 90\%. This thesis utilizes the developed NLP methods to generate a large dataset consisting of 57,155 share buyback announcements.
By analyzing this dataset, this thesis aims to show that the majority of companies, which have a share buyback announced are underperforming the MSCI World. A minority of companies, however, significantly outperform the MSCI World. This significant overperformance leads to a net gain when looking at the averages of all companies. If the benchmark index is adjusted for the respective size of the companies, the average overperformance disappears, and the majority underperforms even greater. However, it was found that companies that announce a share buyback with a volume of at least 1\% of their market capitalization, deliver, on average, a significant overperformance, even when using an adjusted benchmark index. It was also found that companies that announce share buybacks in times of crisis emerge better than the overall market.
Additionally, the generated dataset was used to train a total of 72 machine learning models. These models were compared, combined, and backtested. Through this, it was able to find a large number of strategies that could achieve an accuracy of up to 77\% in backtests and generate great excess returns. A variety of performance indicators could be improved across six different time frames and a significant overperformance was identified. This was achieved by training several models for different tasks and time frames as well as combining these different models, generating significant improvement by fusing weak learners, in order to create one strong learner.

\microtypesetup{protrusion=false}
\tableofcontents{}
\microtypesetup{protrusion=true}

\mainmatter{}

% !TeX root = ../main.tex
% Add the above to each chapter to make compiling the PDF easier in some editors.

\chapter{Introduction}\label{chapter:introduction}
In 2017, there were approximately 41,000 publicly listed companies worldwide \cite{DeLaCruzA.A.MedinaandY.Tang.2019}. To find potential investments, it is impossible for an investor or even a team to analyze each of these companies superficially, let alone in great detail. In order to still gain an overview over the global market, investors filter the large number of companies with the help of screeners. However, these screeners still include a large number of companies and usually only filter by a few characteristics. In this thesis we want to go one step further. We also screen companies for a well-known characteristic that can indicate overperformance, but we do not leave it at that. Additionally, companies that we have found through our screener are pre-analyzed using machine learning. We do this to further narrow our search funnel, letting the machine learning algorithms pre-select the most promising companies. The use of machine learning has the potential to perform a much deeper analysis of these companies compared to complex screeners. Specifically, we screen for companies announcing a share buyback and analyze the characteristics of the company as well as the characteristics of the buyback using a total of 24 trained state-of-the-art machine learning models. This significantly reduce the time required to find promising investments.
\section{Share buybacks}
        \begin{quote}
            "It is our belief that a company’s board has a responsibility to recognize opportunities to increase shareholder value, which includes allocating capital to execute large and well-timed buybacks." - Carl Icahn \cite{CarlIcahnIcahnEnterprises.24.08.2013}
        \end{quote}
Share buybacks - informally defined as a company buying their own shares - can be a well suited method to generate value for current shareholders and a positive signal for new investors.  Unfortunately, buybacks can differ greatly in their effects on the stock and in the incentive behind them. In this thesis we created a large dataset to analyze buybacks from around the world, reaching back to 2005 with the goal to find correlations and to train machine learning models to predict future returns.

\section{Research questions}
    The main purpose of this thesis is to investigate the following research questions. 
    \subsubsection{Is it possible to reliably detect share buyback announcements?}
    Before being able to analyze a buyback one has to identify the buyback as such. For that reason we are going to discuss and propose ways to reliably filter buyback announcements out of a current flow of stock related news.
    \subsubsection{Do share buybacks generate excess performance?}
    Buybacks can have a significant impact on the stock price but the effects are not necessarily significant nor outperforming the market. To see if buybacks are even worth a detailed analysis to enhance ones portfolio, we use a data set consisting of nearly 60,000 share buybacks to identify drivers of stock performance after a share buyback announcement.
    \subsubsection{Is it possible to reliably predict future returns of share buybacks?}
    In general, buybacks should generate value add for shareholders and indicate positive growth. Our goal is to be able to forecast how a company will perform in the short, medium and long term following a share buyback and use this insight to drastically reduce the number of companies being analyzed. For that reason, we try to differentiate and classify buybacks according to their future performance using several machine learning models. 
\section{Approach}
As with our research questions, our approach can be divided into three parts. Each part addresses the corresponding research question. 
First, we develop natural language processing (NLP) models with the task of filtering news about share buyback announcements in a news stream. This allows us to generate a large dataset of share buyback announcements. 
In the second part we enrich this dataset with company and market specific data to find statistical outliers. 
Lastly, we use the dataset to train several machine learning models and compare their performance.
\section{Contributions}
In this thesis we show effective applications of two NLP techniques for the classification of financial markets-related news. 
We show the use of these to create more complex datasets, which in turn can be used to train machine learning models for more difficult tasks.
Within the dataset, we find evidence for buyback-related overperformance and its positive influence in bear markets.
Furthermore, we show the applicability of complex decision trees as well as neural networks for the prediction of future share price developments. In particular, we show how we can significantly improve the performance of independently trained machine learning models with initially poor performance by combining them with each other.
\section{Organization}
In chapter \ref{chapter:Background}, the importance, the advantages and disadvantages, as well as the possibility of abusing of share buybacks are explained. Furthermore, the different machine learning models, concepts and training methods are elaborated. Chapter \ref{chapter:Empirical analysis of share repurchases announcements} is dedicated to data preprocessing, how the dataset was collected and prepared for evaluations. This follows descriptive statistics and machine learning methods to analyse the generated data. The chapter includes the presentation of the results of these models, how they are to be evaluated along with their limitations.

% !TeX root = ../main.tex
% Add the above to each chapter to make compiling the PDF easier in some editors.

\chapter{Background}\label{chapter:Background}

\section{Share buybacks}\label{sec:exbb}
    When a company conducts a share buyback \footnote{Also formulated as a combination of share/stock/equity buyback/repurchase}, it buys its own shares on the open market or from its shareholders. In most cases, the purchased shares are then canceled by the company. This reduces the number of outstanding shares. 
    Share buybacks can have a significant impact on the share price and therefore are subject to a number of regulations, depending on the specific locality of the company. One of these regulations is the obligation to give notice. In order to carry out a share buyback program, a company must announce this to the public in advance \cite{Yallapragada.2014}. Generally, this is done through local news portals like DGAP.de\footnote{News sample: https://kutt.it/U2zVTL} in Germany or by reporting to the local authorities which in return provide the reports to the public, such as the SEC\footnote{SEC-database example search: https://kutt.it/iY0wLw} in the USA.

    \subsection{Advantages and disadvantages of share buybacks}
        Share buybacks have varying positive and negative effects on their respective company.
        \subsubsection{Advantages of share buybacks}
            The most obvious positive effect is directly related to the reduction of outstanding shares. By reducing the number of outstanding shares, various key figures and values of the company can be influenced. Earnings per share (EPS) is one of the most important financial ratios to show the value of a company. EPS is calculated by dividing the earnings of a period by the number of outstanding shares. Buybacks allow companies to increase their EPS without increasing total earings. The same applies to values such as the share price \cite{horan2012buybacks}.
            
            Furthermore, companies can better protect themselves against takeover threats from other companies through a share buyback. Especially companies having a lot of cash on their balance sheet can quickly become an interesting takeover object. By reducing cash and boosting the value of their shares firms can counter takeover threats effectively.
            
            Buybacks also bring an income tax advantage for the shareholders. Companies that want to distribute cash to shareholders typically do so by dividend payments. Dividends received by a shareholder must be taxed directly as income. Share buybacks increase the value of the shares held by shareholders, but they do not have to pay taxes on this profit until they sell it for profit.
            
            Buybacks can be seen as a signal to the market that the company is undervalued. One reason for this can be that information is known internally which the public lacks in order to evaluate the company correctly. Because of this, share buybacks can be an interesting signal for investors to invest in a potentially undervalued company \cite{Yallapragada.2014}.

        \subsubsection{Disadvantages of share buybacks}
            As previously mentioned, values such as EPS or the share price can be directly influenced by a buyback. This brings the potential for manipulation of these figures in one's own interest. For example, share buybacks can be carried out shortly before the earnings report in order to create a more positive image of the company. Also, personnel who are paid contractually agreed bonus payments based on, for example, the share price has a strong motivation to abuse share buybacks for their own interest. It is common for a CEO to receive a bonus payment when the company share price exceeds a certain threshold. Therefore, a CEO is obviously incentivized to carry out a share buyback if this brings the share price above this threshold, no matter how much sense a share buyback makes for the company \cite{Palladino.2020}.
            
            But to be able to benefit from the positive effects of a share buyback, a company does not have to carry it out. The announcement of a share buyback itself has a positive effect on the share price. Unfortunately, there is no obligation for a company to carry out an announced share buyback to the full extent or at all. Thus, the company or personnel can profit from a short-term increase in the share price without having generated any value add for the shareholder. As a result, management personnel holding stock options have a strong motivation to announce a share buyback in order to increase the share price in the short term before exercising these stock options \cite{Yallapragada.2014}.

     \subsection{Historical performance of share buybacks}
    After describing the advantages and disadvantages a share buyback can have for an individual company and its shareholders, we now take a look at a general historical overview of many share buybacks. This shows that companies perform better after announcing their share buyback program than a suitable peer group. The longer the time window, the more insignificant the positive effect of the share buyback \cite{Dann.1981, PaulFruin.2014, Bartov.1991}. In addition, it is found that there is a negative correlation between the size of the company and overperformance. Thus, the smaller the company, the stronger and longer lasting the effect of a share buyback \cite{PaulFruin.2014}.
    For example, companies from the Russell 3k Index, which execute buybacks, overperform on average by 0.6\% over a time frame of one month and by 1.38\% over a time frame of one year, after announcing a share buyback (Analyzed  time frame: Jan 2004 - July 2013) \cite{PaulFruin.2014}.
        
    \subsection{Conclusion}
        A purposeful share buyback is a reasonable way to use excess cash and boost undervalued shares. It can generate value add for shareholders and is a positive signal to the market in general. 
        Unfortunately, share buybacks offer great potential for abuse and can be used for personal interest, often contrary to the interest of the company or shareholders. 
        
\section{Machine learning}
In this thesis many different machine learning algorithms, models, namely from the field of natural language processing (NLP), tabular data regression as well as classification, are used. This chapter presents the models used throughout this work along with methods for training and improving performance.

    \subsection{Text classification}\label{subsec:textclass}
    To be able to filter out any unrelated news before processing them any further, we need to apply classification methods to distinguish between relevant and irrelevant news. In this thesis, we used two methods. A simple Regular Expression Matching and a NLP deep-learning model.
        \subsubsection{Regular Expression Matching}
        Regular Expression Matching (RegEx) is used to find specific patterns in a string or sub-string. These patterns are defined with the help of a predefined syntax, in this case, the syntax of Lib/re.py\footnote{https://docs.python.org/3/library/re.html}, the regular expression operations of Python \cite{Python.19.07.2022}.
        \subsubsection{DistilBERT}
        One of the biggest advancements for NLP in recent years was the introduction of transfer learning. Modern NLP-models take advantage of this by implementing transformers, which enables them to get a deeper understanding of a given input text using attention mechanisms. These attention mechanisms help to identify the most relevant parts in a text sequence \cite{Vaswani.12.06.2017}. A transformer consists of an encoder and a decoder, where the encoder embeds the text sequence to be processed and the decoder generates a new sequence based on the decoder's embedded words. 
        The NLP-Model Bidirectional Encoder Representations from Transformers (BERT) takes advantage of these transformers, which achieved state-of-the-art accuracy on various NLP-tasks and was published by the authors to make pre-trained versions of BERT available for the public. BERT uses the encoders part of these transformers and concatenates several of them \cite{devlin2019bert}.
        DistilBERT is a distilled version of BERT which reduces the size of BERT by 40\%, making it 60\% faster to train while achieving 97\% of accuracy \cite{https://doi.org/10.48550/arxiv.1910.01108}. We take advantage of this to reduce training time and cost.
        
    \subsection{Tabular data regression and classification} \label{subsec:backgroundData}
        \subsubsection{Decision Tree}
\begin{figure}[h]
            \centering
            \includegraphics[scale=0.75]{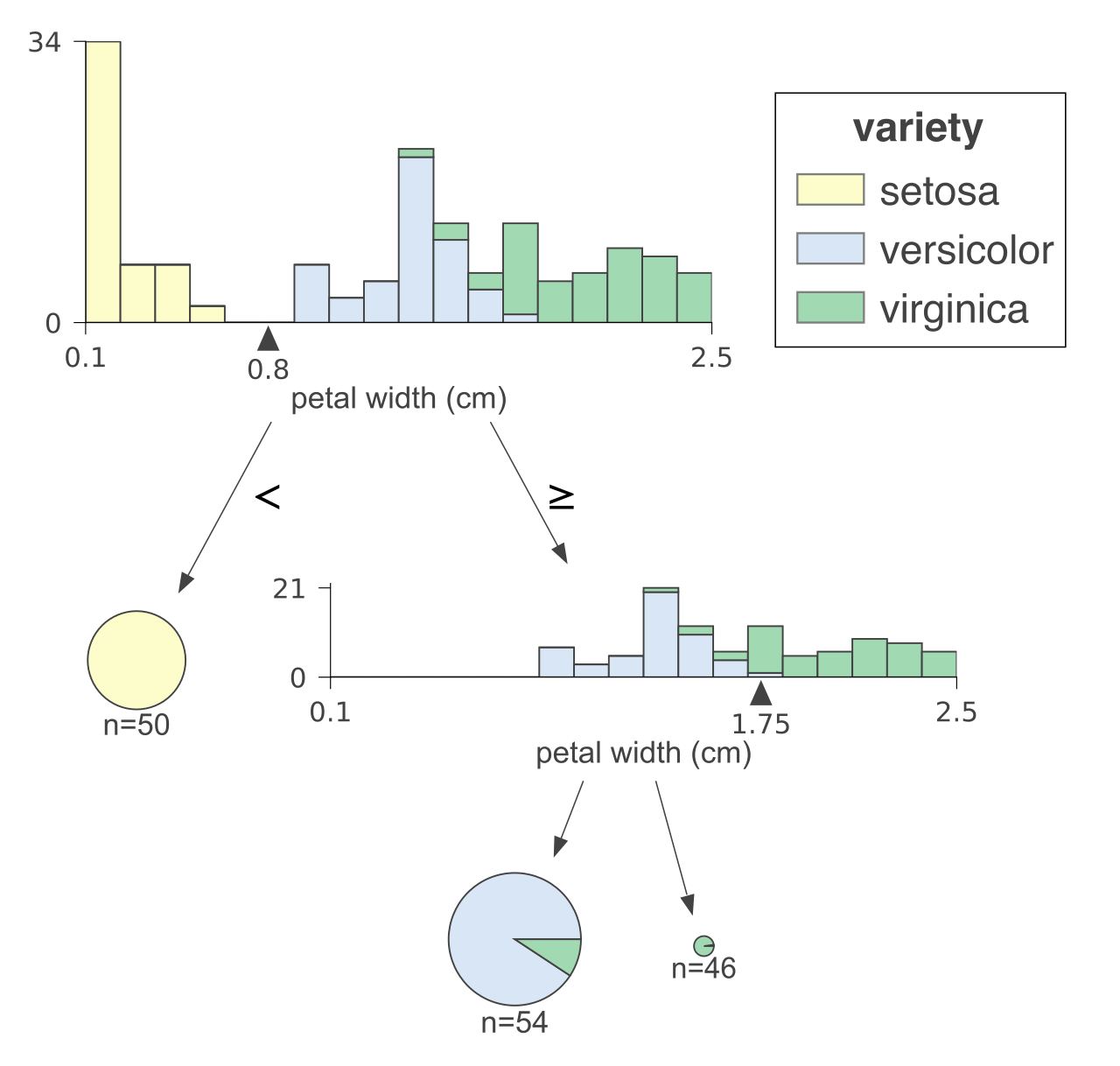}
            \caption{Visualisation of a Decision Tree \cite{dtreeviz.}}
            \label{fig:decivionVis}
        \end{figure}

        A decision tree has a tree-like structure, which is walked through from the root of the tree till a leaf is reached. 
        Starting at the root of the tree, the root node splits into two more nodes. This is true for all nodes except the leaf. Each level of the tree handles an input variable and decides which continuing node to visit based on a threshold. At the end of a walk, you end up with a leaf that represents the class or value predicted by the model.
        Plot \ref{fig:decivionVis} shows a classification decision tree. This tree is based on the Iris Data Set\footnote{https://archive.ics.uci.edu/ml/datasets/Iris}, which is one of the best-known datasets for pattern recognition. It contains three classes, each of which occurs 50 times in the dataset \cite{MachineLearningRepository.}.
        The tree consists of two levels and classifies on the basis of one value, the three types of an iris plant. 
        Each decision node is visualized with the distribution in the dataset over the three classes and the split point. The leaf nodes show the distribution of the results.
        The type setosa, for example, is recognized with an accuracy of 100\%. For the other two leaf nodes, an inaccuracy can be seen since the classes versicolor and virginica cannot be clearly divided by petal width.

            \paragraph{Decision Tree Learning - Classification and Regression Trees (CART)}
            To be able to take a sensible walk through the tree, the variable, and the thresholds, after which it is decided whether to continue at the left or right node, must be adjusted beforehand. This is done with decision tree learning.
            There are different learning algorithms. In this case, we use the CART algorithm. When using CART, the underlying dataset is divided into smaller sub-datasets in each step. The dataset is split based on the variable that has the smallest impurity compared to the other variables. 
            Impurity can be calculated in different ways. CART uses the Gini Impurity. This indicates with a number between zero and one how likely it is to get a misclassification \cite{StevenLoaiza.Gini.}. 
            For example, we assume a regression problem for which we take from a dataset the input data $X_1$, $X_2$, and a continuous label Y. To generate a decision tree now, we start to split the dataset into two sub-datasets by one input variable. Thus, the first split can be at $X_{1\ }=\ t_{1\ }$. With this, we divide the dataset into two "regions":

            \begin{equation}
                R_1=\left\{\left.\left(X_1,X_2\right)\ \right|\ X_1\le t_1\right\}
            \end{equation} and
            \begin{equation}
                R_2=\left\{\left.\left(X_1,X_2\right)\ \right|\ X_1>t_1\right\}
            \end{equation}.
            
            We can further split these regions recursively. So, we can divide $R_1$ into $R_1$and$\ R_3$. If one chooses $X_2=t_2$ so is
            
            \begin{equation}
                R_1=\left\{\left.\left(X_1,X_2\right)\ \right|{\ X}_1\le t_1\ \land\ X_2\le t_2\right\}
            \end{equation} and 
            \begin{equation} 
            R_3=\left\{\left.\left(X_1,X_2\right)\ \right|\ X_1\le t_1\ \land\ X_2>t_2\right\}
            \end{equation}.
            
            This rule-based splitting of the dataset can be continued recursively as desired with the variables $X_1$, $X_2$.
            The corresponding regression model outputs the constant $c_m$ as the mean of $Y$ of the corresponding region, which is defined as:
            
            \begin{equation}
                f\left(X\right)=\sum_{m=1}^{3}{c_mI\left\{\left(X_1,X_2\right)\in R_m\right\}}
            \end{equation}
            
            Or generally with regions $R_1,\ R_2,\ ...,\ R_M$ in $M$:
            
            \begin{equation}
                f\left(X\right)=\sum_{m=1}^{M}{c_mI\left\{X\in R_m\right\}}
            \end{equation}
            To decide which variable is used to split the dataset at which threshold, a greedy algorithm is used. A greedy algorithm searches from its starting position, step by step, for a local optimum. It optimizes a predefined heuristic. Here the heuristic is the impurity, which has to be minimized. The algorithm cannot overview the whole problem space, but always only the next step. Therefore, it always approaches the values that provide the greatest improvement upon one step \cite{bouchet1987greedy}.

            % Assuming that the dataset consists of $N$ datapoints which consists of the inputs $p$ and the output $alpha$ so the $\left(x_i,y_i\right)$ and $i\ =\ 1,\ 2,\ \ldots,\ N$ with $x\ =\ {(x}_{i1}, x_{i2},\ \ldots,\ x_{ip})$. The criterion to determent the best binary partition is:

            % \begin{equation}
            %     {\hat{c}}_m=ave\left\{\left.y_i\right|x_i\in R_m\right\}
            % \end{equation}
            
            % To now find the best variable\ j and threshold t as split points with regions defined as:
            
            % \begin{equation}
            %     R_1\left(j,s\right)=\left\{\left.X\right|X_j\le s\right\}
            % \end{equation} and
            % \begin{equation}
            %     R_2\left(j,s\right)=\left\{\left.X\right|X_j\le s\right\}
            % \end{equation},
            
            % we minimize with the help of a square loss function:
            
            % \begin{equation}
            %     \min\below{j,\ s}\left[\min\below{c_1}\sum_{x_i\in R_1\left(j,s\right)}\left(y_i-ave\left(y_i\ |\ x_i\in R_1\left(j,s\right)\right)\right)^2\ +\ \min\below{c_2}\sum_{x_i\in R_2\left(j,s\right)}\left(y_i-\ ave\left(y_i\ |\ x_i\in R_2\left(j,s\right)\right)\right)^2\right]
            % \end{equation}
            
            Once the best variable is found, each sub-dataset can be split again into further sub-datasets using all variables. 
            Decision trees can be divided into two types. Classification trees and regression trees. Structurally they are the same, but they differ in their leaf and the heuristics used to build the tree. Classification trees are based on a discrete set of values from the dataset class of the corresponding region, whereas regression trees represent continuous values, generated from the average of the respective region. Classification Trees use heuristics such as the Gini-coefficient or log loss. This way one can create an almost arbitrarily large tree. However, this is not very sensible, because this quickly results in overfitting the model \cite{.2009}.          
            \paragraph{Minimal Cost-Complexity Pruning}
            To prevent over-fitting, minimal cost-complexity pruning is used. The goal of minimal cost-complexity pruning is to find a sub-tree of the previously generated tree that minimizes the impurity depending on the heuristic used, by reducing the complexity of the tree. Here, the complexity parameter $\alpha\geq0$ is defined to compute the cost-complexity measure of the tree $R_\alpha(T)$. With $\left|\widetilde{T}\right|$, as the number of terminal nodes of $T$ and $R(T)$ as the total misclassification rate. Minimal cost-complexity pruning finds the sub-tree of $T$ which minimizes $R_\alpha\left(T\right)$ with\cite{scikitlearn.b}:

            \begin{equation}
                R_\alpha\left(T\right)=R\left(T\right)+\alpha\left|\widetilde{T}\right|
            \end{equation}

        \subsubsection{Forests of Randomized Trees}
        To generate and connect multiple trees, we use two ensemble methods, both of which are based on randomized trees.
            \paragraph{Random Forest}aims to improve the accuracy and robustness of never seen data samples by building multiple trees, with a slight variation in all of them. These slight variations are introduced by two factors of randomness. One is the bootstrapping of datasets. A separate dataset is bootstrapped for each tree by randomly pulling samples from the original dataset. It is important that duplicates are allowed. The second source of randomness is how the variables are split at a given node. They are still selected according to the minimum impurity, however not in comparison with all variables, but rather with a randomly selected subset of all variables. This way one can build many slightly different trees, which are all based on the same dataset. To make a prediction, the input data given to each tree is evaluated individually and at the end, the class or value that was predicted most often is used as the final prediction \cite{Breiman.2001}.
            In the technical implementation of this thesis, however, the average of the probabilistic predictions is used instead of the most frequently predicted class as the final prediction, due to the underlying library \cite{scikitlearn.c}. 
            \paragraph{Extra Trees}or extremely randomized trees adds another source of randomness compared to the random forest method. Instead of using the variable-threshold in combination with the smallest impurity from a subset of variables, all variables are compared, but with a randomly selected threshold \cite{scikitlearn.}. 
        \subsubsection{Gradient Boosting}
        Gradient Boosting is a family of models, which can be used for classification, regression, and more complex tasks like ranking. In general, the idea of boosting is to have a “weak learner”, a model which is slightly better than prediction at random, and improve this weak learner by combining its output with the output of other weak learners. Assuming we have a set of $M$ of weak learners $f_m\left(x\right)$ where x is the input variable, we define $F\left(x\right)$ as the final, boosted model as:
        \begin{equation}
            F\left(x\right)=\sum_{m=1}^{M}{f_m(x)}
        \end{equation}
        In the case of gradient boosting, a gradient is used to boost the weak learners. 
        To apply gradient boosting, we start with a weak learner $f_1\left(x\right)$, which receives as input $x_i$, outputs ${\hat{y}}_i$ and tries to predict $y_i$, the ground truth of a given dataset. Here $i$ is the index of datapoint of the given dataset with $i=1,\ \ldots,\ N$. This first weak learner is fitted to predict $y_i$. Once the first model is fitted, we calculate a loss for each prediction using an appropriate loss, depending on the tasks. For this example, we assume a regression task and will choose the mean squared error (MSE) as our loss

        \begin{equation}
            {L_{MSE}\left(y_i,\ {\hat{y}}_i\right)={(y_i-\ {\hat{y}}_i)}^2}
        \end{equation}

        . With a loss function defined, we calculate the gradient of ${\hat{y}}_i$ defined as:
        \begin{equation}
            {\hat{r}}_{m,\ i}=\frac{d{\ L}_{MSE}(y_i,\ {\hat{y}}_i)}{d\ {\hat{y}}_i}
        \end{equation}
        We then use the gradient to fit a new weak learner $f_2\left(x\right)$, which does not predict the ground truth of the dataset anymore but the loss gradient of the model which predicts the ground truth. Once the second model is fitted, a constant $\gamma_2$ is calculated to minimize the overall loss: 
        \begin{equation}
            \gamma_2=\min\gamma{\sum_{i=1}^{N}{L_{MSE}(y_i,\ f_1\left(x_i\right)+\gamma f_2(x_i))}}
        \end{equation}
        The resulting model is now definded as $F(x)$ and constructed as:
        \begin{equation}
            F\left(x\right)=\ f_1\left(x\right)+\gamma_2f_2(x)
        \end{equation}
        To enhance the model $F(x)$ further, one would fit a third model $f_3(x)$ to predict the gradient of the loss of $f_2(x)$. This process can be repeated arbitrarily \cite{Schapire.2014}. Gradient boosting is used in combination with all kinds of machine learning techniques but especially tree boosting, where it achieves state-of-the-art results \cite{Chen.2016}. 

            \paragraph{XGBoost}is a scalable machine learning solution that creates an end-to-end tree boosting system. Introduced in 2014, DMLC\footnote{https://dmlc.github.io/} published an open-source library that made XGBoost available to the public. XGBoost major contributions to tree-based boosting are out-of-core computation, cache-aware, and sparsity-aware learning, and combining all in an end-to-end system, greatly improving training time and efficiency, especially when using large datasets.
            To improve the most time-consuming part of tree learning, which is the sorting of data, XGBoost implements in-memory units, named “blocks”. Each block stores compressed data ordered by the corresponding feature value. This must only be computed once and can be reused in each training iteration, enabling parts of the algorithm to be parallelized. 
            This block structure optimizes the computation complexity. However, this introduces non-continuous memory accesses, which adds an immediate dependency between the accumulation and fetch operation. To solve this, XGBoost uses cache-aware access. Depending on the split-finding algorithm this problem can be fixed by fetching the gradient statistics separately and performing the accumulation in a mini-batch manner. Alternatively, this can also be solved by adjusting the block correspondingly to the cache storage costs of the gradient statistics. 
            To further optimize the algorithm and make use of as much computer resources at once as possible, XGBoost introduces out-of-core computation to take advantage of free disk space for data that doesn’t fit into the main memory. Using the already described block architecture this enables XGBoost to pre-load data from the disk with computation happening in concurrence \cite{Chen.2016}.
            \paragraph{LightGBM}, developed by Microsoft (2016), is another gradient tree boosting implantation which further improves computation time. It builds on the improvements of XGBoost and pGBRT (another gradient boosting tree algorithm). Adding to those previous implementations, LightGBM introduces Gradient-based One-Side Sampling (GOSS) and Exclusive Feature Bundling (EFB) which combined are the core improvements of LightGBM. 
            Earlier implementations must scan for all features of all data instances to maximize the information gain respectively to minimize the impurity for each split. This is very computationally intensive and makes these algorithms very time-consuming. The intuitive approach to reduce the computational effort would be to reduce the number of instances and features. However, such data sampling proves to be a difficult challenge if one does not want to accept a large degree of loss in accuracy. LightGBM tries to solve this challenge with two approaches. 
            GOSS reduces the number of data instances by dropping only instances with a small gradient randomly and always keeping those with a large gradient.
            EFB tries to reduce the number of features in the dataset. Here, features with a sparse feature space are efficiently and safely combined using algorithms based on the graph coloring problem \cite{Ke.2017}. 
            \paragraph{CatBoost,} created by Yandex, is one of the latest improvements for gradient-boosted tree algorithms. As the name suggests, CatBoots (categorical boost) introduces support for categorical features, without having to convert them to numerical features, unlike earlier algorithms. That being the biggest contribution, CatBoost also introduces a new schema to calculate the leaf values, which reduces overfitting \cite{Dorogush.24.10.2018}.

        \subsubsection{Neural Network} \label{subsec:backgroundNN}
\begin{figure}[h]
            \centering
            \includegraphics[scale=0.23]{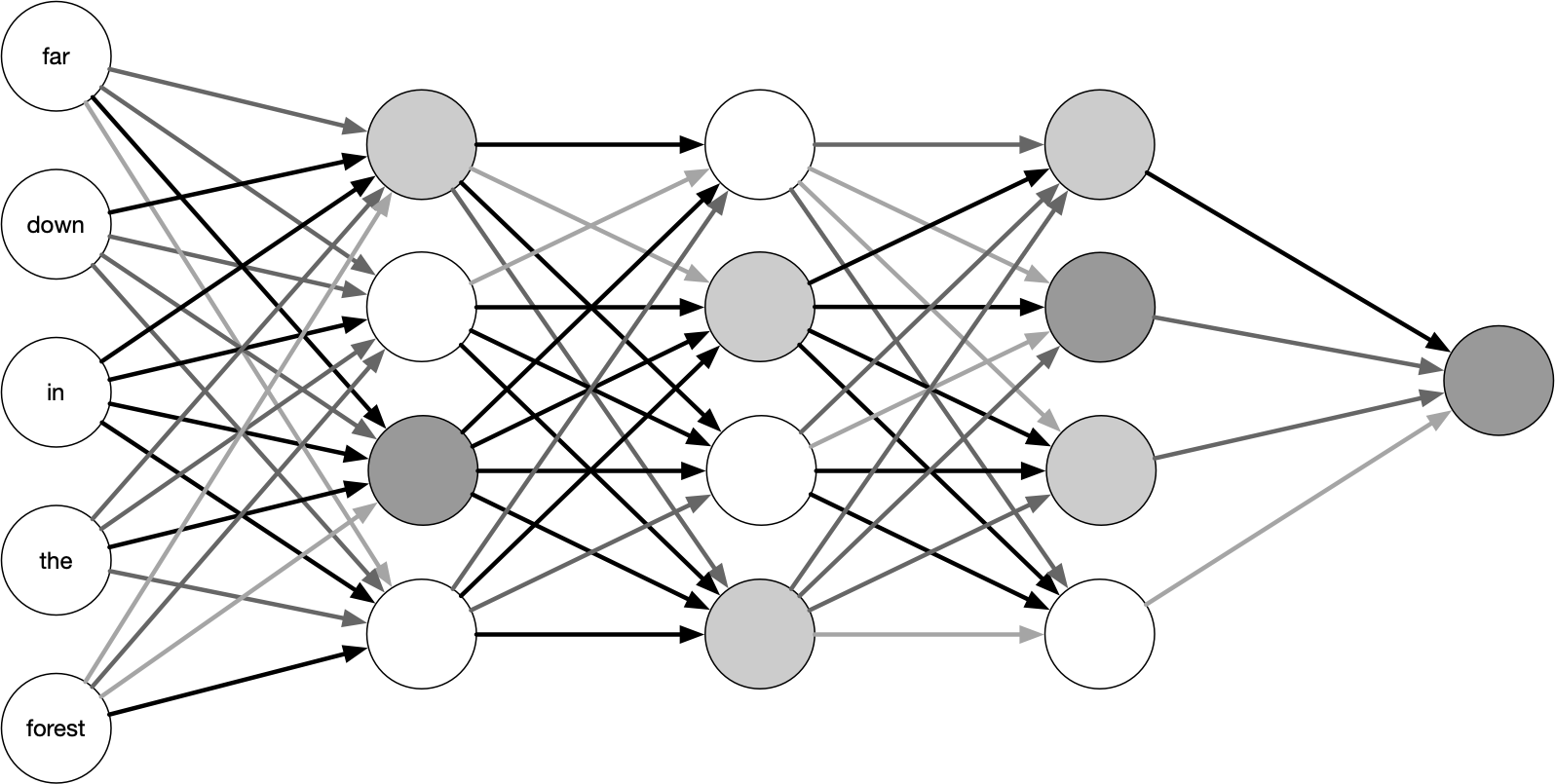}
            \caption{Visualisation of a FC-NN \cite[Figure~8.1]{Hvitfeldt.2022}}
            \label{fig:FCNN}
        \end{figure}

        A neural network is a network of computation nodes or artificial neurons. Each node has weights and a bias that can be applied to a given input, in form of mathematical operations. Each node can forward the input, once the weights and bias are applied. A node on its own is not very helpful in solving complex tasks, but having many nodes connected can enable the network to master complex challenges. While learning the weights and biases of each node are adjusted using backpropagation to fit the neural network to a dataset. Backpropagation is a supervised learning technique using gradient descent\footnote{\cite[Chapter~5]{Goodfellow.2016}}.
        Neural Networks come in various forms and shapes. For the regression and classification of tabular data, we choose a fully connected neural network. Fully connected neural networks are one of the simplest and most used neural networks. They consist of at least an input layer and an output layer while having an arbitrary number of hidden layers in between. The size of a layer describes how many neurons they have. The size of the input and output layers are dependent on the shape of the input data and on the desired shape of the output data. The size, as well as the number of the hidden layers, can be set to any size and are impacting the models performance heavily. The neurons inside of a layer are not connected with each other, but each neuron of a layer is connected with each neuron of the previous layer. Each connection between two neurons is associated with a weight and each neuron is associated with a bias \cite{Goodfellow.2016}.
        Figure \ref{fig:FCNN} shows a visual representation of a fully connected layer with three hidden layers, a hidden layer size of four, an input size of five and an output size of one.

        \subsubsection{Ensemble Techniques} \label{subsec:ensebmle-tec}
        In the following, we are describing the different ensemble techniques, which were used in this thesis to combine the output of different models (different algorithms but also just different parameter settings), in order to achieve a performance that a single model couldn't achieve. After an ensemble technique is in place, the combination of all used models can be seen as one big model. It is important to note that all models combined were trained to predict the same label.
            \paragraph{Ensemble Selection}comes into play once all models were trained on the same dataset. The ensemble algorithm is based on forward step wise selection\footnote{Draper, N. and Smith, H. (1981) Applied Regression Analysis, 2d Edition, New York: John Wiley \& Sons, Inc.} and starts with an empty ensemble set. To this set, the best-performing model is added. Then, the next best model is added to the set. If the combination of both models is better than just the first model, the second model is kept in the set. Otherwise, it is dropped. This process is repeated with all remaining models. After each step one checks if the ensemble of models inside the set can outperform the ensemble before that step. If a set cannot outperform the previous set, the model that got added is dropped. The models inside a set are combined by simply averaging their output \cite{Caruana.2004}. 
            In addition, weights are added to the selected models using a greedy search algorithm to adjust the influence on the resulting output, depending on their goodness \cite{fahmi2020implementation}.
            \paragraph{With Stacking}or Stacked Generalization several models from the same type (e.g. XGBoost) are fitted independently of each other on slightly different datasets. These models are called level-0 models. The predictions of these level-0 models are then added to the original input data to fit a single new model, which is named level-1 model. Through this the level-1 model benefits from the predictions of the level-0 models.  K-fold cross-validation\footnote{https://towardsdatascience.com/cross-validation-in-machine-learning-72924a69872f} methods are used to alter the datasets for the level-0 models and to make sure that the level-1 model is only trained with data that the level-0 models have not seen before \cite{Brownlee.2021}. 
            \paragraph{Ensemble Stacking}combines both Ensemble Selection and Stacking by running the Stacking algorithm on different model types first and then running Ensemble Selection with all level-0 models but also with different level-1 models \cite{PiotrPlonski.https:supervised.mljar.comfeaturesstackingensemble}.
        \subsubsection{Feature engineering} \label{subsec:backgroundFeature}
        Feature engineering is an umbrella term for any process that deals with the acquisition, aggregation, and manipulation of data, which is used as a subsequent step for training machine learning models \cite{HarshilPatel.}. The techniques we use to manipulate and extend the feature space we aggregate are described below.
            \paragraph{Golden Features} is a name for features that add great predictive power to a model. To generate those Golden Features, we generate unique pairs of all features from the original dataset. These unique pairs are then combined once with subtracting and once with divisions, generating a set of features. Each newly generated feature is then used to fit a Decision Tree. The Decision Tree is then used to calculate a loss for each new feature. The top 5\% with the best (lowest) loss are then carried over to the original dataset and used for training \cite{PiotrPlonski.19.07.2022}.
            \paragraph{Feature selection}is a process in which input features that have little or no useful influence on the predictions of a model are identified and removed. The goal is to remove unnecessary complexity and noise from the dataset \cite{JasonBrownlee.}.
            To find the features that have no predictive power, we introduced a new random feature as an additional input for the model that we want to improve. This random feature has a uniform distribution from 0 to 1.
            The model is then trained with the dataset containing the original data plus the random feature. 
            The permutation-based feature importance\footnote{https://scikit-learn.org/stable/modules/permutation\_importance.html} of all features is calculated for the trained model. This is a model inspection technique which can externally estimate the importance of a single feature for the predictions of a model \cite{scikitlearn.feature.}.
            If a feature has a lower relevance than the random feature it is dropped. 
            Once all irrelevant features have been removed from the dataset, the model is trained again \cite{PiotrPonski.Features.}.
        
        \subsubsection{Hyperparameter optimization} \label{subsec:backgroundHyperparameter}
        All the described machine learning models bring a set of hyperparameters with them. Finding the optimal hyperparameters can greatly improve the performance of a model but often feels like groping in the dark for researchers. A model which a researcher wishes to optimize has to be trained several times with slightly different hyperparameters till a satisfying performance is reached and even then the researcher can't be sure to have reached an optimal combination of hyperparameters. To enhance and automate this process, we use diffident hyperparameter optimization techniques, which are less work intensive for us but require more resources \cite{Akiba.07252019}. 
            \paragraph{Hill Climbing}does improve one hyperparameter per run. It improves this hyperparameter by selecting a default value for this hyperparameter, training the model, and then rerunning the training twice. Once with the hyperparameter default value increased and once with the hyperparameter default value decreased. For example, we want to improve the max\_depth parameter of a XGBoost model. The hyperparameter space of the max\_depth parameter is defined as a range of integers between one and nine. We select five as our default value and train the XGBoost model with max\_depth set to five. Once that is finished, we retrain the model twice. Once with max\_depth set to four and once max\_depth set to six. After the two runs are finished, we compare the results and use the max\_depth value which resulted in the best result as our new default value and all values below 6 are dropped from the hyperparameter space. This process can then be repeated with all parameters as long as there are unused values in the space. This can be very computationally expensive, depending on the model used \cite{selman2006hill, PiotrPlonski.03.08.2022}.
            \paragraph{Optuna}offers a define-by-run framework using relational sampling as well as an efficient pruning algorithm. With Optuna a model is trained several times with a different set of hyperparameters. Define-by-run methods assemble the search space of the hyperparameters dynamically and the user does not have to define the parameter space explicitly in advance. This comes with the downside that calculating correlations between hyperparameters are not possible like with older "define-and-run" methods. To still take advantage of correlations between hyperparameters runs Optuna trial-runs at the beginning and uses the knowledge gained in the trial runs for the actual define-by-run search. This way the hyperparameters used to train the models do not need to be defined specifically in advance but can still be selected based on their correlations.
            Furthermore, Optuna does use pruning to greatly improve the cost of its searches. Pruning can be seen as an early stopping mechanism. If a run does not meet predefined criteria and the improvements are insufficient, the run is stopped \cite{Akiba.07252019}.

% !TeX root = ../main.tex
% Add the above to each chapter to make compiling the PDF easier in some editors.

\chapter{Empirical analysis of share repurchases announcements}\label{chapter:Empirical analysis of share repurchases announcements}

\section{Hypotheses}
The hypotheses developed in this paper can be broken down into the following three parts. First, we hypothesize that automated detection of share buyback announcements in a stream of news is applicable to NLP techniques. 

The two other hypotheses relate to share buybacks themselves. In particular, we suppose that on average, share buybacks will generate short-term overperformance relative to the global market, but that these positive effects will fade and quickly become insignificant over time. The basis of this assumption is that share buybacks fundamentally entail a reduction in the number of outstanding shares. As in discussed the section \ref{sec:exbb}, this inevitably increases the share price. Once a share buyback has been priced in by the market, on average, no significant effects should be discernible.  

Our final hypothesis is that it is possible to use machine learning to automatically analyze share buybacks and the company executing them when they are announced in order to forecast future returns.
\section{Methodology}
    \subsection{Overview}
    In this section, we give an overview of the methodology applied in this thesis. 
    The applied steps can be divided into three sub-steps. The creation of a dataset, the statistical analysis of this dataset, and the fitting of machine learning models to this dataset.
    
    In order to be able to analyze share buybacks and automatically detect them in the future, we need a corresponding dataset on which the subsequent steps are also based.  For this, we performed a large internet crawl, from which we created a sub-dataset and labeled it by hand. We used this sub-dataset to develop two NLP algorithms for the detection of share buyback announcements. Using these two NLP techniques, we were able to filter the remaining share buyback announcements from the crawled data and are also able to detect future share buybacks quickly and automatically. 
   
    The next step was to statistically analyze this dataset of share buyback announcements in order to be able to make general statements about the performance of share buybacks. 
    
    Finally, also using the initially generated dataset of share buyback announcements, we trained 24 state-of-the-art machine learning models to forecast the performance and overperformance of a company after a share buyback announcement on 6 different time horizons.
    
\begin{figure}[t]
        \centering
        \includegraphics[scale=0.6]{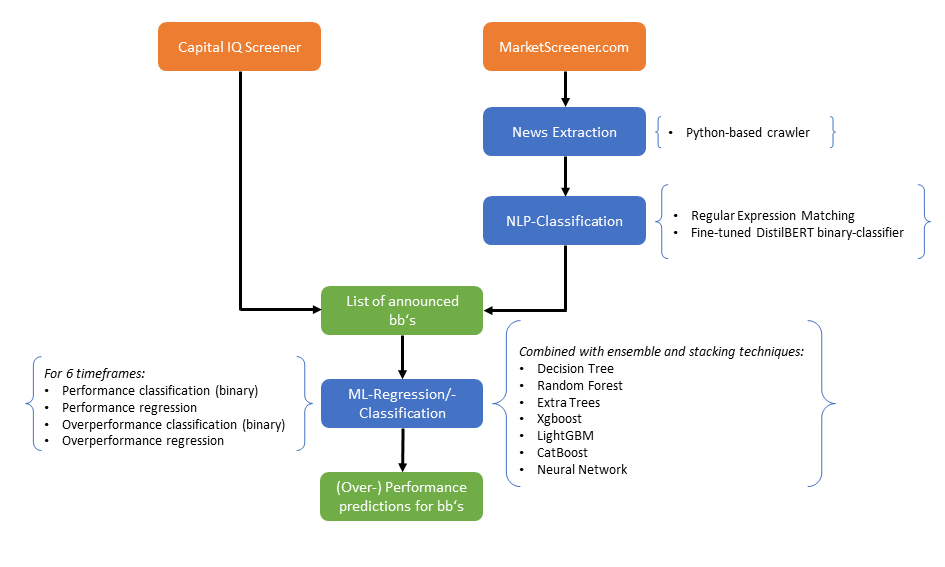}
        \caption{Flowchart of share buybacks pipeline}
        \label{fig:Pipeline}
    \end{figure}

    We were able to build the following pipeline for analyzing share buybacks using marketscreener.com, S\&P Capital IQ, and the models we developed.
    The pipeline built in this thesis to analyze share buybacks essentially consists of two tasks. The finding of share buyback announcements and the analysis of these buybacks as well as the corresponding company. Share buybacks that are fed into the pipeline via Capital IQ can be processed without pre-processing. But share buybacks that are fed into the pipeline via the news aggregator marketscreener.com must first be recognized as buybacks. These potential share buybacks will be classified as buybacks or non-buybacks using our developed two NLP-algorithms.
    Thereupon, each individual share buyback and its associated company can be analyzed by our 24 machine learning algorithms. For this purpose, a variety of models are used to solve not only classification but also regression tasks.
    The flowchart \ref{fig:Pipeline} is a visualization of the described pipeline.
    In the following, each step that leads to the final pipeline is described including the gathering of the dataset, the analysis of the said dataset, and the development of the various trained models.
    
    \subsection{Dataset creation}
    In this subsection, we will show which sources and techniques were used to generate a dataset, with the goal to get a representative overview of stock buybacks in various market environments, on which this thesis is based.
        \subsubsection{Data gathering}
        We used three different data sources to create the dataset, marketscreener.com, S\&P Capital IQ and gurufocus.com.
            \paragraph{The news aggregator marketscreener.com}combines a variety of different news sources, including Associated Press News, EQS Group, Reuters, Publicnow, Dow Jones Newswires, S\&P Capital IQ, and GlobalNewswire. The website allows the user to search among all news \cite{MarketScreener.}. We take advantage of this by first searching for news with share buyback references, then crawling the URLs of this news, and finally scraping the information from these news. The search terms used to find the news to crawl were "stock buyback", "share buyback", "stock repurchase", "share repurchase" and "equity buyback". As a result, we collected a total of 682,361 news links. These were then filtered before the scraper pulls more information about the news. Filtered were duplicate URLs, news that does not directly refer to a company (can be recognized by analyzing the URL), and news we can't associate a time and date to.  The scraper itself also filters companies for which marketscreener.com could not show us usable identifiers, like ISIN, WKN, or Ticker and Exchange. This reduced the amount of news by 51\%. The remaining 49\% of the news are then filtered according to their content. We used RegEx filters and a BERT model to distinguish share buyback announcements from all other news. A detailed description can be found in the following subsection.
            The news now classified as share buyback announcements were additionally filtered for obvious duplicates since multiple news agencies may report on the same share buyback and this may also be slightly delayed. Thus, all share buyback announcements were removed if a share buyback about the same company had already been announced within 30 days before.
            That left us with just under 7\% or 45,445 of the initial crawled links.

            % The flowchart \ref{fig:Pipeline-Data} gives an overview over the creation process of our dataset.
            
            % \begin{figure}[h]
            %     \centering
            %     \includegraphics[scale=0.6]{images/Buyback-Pipeline.drawio.pdf}
            %     \caption{Flowchart of share buybacks pipeline}
            %     \label{fig:Pipeline-Data}
            % \end{figure}
            
            \paragraph{The market intelligence platform S\&P Capital IQ}also provides data about share buyback announcements and we used it to expand our dataset with share buyback announcements that we missed but more importantly, to enhance our dataset with company-specific data at the time of the announcement, utilizing the S\&P Capital IQ Excel Plug-in\footnote{https://www.capitaliq.spglobal.com/apiservices/office-tools-service/download?lang=en-us\&platform=x64} \cite{S&PGlobalMarketIntelligence.}. This left us with 57,155 share buyback announcements.
            
            \paragraph{The financial news site gurufocus.com}was used by us to get data regarding the ownership of a company. Specifically, we are interested in insider and institutional ownership data, meaning how many shares of a company are held by insiders and how many by institutional investors \cite{gurufocus.}. While using  S\&P Capital IQ we found that the historical ownership data available is unusable and obviously wrong. We determined that this was due to a bug and reported it. Therefore, we have used a scraper to locally save the entire ownership history of all companies in our dataset, which we could find on gurufocus.com for free, and added the percentage and number of shares held by insiders and institutional investors at the time of announcing the share buyback to our dataset. Unfortunately, using the free version of gurufocus.com limits us to companies traded in the US.

        \subsubsection{Classification of share buyback announcements}
        To run meaningful descriptive statistics and to train machine learning models, we need a dataset of share buyback announcements as large as possible.
        To get to this dataset we first need to find a way to classify the large number of news items that could be share buyback announcements. 
        Our goal is to classify the large amount of financial news to generate a large dataset for later analysis.
        The reasoning behind this approach is to minimize manual work and still generate a very large dataset. Therefore, we label manually only a small sub-dataset. Using this labeled sub dataset, we create algorithms to automate this manual labeling and finally generate a much larger dataset using these algorithms. This large dataset can then be used to train machine learning models for more complex tasks.
        As mentioned in the subsection above, we need to find a way to distinguish share buyback related news from actual share buyback announcements. For this, we use two NLP techniques. One is a RegEx filter and the other is the classification model DistilBERT. Both techniques are described in the subsection \ref{subsec:textclass}.

            \paragraph{Training dataset}
            To train the model, as well as to create the RegEx-filter we need a dataset with labels to work from. Since there is no dataset of share buyback news yet, we must create one ourselves. We did this by randomly selecting 503 samples from the scraped news, loading them into a separate data frame, and labeling these 503 news by hand. Then we divided the dataset into a training and an evaluation dataset. The training dataset consists of 402 data point and the evaluation dataset consists of the remaining 101.
            \paragraph{A RegEx filter} was created using the dedicated Python library re.py. The approach was to classify the news based on the title. By manually adapting the RegEx filter to the training dataset we generated a simple but already very effective algorithm. The accuracy of the algorithm can be seen in table \ref{tab:bbann}. Algorithm \ref{alg:regex} shows in pseudo code how the algorithm is constructed.

            \begin{algorithm}
            \caption{Share buyback announcement RegEx-filter}\label{alg:regex}
            \begin{algorithmic}[1]
            \Procedure{isBuyback}{s}
            \State $\textit{patterns} \gets \text{\{set of patterns that indicate an announcement\}}$
            \State $\textit{antiPatterns} \gets \text{\{set of patterns that indicate against an announcement\}}$
            \State $\textit{phrases} \gets \text{\{set of phrases that indicate an announcement\}}$
            \State $\textit{antiPhrases} \gets \text{\{set of phrases that indicate against an announcement\}}$
            \State
            \If {$s \textit{ matches antiPatterns } \lor \textit{ substring of } s \textit{ in antiPhrases}$} 
                \State \Return false
            \EndIf
            \State
             \If {$s \textit{ matches patterns } \lor \textit{ substring of } s \textit{ in phrases}$} 
                \State \Return true
            \EndIf
            \State
            \State
            \Return false
            \EndProcedure
            \end{algorithmic}
            \end{algorithm}
            
            \paragraph{The model DistilBERT}was used as a second approach to this classification problem. We used a version of DisitlBERT, which was pre-trained on the English binary sentiment analysis dataset sst-2\footnote{https://huggingface.co/datasets/sst2} \cite{HuggingFace.}. The model was fitted to the training dataset using the optimizer AdamW \cite{PyTorch.}. The default settings were used. To compensate for the imbalanced train dataset we have adjusted the class weights because there are about 2.5 times as many non-buyback announcements in the train dataset as real announcements. This resulted in class weights of 1 (for non-buybacks) and 0.38 (for buybacks). The batch size used is 256 and the training is terminated after two consecutive epochs, without improvement of the evaluation loss. The selected loss is Cross Entropy Loss \cite{PyTorch.CrossEntropyLoss}. The training run started with a training loss of 0.65 and an evaluation loss of 0.47. The training was terminated after 12 epochs. The best snapshot of the model by evaluation loss was used as the final model. This resulted in a model with a training loss of 0.067 and an evaluation loss of 0.18. Due to the small dataset, no hyperparameter tuning was done, and overfitting is clearly applying.
            
            \paragraph{The performance}of the two approaches is very similar. The RegEx filter has zero false positives. DistilBERT unfortunately has two false positives. Most share buyback announcements are worded very similarly. The deep learning model DistilBERT uses embeddings to encode words in a way that machines can understand. This makes it possible to understand more complex contexts. Furthermore, the model is pre-trained on a large language dataset and can apply the learned knowledge to recognize stock buyback phrases, even though the model has not seen these phrases before. The same applies when other words or synonyms are used. This makes DistilBERT much more robust \cite{DiJin.2020}. 
            A RegEx filter, on the other hand, has a fixed set of patterns and phrases to which it falls back. If even the smallest deviations occur, the filter can no longer recognize the news as a share buyback. This ensures a very small number of false positives but makes the filter very inflexible. If you want to achieve a very large coverage with a regex filter, you need an almost infinite space of wording, because you have to cover any small deviations in formulations and wording \cite{PieterLuitjens.}.
            Although the used dataset is clearly too small for even the scaled-down model DistilBERT, it achieves acceptable performance. Thus, we are sure that with a larger dataset and hyperparameter tuning, human-level accuracy can be achieved.

            \begin{table}[h]
                \centering
                 \begin{tabular}{|c|c|c|c|}
                \hline
                    Method       &   \textbf{Acc} &   \textbf{FP}  &   \textbf{FN}  \\
                \hline
                    \textbf{RegEX}       &   0.88&   0   &   12  \\                    \textbf{DistilBert}  &   0.90&   2   &   8   \\
                \hline
                \end{tabular}
                \caption{Share buyback announcement classification results}
                \label{tab:bbann}
            \end{table}
            
            \paragraph{In conclusion}we were able to show in this section that a classification of share buyback announcements can be well automated with already simple methods and that also more complex NLP models can be well suited for this task. Despite a really small dataset, models can be trained with relatively high accuracy. Thus, we can detect future share buybacks immediately after they are announced. However, since we also have access to Capital IQ for historical share buyback announcements, we used this data to validate our share buyback announcements crawled from marketscreener.com and classified via RegEx and DistilBERT. In this way, we avoid contaminating our main dataset with few but still potentially significant false positives and keep noise at a minimum.
            
    \subsection{Descriptive statistics}  \label{subsec:statistics}
    In this chapter, we will describe the structure and composition of the dataset we generated, consisting of share repurchase announcements and enriched with company-related data. In addition, we will segment and analyze the dataset according to various aspects. The completeness of the data is dependent on the data available to us through Capital IQ. Not every data point is available to the same extent for every company in their database. To minimize noise, or missing historical data, we have only included companies in the analysis for which Capital IQ could provide us with at least a market cap. This left us with a total of 55,050 share buyback announcements.
        % \subsubsection{Definition of variables}
        % \subsubsection{Dependent variables}
        \subsubsection{Distribution}
\begin{figure}[p]
                \centering
                    \includegraphics[scale=0.33]{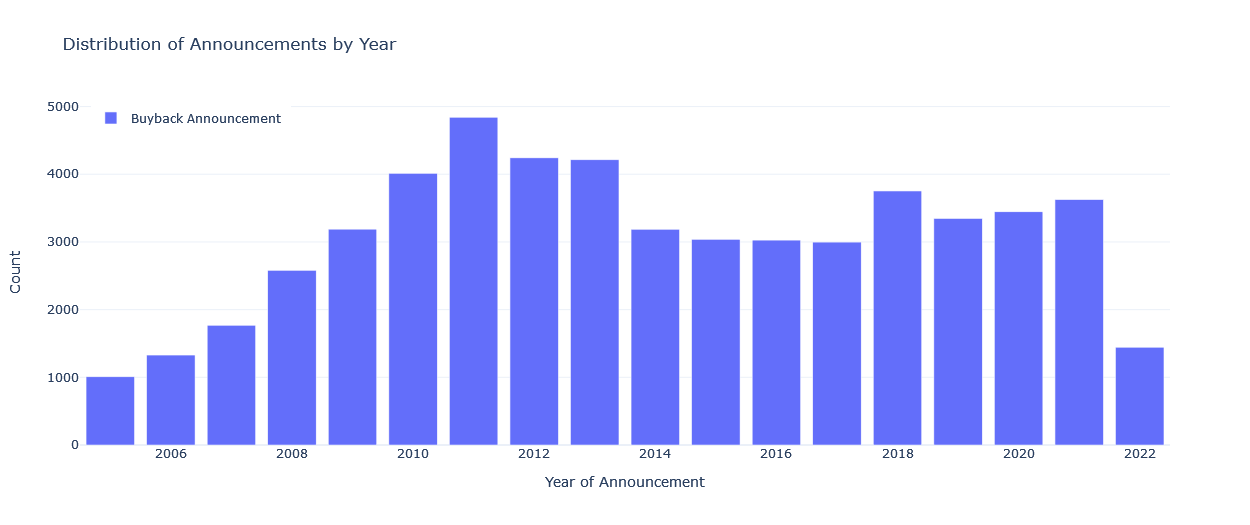}
                    \caption{Distribution of Announcements by Year}
                    \label{fig:distYear}
            \end{figure}

\begin{figure}[p]
                \centering
                    \includegraphics[scale=1]{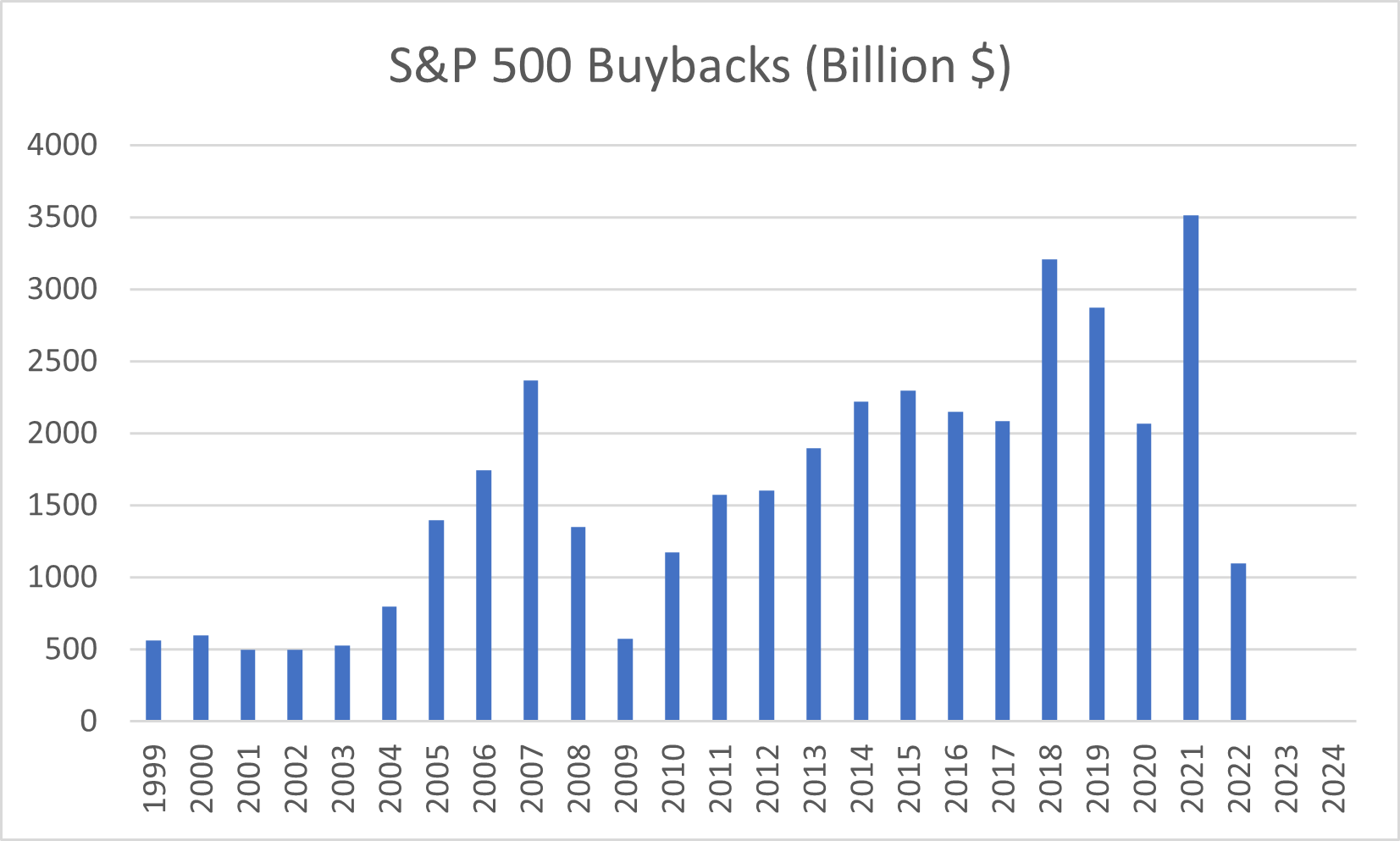}
                    \caption{S\&P 500 Buyback Volume in Billion Dollars \cite{Dr.EdwardYardeniJoeAbbottMaliQuintana.20220812}}
                    \label{fig:vol_bb_sp}
            \end{figure}

        In this subsection, we show the distribution of different features of the dataset to give the reader a rough overview of the composition of the dataset we generated.
        \paragraph{The distribution by year of announcement}can be seen on the chart \ref{fig:distYear}. A sharp increase can be seen at the beginning of the observed time period namely 2005 to 2011. After the peak in 2011, share buyback announcements initially fall and then stabilize. The share buyback announcements from 2022 only go until April and are therefore not complete. But at least for the US economy, 2022 is a year with already strongly increasing numbers of share buybacks and appears to be developing into a record-breaking year \cite{BobPisani.}.
        The chart \ref{fig:vol_bb_sp} shows the amount of money that was invested in share buybacks by companies out of the S\&P 500. From 2004 to 2007, there was a sharp increase in money invested in share buybacks from companies in the S\&P 500. This is also reflected in our dataset. The sharp drop in volume as seen in the financial crisis of 2008 or the Covid-19 crisis is not reflected in our dataset. This indicates that in crises, share buybacks are announced in equal measure, but are hardly if at all, executed to the same extent as in bull markets.

            
\begin{figure}[h]
                \centering
                    \includegraphics[scale=0.4]{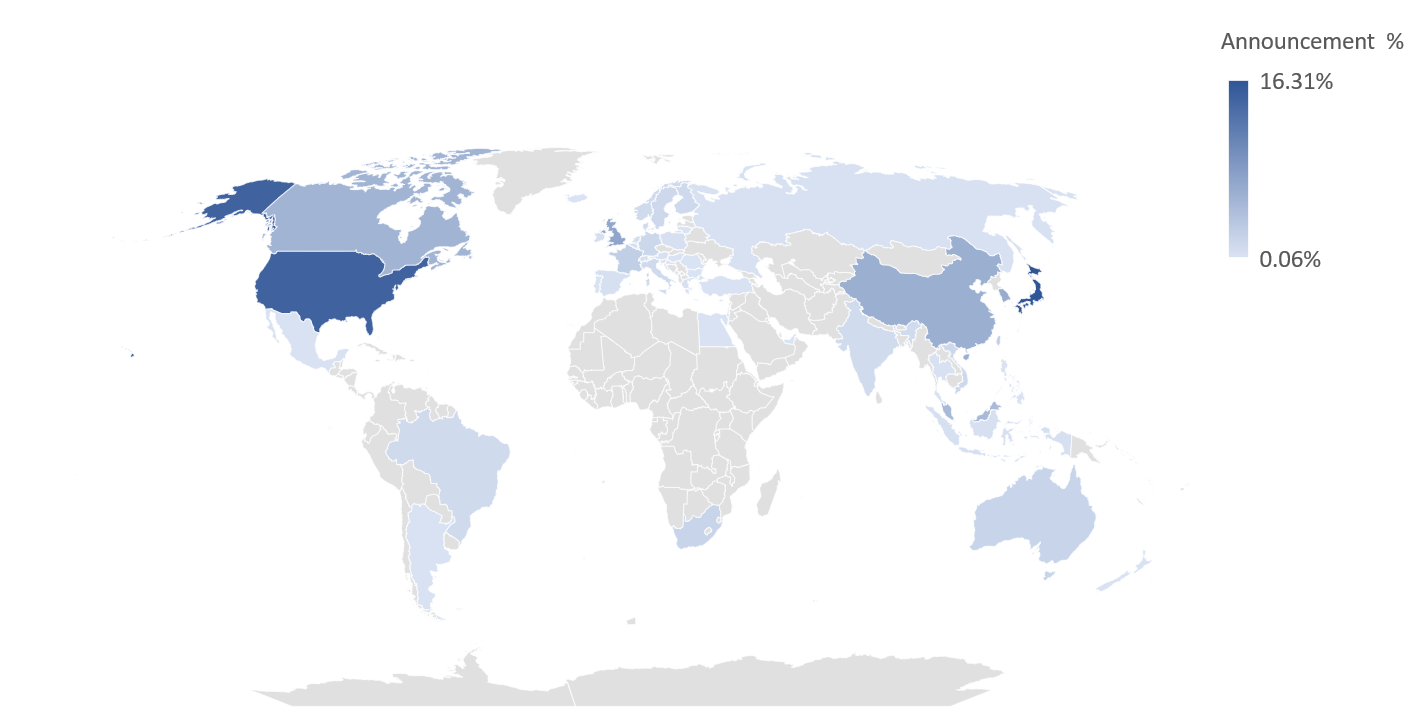}
                    \caption{Distribution of Announcements by Country}
                    \label{fig:distCountry}
            \end{figure}

            % \paragraph{Distribution month of announcement}
            \paragraph{The distribution of countries}can be seen on the chart \ref{fig:distCountry}. The chart shows the distribution of share buyback announcements by country. Most share buyback announcements come from Japan with 16.3\%, closely followed by the USA with 14.7\%. The following countries are Hong Kong, the United Kingdom, and South Korea. The countries plotted on chart \ref{fig:distCountry} are the top 59 countries by number of announcements. After the 59 countries, the number of announcements per country becomes very low. For example, Morocco is in 60th place with 0.04\% of all announcements, which corresponds to 23 announcements since 2005. The developed countries, as well as most of the East Asian countries, are represented in the dataset. Africa and South America, as well as West Asia, are hardly or not at all to be found in the dataset. In total there are 92 countries in the dataset. The top ten countries and their figures are shown in appendix table \ref{tab:country}.
            Noteworthy is the prominence of Japanese companies in the dataset. Share buybacks were prohibited in Japan until 2001. Under the administration of Shinzō Abe, the number of share buybacks in Japan has risen rapidly.   However, these share buybacks are currently coming under increasing criticism, as some companies are using them exclusively to rectify share value losses and are even going into debt for these buybacks which can risk the health of the company and do not benefit their employees \cite{GearoidReidy., MinJeongLeeKeikoTabata.}.
            
            \paragraph{The distribution of primary industries}is much more evenly\footnote{The top ten industries by number of announcements in the appendix \ref{tab:primany}}. In total are 159 different primary industries included in the dataset. The industry with the most announcements is Industrial Machinery, with only 3.51\% of announcements.
            
            \paragraph{We also analyze the distribution market capitalization}, because market capitalization is one of the most fundamental financial metrics of a publicly traded company.  It reflects the market valuation of the shares available to the public and is calculated using the current share price times the outstanding shares. Companies can be divided into different classes depending on the size of their market cap. Typically, companies are divided into small-, mid-and large-cap. Since companies with a market cap of less than \$300 million are already too small for small-cap, we have included two additional classes for very small companies. This gives us a much better picture of the company composition since we have a large number of very small companies in our dataset. In addition, we have also introduced the mega-cap class to separate the significantly largest companies from the large-caps. In total, we have the classes nano-, micro-, small-, mid-, large- and mega-cap, which are defined as described in the table \ref{tab:marketcapRanges}.

            \begin{table}[h]
            \centering
            \begin{tabular}{|c|c|c|}
                \hline
            Class & Bottom (>=) & Top (<) \\
                \hline
            Nano  & \$0 M                    & \$50 M            \\
            Mirco & \$50 M                   & \$300 M           \\
            Small & \$300 M                  & \$2 B             \\
            Mid   & \$2 B                    & \$10 B            \\
            Large & \$10 B                   & \$200 B           \\
            Mega  & \$200 B                  & inf               \\
                \hline
            \end{tabular}
            \caption{Market Capitalization Ranges}
            \label{tab:marketcapRanges}
            \end{table}
            
\begin{figure}[h]
                \centering
                    \includegraphics[scale=0.2]{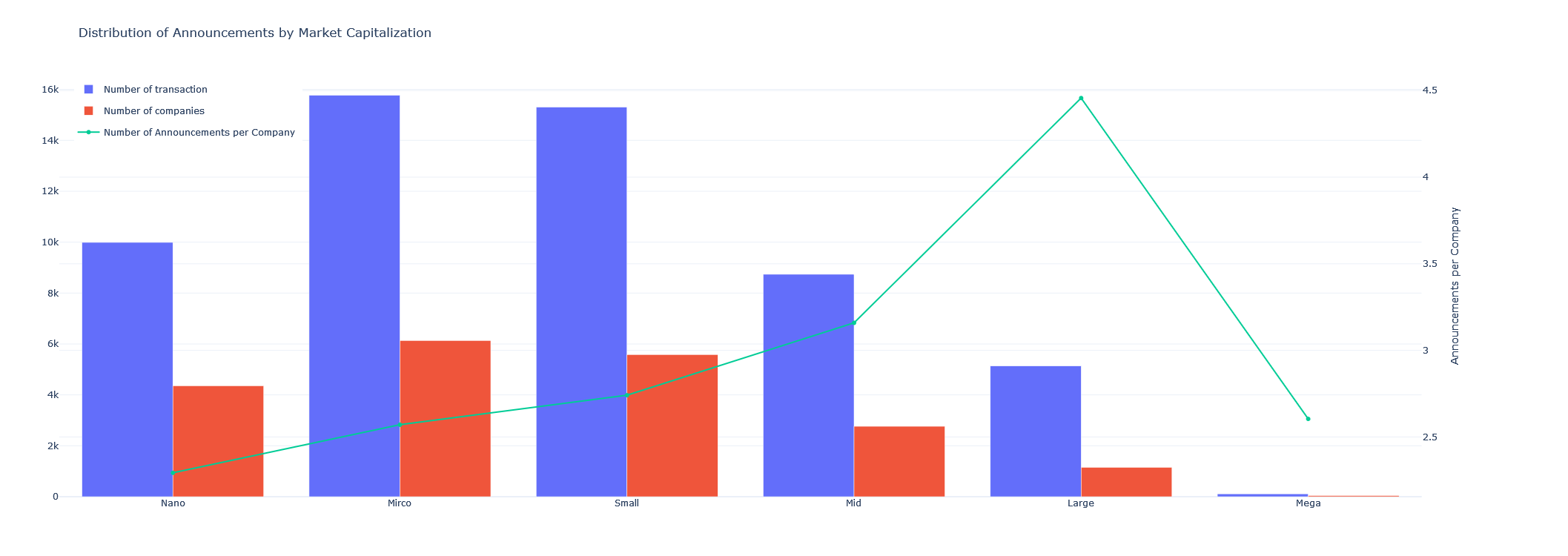}
                    \caption{Distribution of Announcements by Market Capitalization}
                    \label{fig:distMarketCap}
            \end{figure}

            Bar chart \ref{fig:distMarketCap} shows the distribution of companies in our dataset, divided into the market cap classes just defined. 
            Most announcements come from small-cap companies with 15,300 announcements. However, most companies are in the micro-cap class with 6,100 companies. It is interesting to see that the number of announcements a company makes increases with the market cap. 
            % One reason for this may of course be that larger companies are on the market longer and therefore have the opportunity to carry out more share buybacks. (TODO: Source for last statement plus small calculation why this is not enough) 
            The number of share buyback announcements per company is surprisingly falling for mega-caps. Unfortunately, the mega-cap class is not significantly large. It only consists of 43 companies with a total of 112 announcements\footnote{More detailed description of the distribution by market cap in appendix \ref{tab:marketCapDist}.}.
            
        \subsubsection{Performance and overperformance of share buybacks} \label{subsec:perfAndOver}
        In this subsection, we discuss the price performance of the companies included in the dataset after the announcement of a share buyback. 
         The observation periods we chose are one week, one month, six months, one year, two years, and five years.
        The main focus of this thesis is on the overperformance compared to the MSCI World, however, we will partially discuss the general performance, meaning the performance of the stocks themselves, as well. 
        In the following paragraphs, we will first discuss the performance and overperformance of the companies included in the dataset over the defined periods.
        Then we will break down the overperformance according to different criteria and examine them in more depth. Specifically, we will look at overperformance in terms of market capitalization, as well as the differences between share buybacks in a bull market and in a bear market.

            \paragraph{A general definition and explanation of the charts and figures shown}is provided in this paragraph to avoid potential misinterpretation.
            In general, we base the price performance on the last daily close price seen from the time of the announcement and compare this with the daily close price on the day of the announcement with the respective period added on top.
            
            It is important to mention that the overperformance is given in percentage points (pp) and not in percent. So if we discuss a stock performance of 0.15pp and a stock overperformance of 0.05pp in the same period, it means that the stock has gone up by 15\%, but the MSCI World has only gone up by 10\%, resulting in an overperformance of 5\% or 0.05pp.
            
            Also worth mentioning is that unless otherwise stated, all returns are not annualized. We have deliberately chosen not to do so, as our period under review is initiated by a single event and the stock performance is measured from that event. We are not particularly interested in the average annual return of companies that carry out share buybacks, but rather in how the returns changes, and this in comparison with very short but also longer periods of time.
            
            You will also notice that the count-value, the number of samples on which the figures shown are based, becomes smaller with longer time frames. This is simply due to the fact that for the time frame of for example 5 years, only samples that are older than 5 years can be used. This significantly reduces the number of usable samples.
            
            On some graphs, you will notice that we denote the removal of outliers with "outliers removed (min: 1\%-, max: 99\%-percentile)". This means that on the graph we display the top 99\%-percentile as the maximum value and the 1\%-percentile as the minimum value. This greatly improves the readability of the graphs.
            
            Furthermore, to avoid confusion, we want to give a short explanation for negative returns smaller than 1. You will find in some box plots a minimal overperformance of for example -1.53, or -153\%. The reason behind this is not erroneous data, but that in this particular case the reference index has risen by, for example, 60\% in 5 years, but the stock under consideration has fallen by 93\%. This results in a delta and thus in underperformance of -153\%.
            
\begin{figure}[h]
                \centering
                \includegraphics[scale=0.36]{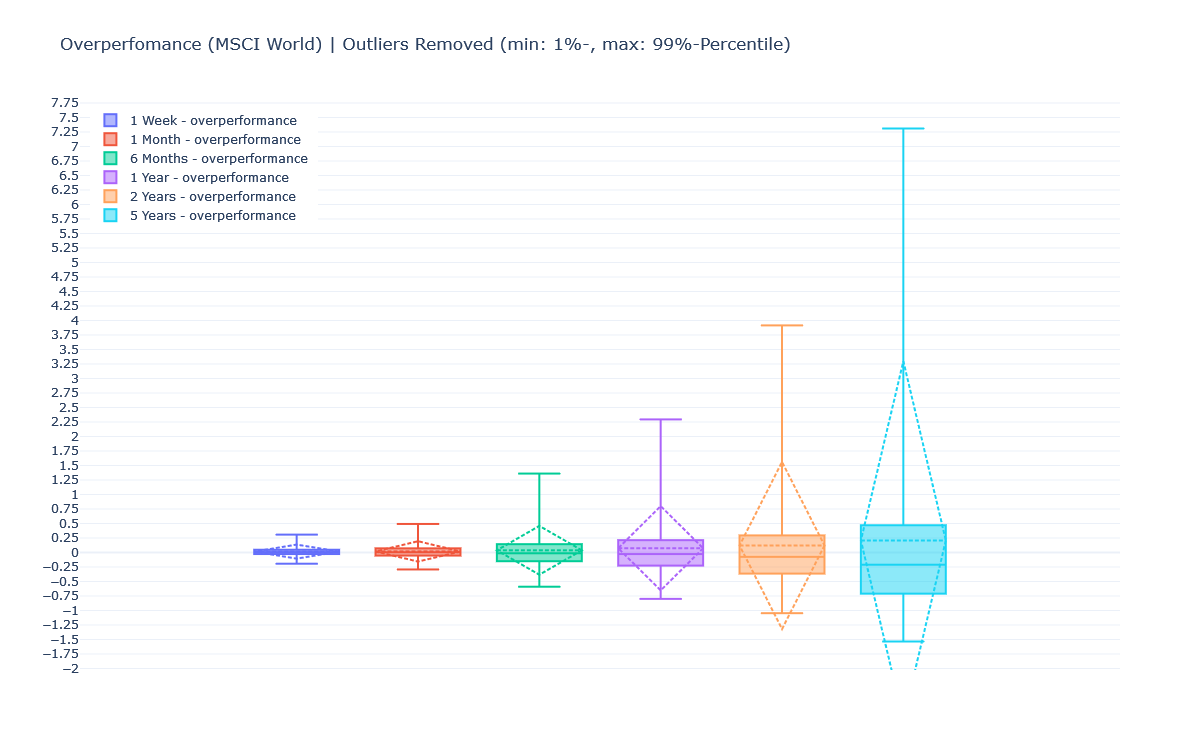}
                \caption{Box-Plot of Overperformance After Announcement}
                \label{fig:overOverall}
            \end{figure}

            \paragraph{The overall performance and overperformance}is shown respectively in the two tables \ref{tab:perfOverall} and \ref{tab:overOverall}. The overperformance is also shown as a box-plot in chart \ref{fig:overOverall}. The performance and overperformance are almost identical in short time frames. For the period of one month, the two values differ by just 0.0046 percentage points. The reason for this is the benchmark index with which the overperformance is calculated. The MSCI World had a compound annual growth rate (CAGR) of 9.38\% from January 2005 to April 2022 \cite{.20220818}. This corresponds to a monthly rate of 0.75\% or 0.0075 percentage points, which is somewhat inline with the 0.0046 difference.

            \begin{table}[p]
            \centering
            \begin{tabular}{|c|c|c|c|c|c|c|}
            \hline
            \textbf{time frame} &
              \textbf{Count} &
              \textbf{Mean} &
              \textbf{\begin{tabular}[c]{@{}c@{}}Standard\\ Deviation\end{tabular}} &
              \textbf{\begin{tabular}[c]{@{}c@{}}25\%\\ Percentile\end{tabular}} &
              \textbf{Median} &
              \textbf{\begin{tabular}[c]{@{}c@{}}75\%\\ Percentile\end{tabular}} \\ \hline
            \textbf{1 Week}   & 54,945 & 0.0165 & 0.1239 & -0.0266 & 0.0076 & 0.0497 \\ 
            \textbf{1 Month}  & 54,760 & 0.0238 & 0.1843 & -0.0518 & 0.0100 & 0.0806 \\ 
            \textbf{6 Months} & 52,822 & 0.0797 & 0.4524 & -0.1252 & 0.0290 & 0.2052 \\ 
            \textbf{1 Year}   & 50,974 & 0.1602 & 0.7630 & -0.1738 & 0.0519 & 0.3243 \\ 
            \textbf{2 Years}  & 48,186 & 0.2874 & 1.4720 & -0.2421 & 0.0685 & 0.4845 \\ 
            \textbf{5 Years}  & 37,276 & 0.6091 & 3.1105 & -0.3394 & 0.1453 & 0.8668 \\ \hline
            \end{tabular}
            \caption{Table of Performance After Announcement}
            \label{tab:perfOverall}
            \end{table}
            
            \begin{table}[p]
            \centering
            \begin{tabular}{|c|c|c|c|c|c|c|}
            \hline
            \textbf{Time frame} &
              \textbf{Count} &
              \textbf{Mean} &
              \textbf{\begin{tabular}[c]{@{}c@{}}Standard\\ Deviation\end{tabular}} &
              \textbf{\begin{tabular}[c]{@{}c@{}}25\%\\ Percentile\end{tabular}} &
              \textbf{Median} &
              \textbf{\begin{tabular}[c]{@{}c@{}}75\%\\ Percentile\end{tabular}} \\ \hline
            \textbf{1 Week}   & 54,945 & 0.0159 & 0.1211 & -0.0266 & 0.0070  & 0.0474 \\
            \textbf{1 Month}  & 54,760 & 0.0192 & 0.1753 & -0.0531 & 0.0046  & 0.0716 \\ 
            \textbf{6 Months} & 52,822 & 0.0382 & 0.4211 & -0.1503 & -0.0134 & 0.1436 \\ 
            \textbf{1 Year}   & 50,974 & 0.0737 & 0.7261 & -0.2264 & -0.0272 & 0.2127 \\ 
            \textbf{2 Years}  & 48,186 & 0.1214 & 1.4407 & -0.3639 & -0.0751 & 0.2945 \\ 
            \textbf{5 Years}  & 37,276 & 0.2078 & 3.0906 & -0.7098 & -0.2108 & 0.4698 \\ 
            \hline
            \end{tabular}
            \caption{Table of Overperformance After Announcement}
            \label{tab:overOverall}
            \end{table}
            
            Throughout all time frames, the average overperformance is steadily increasing. Starting with 1.59\% after one week and ending with an increase of 20.78\% after 5 years. However, this increase is slowing down over time. Therefore, we annualized the returns. For periods of less than one year, we have annualized the average values. For periods longer than one year, we annualized each sample and took the average. For short time periods, especially for one week period, sometimes enormously high annual returns have been calculated, which overshadowed any sense because, so few samples completely distort the average values. If we look at the annualized average and median values in Table \ref{tab:overallAnn}, we see that they are falling steeply. The initial surge at the announcement of the share buyback is several times greater than any other change. After that, there is a rapid fall and after one month a continuous decline. The biggest influence has the share buyback at its announcement and the effects decreased from then on until the 5-year return is almost completely aligned with the market average. From a market efficiency perspective, it can be assumed that in the first week the information about the share buyback announcements is processed and priced in. Thereafter the effects of the announcement itself only contribute to a small extent to the performance of the share.
            
            \begin{table}[h]
            \centering
            \begin{tabular}{|c|c|c|c|c|c|}
            \hline
            \textbf{Time frame} &
              \textbf{Count} &
              \textbf{\begin{tabular}[c]{@{}c@{}}Mean \\ Performance\end{tabular}} &
              \textbf{\begin{tabular}[c]{@{}c@{}}Median \\ Performance\end{tabular}} &
              \textbf{\begin{tabular}[c]{@{}c@{}}Mean \\ Over-\\ performance\end{tabular}} &
              \textbf{\begin{tabular}[c]{@{}c@{}}Median \\ Over-\\ performance\end{tabular}} \\ \hline
            \textbf{1 Week}   & 54945 & 1.3437 & 0.4851 & 1.2664 & 0.4409  \\ 
            \textbf{1 Month}  & 54760 & 0.3268 & 0.1271 & 0.2562 & 0.0562  \\ 
            \textbf{6 Months} & 52822 & 0.1657 & 0.0589 & 0.0779 & -0.0267 \\ 
            \textbf{1 Year}   & 50974 & 0.1602 & 0.0519 & 0.0737 & -0.0272 \\ 
            \textbf{2 Years}  & 48186 & 0.4841 & 0.4382 & 0.0265 & -0.0369 \\ 
            \textbf{5 Years}  & 37276 & 0.1829 & 0.1649 & 0.0014 & -0.0380 \\ \hline
            \end{tabular}
            \caption{Table of Performance After Announcement Annualized}
            \label{tab:overallAnn}
            \end{table}
            
            As already mentioned, on average, companies that execute share buybacks clearly outperform the market. Upon closer examination, it becomes apparent that such a general statement would be a fallacy. A detailed look at the box-plot \ref{fig:overOverall} supports the doubt. Even if the average of the overperformance is consistently positive, this does not necessarily argue in favor of share buybacks. If we look at the median, we see that it mirrors the average almost along the 0-y line.
            Up to the period of one month, the median is still positive, but from then on it drops into the negative and continues to drop until it reaches an underperformance of 21.07\% at the five-year time frame.
            Furthermore, the standard deviation is very high, so there are large differences in the performance after a share buyback. Therefore, it is apparent that a minority of shares perform very well after a share buyback and pull the entire average up with them. 
            In addition, these very well-performing companies bring a theoretically endless upside with them, while the very poorly performing companies only have a limited downside, which also pushes the average further upwards. The maximum loss of a company can be 100\%, but there are no limits on the upside.
            
            In addition, the weighting of the companies in the MSCI World depends on the market cap. Therefore, mainly large companies are represented in this index. However, our dataset consists to a large extent of small companies. Smaller companies have on average a higher return, but also bring a higher risk with them \cite{bares2011small}. Therefore, part of the seen overperformance is also due to the risk-premium that small companies bring with them. Furthermore, small companies also bring a factor-premium\footnote{https://www.invesco.com/ch/en/insights/introduction-to-factor-investing.html} with them, which again has a small positive influence on average \cite{Kommer.2018}. The influence of market capitalization is discussed in more detail in the following paragraphs.

\begin{figure}[p]
                \centering
                \includegraphics[scale=0.55]{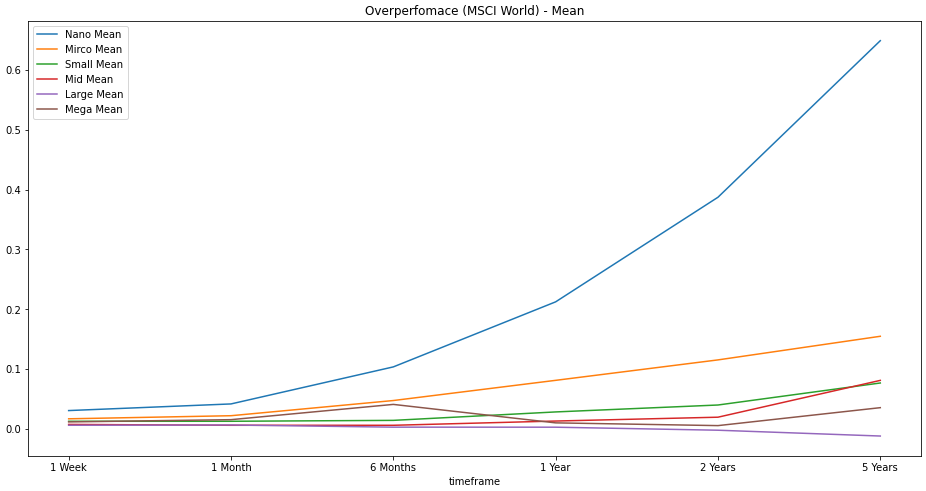}
                \caption{Line-Plot of Mean-Overperformance by Market Capitalization}
                \label{fig:marketcap_mean}
            \end{figure}

\begin{figure}[p]
                \centering
                \includegraphics[scale=0.55]{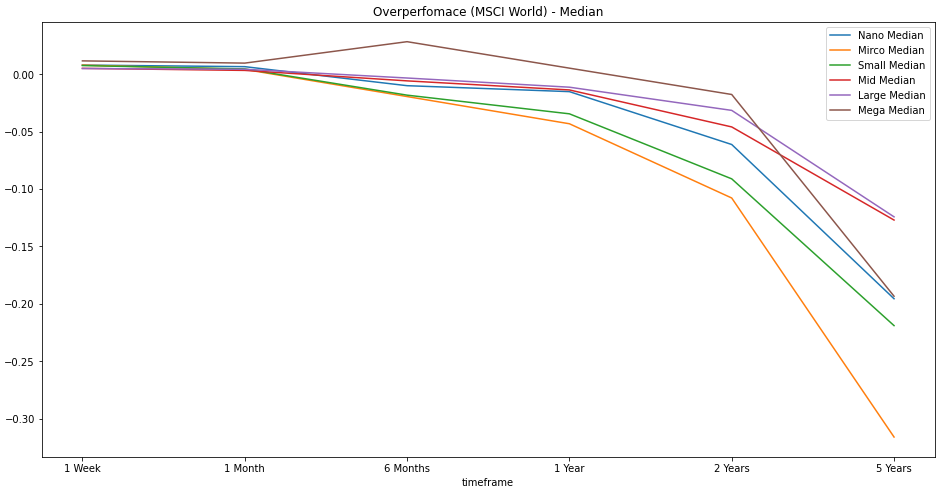}
                \caption{Line-Plot of Median-Overperformance by Market Capitalization}
                \label{fig:marketcap_median}
            \end{figure}

            \paragraph{To find differences regarding the market capitalization}we split all samples of our dataset into the same classes that were previously described in the table \ref{tab:marketcapRanges}.
            The graph \ref{fig:marketcap_mean} shows the time trend of the average excess return of the six market cap classes and the graph \ref{fig:marketcap_median} the trend of the median.
            As expected, the smaller market caps perform better, with the nano-caps significantly outperforming all others, with an annual excess return of 21.25\%. Mid-caps are able to catch up with small-caps over the 5-year period. From the second year on, large-caps are already negative on average. After five years, the large-caps have an average underperformance of 12.42\%. The mega-caps actually outperform the large-caps, but the sample size of the mega-caps is too small to make a meaningful statement.

            Looking at the median, all cap sizes show a consistent picture. As in the previous overall analysis, the median decreases sharply over time. After six months, all categories are already in the negative, with the exception of the mega-caps, which will not be discussed further for already mentioned reasons.  However, the large-caps underperform on median by only 3.14\% and are after five years together with the mid-caps at about -12\%, in comparison with the -21.07\% overall market caps.
            Additionally, it is noteworthy that over all time periods, the nano-caps have a smaller underperformance than the micro- and small-caps. This could speak for a stronger factor premium of the nano-caps. This further supports our thesis that only a few companies contribute to the overall average overperformance.

\begin{figure}[p]
                \centering
                \includegraphics[scale=0.55]{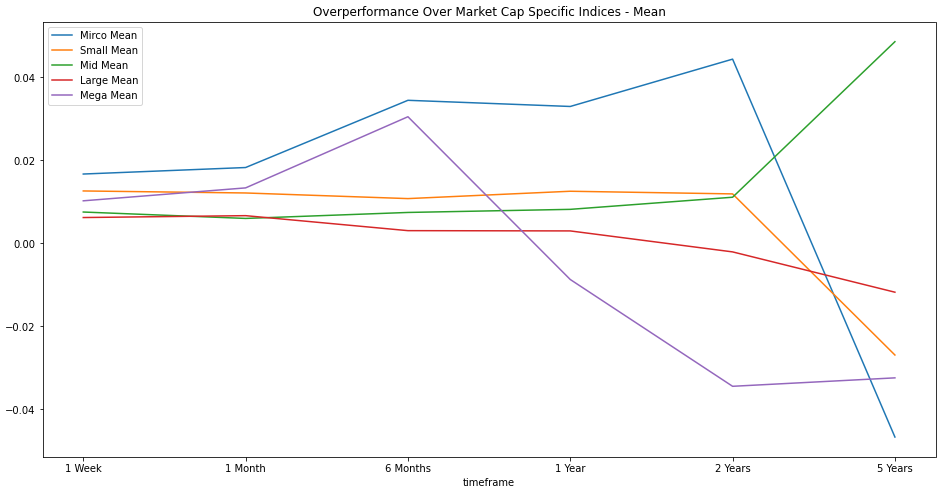}
                \caption{Line-Plot of Size Specific Mean-Overperformance by Market Capitalization}
                \label{fig:marketcap_mean_specific}
            \end{figure}

\begin{figure}[p]
                \centering
                \includegraphics[scale=0.55]{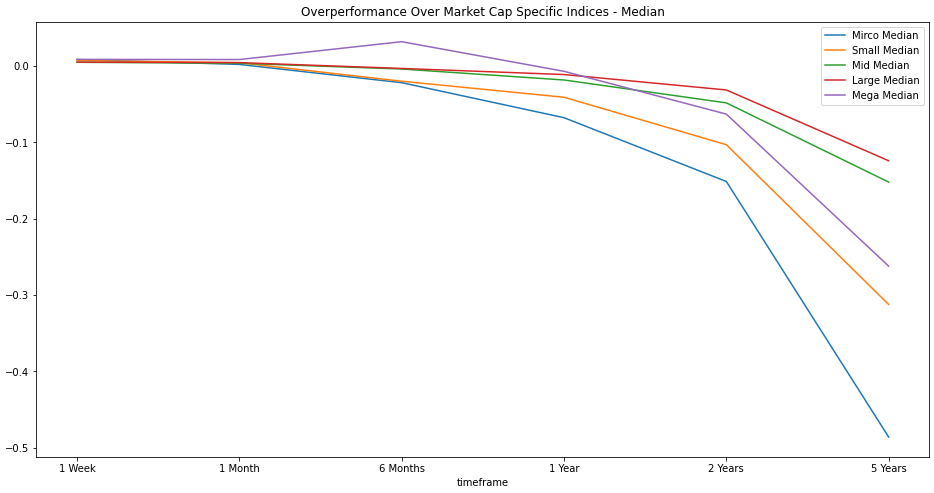}
                \caption{Line-Plot of Size Specific Median-Overperformance by Market Capitalization}
                \label{fig:marketcap_median_specific}
            \end{figure}

            \paragraph{To address market capitalization specific overperformance}, we compared each market cap class individually with a fitting index. Nano-caps are unfortunately omitted from the following. The smallest index available to us is the Russell Micro-cap Index, which tracks the US micro-cap equity market \cite{FTSERussell.}. We use it as a benchmark for the micro-caps from our dataset.  
            For the small caps, we use the MSCI World Small Cap Index, which tracks small caps from 23 developed countries \cite{MSCI.}.
            The MSCI world mid-cap index covers the mid-caps of the same 23 countries \cite{MSCI.mid}.
            For the large-caps, we use the MSCI World.
            Out of curiosity, we compared the mega-caps with the iShares Dow Jones Global Titans 50, which consists of the 50 largest companies on the major stock exchanges, even though we don't expect any meaningful learnings \cite{iShares.}. 
            To make a direct comparison, we have plotted the same two charts as in the previous paragraph but using the individually chosen indices to calculate the overperformance. In plot \ref{fig:marketcap_mean_specific} and \ref{fig:marketcap_median_specific}. The results are most certainly interesting. The plot with the average overperformance no longer shows a clear picture. The size of the returns only exceeds the 5\% mark over the five-year period and once over the 2-two-year period, and then only barely. Across all market cap classes, it can be said that no overperformance is discernible, except for the micro- and small-caps in the first week of just under 1-2\%, which extends slightly into the future. The movements remain very small even after long periods of time and we do not see a trend here that supports a fundamental overperformance.

            Interestingly, the graph with the median still shows a clear picture. 
            The median is consistently negative, decreasing over time. However, the average was still neutral, which again speaks for the fact that a few companies are performing better than the average, which may no longer generate excess returns, but still are pushing the average up. 
            % However, it is debatable whether share buybacks have any influence at all or whether most companies generally underperform their benchmark index and only a few drive the index up \cite{BryanTing.}.

\begin{figure}[p]
                \centering
                \includegraphics[scale=0.55]{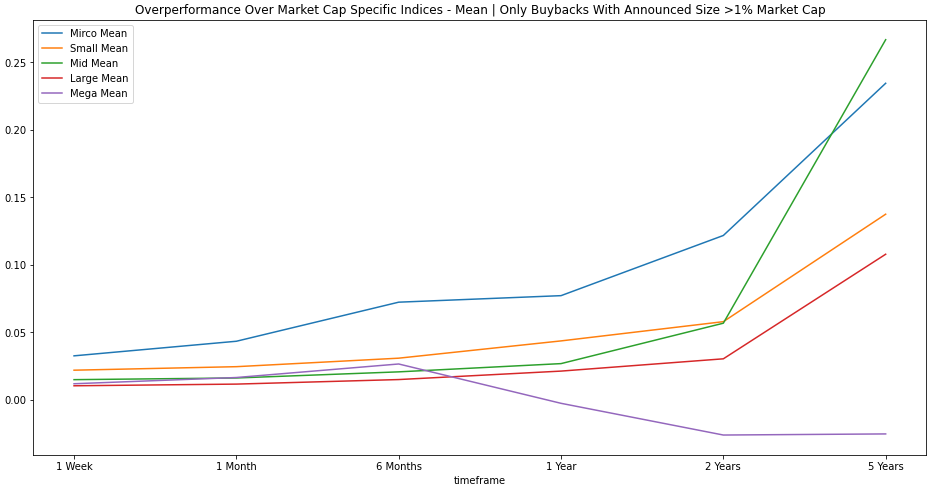}
                \caption{Line-Plot of Size Specific Mean-Overperformance by Market Capitalization with Announced Transaction Size >1\%}
                \label{fig:marketcap_mean_specific_big_transactions}
            \end{figure}

            \paragraph{The announced transaction size}of the buyback should be at least in the short-term one of the most important factors to evaluate the effects of a buyback since this measure indicates how many shares will be bought back and thus by how many the number of outstanding shares get decreased like already explained in section \ref{sec:exbb}.
            Simply put, the greater the volume of the buyback the greater the initial price bounce. 
            Therefore, we have kept only the data points that have a transaction size greater than 1\% of their own market cap for the following charts. This has obviously drastically reduced the number of data points, but the distribution across the different market-cap classes has not changed significantly. The only thing worth mentioning is that the mode is now small-caps. The exact distribution can be seen in the attached graph \ref{fig:dist_marketcap_big_transactions}.  
            What this surprisingly shows us is that despite using the individual comparison indexes, we can still see a mean overperformance, plotted on \ref{fig:marketcap_mean_specific_big_transactions}. 
            Across all market caps, share prices develop positively over time and generate an excess return, the smaller the higher the average return. However, the mid-caps manage to surpass all others at the five-year window.
            
\begin{figure}[p]
                \centering
                \includegraphics[scale=0.55]{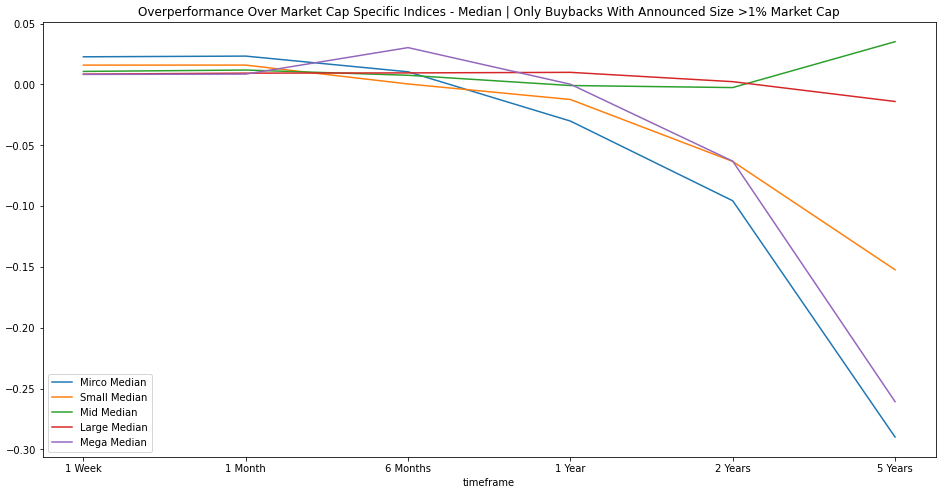}
                \caption{Line-Plot of Size Specific Median-Overperformance by Market Capitalization with Announced Transaction Size >1\%}
                \label{fig:marketcap_median_specific_big_transactions}
            \end{figure}

            The median has also improved significantly across all time frames and classes. Up to the period of six months, the median is positive or only slightly negative for all classes, more or less neutral overall. This can be seen on plot \ref{fig:marketcap_median_specific_big_transactions}.
            Over longer periods, the median remains negative for most, but less strongly, and both mid- and large-caps remain positive or only slightly negative. 
            
            This shows the transaction size as a good indicator to interpret the impact of a share buyback. However, most share prices are still declining in the long term compared to the market, even with large transaction sizes, and it is still necessary to find out how to identify the minority of very good-performing companies.
            
\begin{figure}[h]
                \centering
                \includegraphics[scale=0.3]{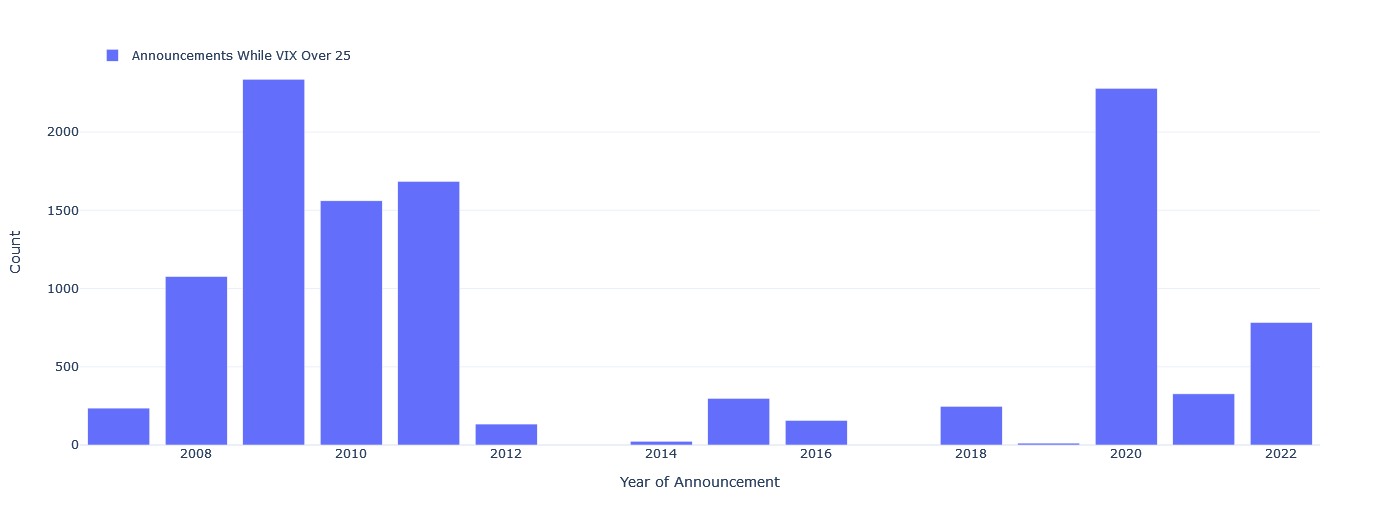}
                \caption{Distribution of Announcements by Year While VIX over 25}
                \label{fig:bullbear_dist_year}
            \end{figure}

            \paragraph{The difference in bull and bear markets}is another criterion under which we examine our dataset. One of the questions we initially had to clarify is how we define a bull or bear market. We decided to use the Cboe Volatility Index (VIX index) for this purpose. The VIX is designed to reflect the estimated volatility of the S\&P 500. Specifically, it reflects the market's estimate of the fluctuation of the S\&P 500 over the next 30 days \cite{Cboe.}. In phases with little uncertainty in the market, the VIX moves in a low range, while in phases with a lot of uncertainty in the market, it rises again. Values between 12 and 20 are considered normal. Everything below is a comparatively low value, and everything above is a comparatively high-value \cite{TimEdwards.}). In order to get a clearer demarcation between the Bull and Bear markets, we have decided to use the value 25 as the border between the two market directions. Hence, we see buybacks announced while the VIX is above 25 as announced during a bear market and all others as announced in a bull market. The bar chart \ref{fig:bullbear_dist_year} shows the distribution of announcements while the VIX is above 25. Clearly standing out are the times around 2008, as well as the initial phase of the Covid-19 crisis.
            
\begin{figure}[h]
                \centering
                \includegraphics[scale=0.3]{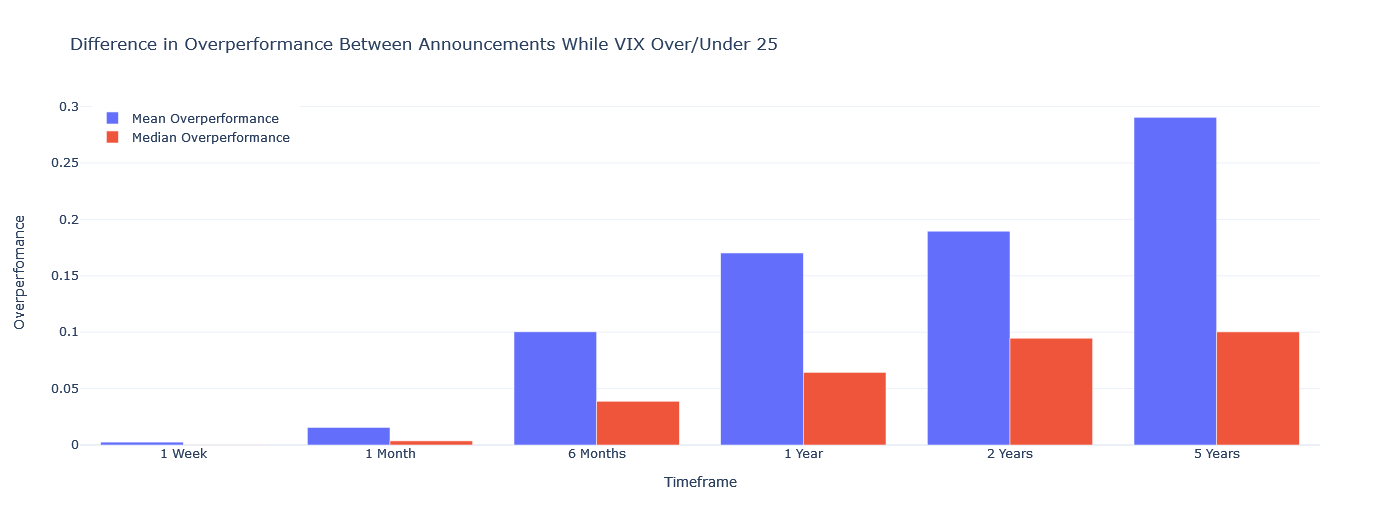}
                \caption{Difference in Overperformance Between Announcements While VIX Over/Under 25}
                \label{fig:bullbear_over_vs}
            \end{figure}

            If we now compare the average overperformance as well as the median of these two periods we find an interesting trend. The bar chart \ref{fig:bullbear_over_vs} shows the delta between the overperformance of all announcements made in the Bear market against all announcements made in the Bull market. The exact figures can also be seen in the attached table \ref{tab:bullbear_over}. 
            The delta can be read as the excess of percentage points of average/median overperformance of share buybacks in a bear market compared to the average overperformance during a bull market.
            Over every time period, we see a clear, positive, steadily increasing delta. 
            For example, after one year we have an average delta of 20.98\%. This does not mean that share buybacks in a bear market beat the MSCI World by 20.98\%, but that they deliver at least 20.98\% percentage points more overperformance than share buybacks during a bear market. Thus, theoretically, if an underperformance of 25\% is generated in a bull market and underperformance of 4.02\% in a bear market, the return still underperforms the benchmark index but is resulting in an overperformance delta of 20.98\%. This is especially true for the median, which also increases steadily throughout. Nevertheless, it must be clearly stated that it seems that share buybacks announced in a bear market have an advantage over other share buybacks.
            
\begin{figure}[h!]
                \centering
                \includegraphics[scale=0.38]{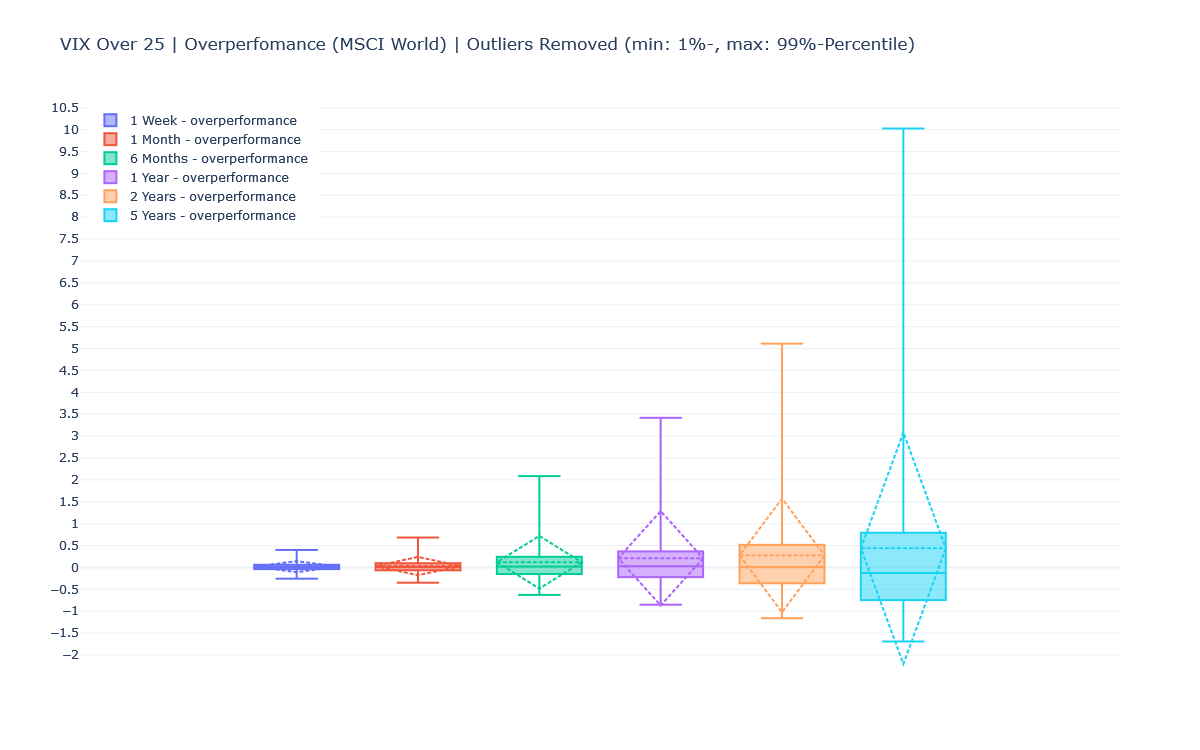}
                \caption{Box-Plot of Overperformance After Announcement While VIX Over 25}
                \label{fig:bullbear_over_box}
            \end{figure}

            If we now look at the general overperformance of share buybacks during a bear market, we see a positive picture. The graph \ref{fig:bullbear_over_box} is a box plot and shows the overperformance in the known period. The exact numbers can be seen in the attached chart \ref{tab:bullbear_over}.
            As with the overperformance of all announcements, we see a consistently increasing overperformance on average. However, the overperformance is significantly larger in the bear market. In the very short time frames (one week and one month), the average overperformance is just 1.3 to 1.6 times higher. After that, however, a difference of slightly more than twice to 3.1 times the overperformance can be seen. 
            Given that recessions last between eight and 18 months \cite{KellyAnneSmith.}, it is clear that companies that buy back shares in volatile times recover significantly better than the overall market.
            If you also look at the median, you can see that this statement even speaks for the majority of companies. The median only becomes negative after 5 years, unlike the overall median where it already becomes negative after 6 months.

        % \subsubsection{Robustness checks}
    \subsection{Classification and regression of future returns using machine learning}
    The last part of our methodology aims at identifying the minority of companies that generate significant excess returns after a share buyback. To achieve this, we created a total of 24 machine learning tasks and trained models for them in three different modes resulting in 72 machine learning models to categorize companies according to their forecasted performance after the announcement of a share buyback. We forecast performance and overperformance for six different time periods. 
    From these 72 models, we selected the best 24, four for each time frame.
    Our goal is to have four models for each time frame, all trained on the same data set, but all predicting something different. To end up with a final prediction for one time frame, we would not rely on the prediction of a single model, but on the prediction of four different models. The goal of doing this is to make the predictions more robust and therefore hopefully more accurate by using multiple models. In addition, we can take this a step further and will combine the predictions from different time frames, to enhance the accuracy for one time frame, even though most of the models used weren't trained for this specific time frame. 
    
    In the following, we will discuss the entire training process, the composition of the machine learning models, and their performance.
        \subsubsection{Overview}
           \paragraph{The time frames}as prediction targets are the same as those that have been used throughout the thesis.
            For each time period, one week, one month, six months, one year, two years, and five years, we generated four different machine learning tasks from our dataset.

            \paragraph{The four different tasks} are divided in classification and regression of performance and overperformance. This results in four tasks per time slot. The classification of the performance, the classification of the overperformance, the regression of the performance, and the regression of the overperformance. Classification aims at binary splitting the predicted (over-)performance into greater than or equal to 0 and less than 0. Regression aims to predict the actual percentage points of (over-)performance that a company will generate over the respective period. Through this, we have a total of 24 different tasks.
            
            \paragraph{The dataset}used is the same as described in the last chapter descriptive statistics \ref{subsec:statistics}, consisting of announcements of share buybacks and additional company data\footnote{A detailed list and explanation of the input data used can be found in Appendix \ref{tab:inputValues}}. The data consists for the most part of company-specific data, which can be found for instance in the income statement or balance sheet. In addition, market-related data such as share price or market cap were added to the dataset. Furthermore, macroeconomics-related data such as the VIX were included.  
            
            \paragraph{To reduce noise}in the dataset we applied a few filters. We have only taken samples for which we have the market cap, the last close price before the announcement, a total enterprise value, and an EBIT. We have done this to set a minimum requirement of information present. Most of the input features used for training are based on at least one of these four figures and if one or more of these are missing, a lot of potentially valuable information is lost and only adds noise to the dataset. In addition, we removed the lower 1\% and upper 99\% percentiles for the regression tasks. With this, we also are trying to keep the noise of the dataset low. The top percentiles include the most extreme samples. Excluding those, improved our losses significantly, as our models seem to struggle to find a pattern for the occurrences of such events. In total, this resulted in a dataset of 51,357 samples. 
            
            \paragraph{To standardize}the data, we subtracted the mean of each feature and divided it by the standard deviation of the respective feature, except for country, primary industry, month of the announcement, and day of the announcement, using sklearn's StandardScaler \cite{scikitlearn.StandardScaler.}. 
            
            \paragraph{We split the dataset}into a training and a test subset with a distribution of 80\% of the samples in the training dataset and the remaining 20\% in the test dataset. We divided the datasets randomly, but always in the same way, so each model trained in the following is based on the same training dataset and was evaluated on the same test dataset. In the following, the train dataset is partially further divided into an evaluation dataset. This depends on the training process and is mentioned accordingly. It is important to note that this is completely independent of our test dataset and that the test dataset was only used at the end to evaluate the trained models.
            
            \paragraph{The computing power}used to train the model were four identical virtual machines. Each virtual machine is equipped with 16 CPU-cores and 48GB RAM. The library to create the training pipeline used by us was optimised for CPU-based training \cite{PiotrPonskiURL.}. Therefor, we had no sensible usage for GPU resources.

        \subsubsection{Differences in training and hyperparameter optimization with AutoML mljar-supervised}
        The library we use to develop the machine learning models is AutoML mljar-supervised\footnote{https://github.com/mljar/mljar-supervised}.
        AutoML mljar-supervised allows us to build a powerful training pipeline for model fitting and hyperparameter optimization with very little code. It can also perform ensemble techniques like described in subsection \ref{subsec:ensebmle-tec}.
        The library has a selection of different modes which all have unique characteristics. We have taken advantage of a total of three different modes, which will now be described in detail \cite{PiotrPonskiURL.}.
        
            \paragraph{The explain} mode is the simplest mode. The purpose of the explain mode is to gain basic knowledge about the dataset using the simplest machine learning models, as well as to recognize the usefulness of applying machine learning to this dataset.  The training dataset is divided into 75\% training and 25\% evaluation datasets. Then a predefined set of models is trained. A baseline model, a linear model, a decision tree, a random forest, a XGBoost model, and a neural network.
            A baseline model simply returns the mean of the labels for regression tasks and the most frequent label for classification tasks. If models perform worse than the baseline, the dataset is most likely not suitable for machine learning. The other models are as described in the subsection \ref{subsec:backgroundData}. No hyperparameter optimizations are performed and the models are trained with their default parameters until they are fully optimized, such as the decision tree or linear model, or until they do not improve their evaluation loss. Each training run will be interrupted after 360 seconds, no matter if the model is fully optimized or not. This is necessary to keep the training time at a manageable length.
            In the end, all models are tried to be combined with an ensemble algorithm, also described in \ref{subsec:ensebmle-tec}.
            Despite the simplistic nature of the mode, we have had a good experience with its performance, as it does not allow much overfitting and can still draw correlations from the dataset \cite{PiotrPonskiURL.}.
            
            \paragraph{The perform} mode trains the models to a much greater extent. The mode uses 5-fold cross-validation to split the training set. The models trained are a linear model, random forest, LightGBM, XGBoost, CatBoost, and a neural network. Again, ensemble is used to combine the models at the end. The training time is the same as with the explain mode. Thus, all models are trained till no improvement is noticeable, they are fully optimized, or the time limit of 360 seconds is reached.
            The models were initialized and trained once with their default values, just like in the explain mode. Additionally, all models were trained 5 times with randomly chosen hyperparameters. 
            In addition, we have enabled the feature engineering methods golden features, and feature selection. Both techniques are explained in \ref{subsec:backgroundFeature}. Golden feature generates 10 new features by applying math operations to the original input features. Feature selection on the other hand drops features that don't add any predictive power.
            Next, all six model types are trained again with the best hyperparameter found so far, but with the golden features and feature selection separately, as well as combined, added additional as input.
            Then the best two of all six model types are taken and optimized with the help of two hill-climbing steps \cite{PiotrPonskiURL.}.

           % All the models generated in this run are then combined with the ensemble algorithm. Then all six model types are combined individually using stacking and also added to the pool of models used for ensemble algorithm and combined . 

            \paragraph{The Optuna}mode uses the in \ref{subsec:backgroundHyperparameter} described optimization framework for hyperparameter optimization. In Optuna mode, 10-fold cross-validation is used to split the dataset. The models used are Random Forest, Extra Trees, LightGBM, XGBboost, CatBoost, and neural network. The hyperparameters of each of these models are optimized for 800 seconds using the Optuna framework. 
            Optuna has as well golden features and feature selection enabled. 
            All these models are then combined again with ensemble. In addition, we have activated stacking at the end. This creates stacked models of the best 15 models, adds them to the pool of models, and combines all models again with ensemble \cite{PiotrPonskiURL.}.
            
        \subsubsection{Hyperparameter}
        The multitude of models that are trained and optimized in this thesis implies an even larger number of hyperparameters. We have adopted the default hyperparameters as well as the allowed value ranges for hyperparameter optimization from AutoML mljar-supervised\footnote{Detailed description and enumeration of all hyperparameters can be found under the following link: https://supervised.mljar.com/features/algorithms/}.
        
        \subsubsection{Training and hyperparameter optimization}
        With all this predefined, we can now start to fit the broadband of machine learning models to the 24 different tasks using the modes explain, perform, and Optuna. In this way, we have generated a total of 24 times three, thus 72 machine learning models. These 72 models are the result of the above-described training modes and can therefore mean that one model consists of many different sub-models. Thus, one model can consist of a tremendous number of individual decision trees and neural networks, which have been combined in various ways. In total, 323,501 seconds or just under 90 hours of training on 16 CPU-cores produced these models. We have combined the individual training parts just described into four very similar training pipelines. One each for the classification of performance, classification of overperformance, regression of performance, and regression of overperformance.
        All these pipelines have the same components and differ only in a few details. In the calculation and creation of the labels and in the standardization. To create the overperformance labels for classification and regression, we first had to calculate the overperformance. However, we had already saved this in our dataset, thanks to the analysis in the previous chapter. 
        For classification tasks, we also had to binary assign labels greater than or equal to zero and less than zero. For the regression tasks, we had to standardize the labels and save the values for later to scale back the values predicted by the model to real values.
        In each pipeline, six models are trained sequentially for the six time periods.
        We start by loading, cleaning, standardizing, and splitting the dataset, as well as generating the labels. Then the training library is loaded, the training settings are defined, and the models are adapted to the passed and standardized dataset as defined by the mode. 
        Starting with the one-week time frame, each time frame is processed. When the best model for the given time frame is found, it is evaluated on the separately defined test dataset and saved. 
        When all six time periods have been run through, the training run and the pipeline are finished.
        Each pipeline was run three times, for the three modes respectively, resulting in a total of 12 runs with four runs for each task of regression and classification of performance and overperformance.  
        
    \subsubsection{Training results}
    In this subsection, we will compare the results of the 72 machine learning models and their performance. Each of these 72 models maps to one of the four tasks\footnote{Performance classification, overperformance classification, performance regression, overperformance regression}, one of the six time frames\footnote{One week, one month, six months, one year, two years, five years} and one of the three training modes\footnote{Explain, perform, Optuna} and can consist of several sub-models. For each model, the respective loss function, the accuracy, the type of model, and the time how long it was trained are given.
    The loss function is useful to compare the models with each other, while the accuracy can give us a feeling for the results in the real world.
    The loss function for classification tasks is the log loss\footnote{https://scikit-learn.org/stable/modules/generated/sklearn.metrics.log\_loss.html} and for regression tasks the root-mean-square error (RMSE)\footnote{https://scikit-learn.org/stable/modules/generated/sklearn.metrics.mean\_squared\_error.html}. It is noteworthy that the RMSE is based on the standardized values, so it cannot be directly transferred to an error of percentage points. But this fact enables us to compare the regression models over the different time frames because without standardization the RMSE would be larger for longer time frames because the values to be predicted are larger. With standardization, they are all scaled to similar feature spaces and therefore comparable.
    The accuracy is self-explanatory for binary classification. For regression, it is to be interpreted as follows. To calculate the accuracy of the regression models, we have mapped the predicted performance and overperformance of the models binary to an (over-)performance greater than or equal to 0 and less than 0. Based on this, we could calculate the accuracy in the same way as for the classification. 
    The type indicates the type of model. Default XGBoost stands for a trained XGBoost model with standard hyperparameters. Ensemble stands for a model that consists of several models combined using the already often mentioned ensemble algorithm. Ensemble stacked also stands for an ensemble model, which additionally contains stacked models.
    The run time tables shown in the following display in seconds how long the individual models were trained on a 16 CPU-core VM.
    For the best model types and run times, the models with the best accuracy are highlighted by bold writing.

        \paragraph{Classification of performance}
        If we look at the log loss table \ref{tab:lossClassPerf} for the classification of performance, we see a picture that will be common to all other models. The Optuna trained models are generally the best performing, but the perform trained models do not compare badly. The accuracy is actually better from a yearly perspective when the models are trained in perform mode than in Optuna mode, but only by 0.0029pp, which is only a difference by 30 correct samples.
        
        \begin{table}[h]
        \centering
        \begin{tabular}{|c|c|c|c|c|c|c|}
        \hline
        \textbf{Time Frame} & \textbf{1W}    & \textbf{1M}    & \textbf{6M}    & \textbf{1Y}    & \textbf{2Y}    & \textbf{5Y}   \\ \hline
        \textbf{Explain} & 0.6453 & 0.65          & 0.6358 & {0.6257} & 0.6162 & 0.5593 \\ 
        \textbf{Optuna}   & \textbf{0.6309} & \textbf{0.6246} & \textbf{0.6057} & \textbf{0.5999} & \textbf{0.5851} & \textbf{0.5029} \\ 
        \textbf{Perform} & 0.6367 & {0.64} & 0.6257 & 0.6101          & 0.5953 & 0.521  \\ \hline
        \end{tabular}
        \caption{Log Loss for Classification of Performance}
        \label{tab:lossClassPerf}
        \end{table}
        
        With increasing time frames the loss decreases steadily and the accuracy increases steadily from 63.85\% to 75.88\%, as shown in table \ref{tab:AccClassPerf}. 
        Interestingly, the difference between the models trained in Explain mode, which can be seen as some sort of baseline, and the other two increases over time. The longer the time frame the more worthwhile hyperparameter optimization.
        
        \begin{table}[h]
        \centering
        \begin{tabular}{|c|c|c|c|c|c|c|}
        \hline
        \textbf{Time Frame} & \textbf{1W}     & \textbf{1M}   & \textbf{6M}     & \textbf{1Y}     & \textbf{2Y}     & \textbf{5Y}     \\ \hline
        \textbf{Explain}     & 0.6214 & 0.615  & 0.6299 & 0.6359          & 0.6583 & 0.7088 \\ 
        \textbf{Optuna}   & \textbf{0.6385} & \textbf{0.64} & \textbf{0.6558} & {0.6559} & \textbf{0.6861} & \textbf{0.7588} \\ 
        \textbf{Perform} & 0.6378 & 0.6274 & 0.6431 & \textbf{0.6588} & 0.6761 & 0.7465 \\ \hline
        \end{tabular}
        \caption{Accuracy for Classification of Performance}
        \label{tab:AccClassPerf}
        \end{table}
        
        The best model types are also very uniformly placed, as shown in table \ref{tab:BestClassPerf}. In explain mode, the default XGBoost is the best model, except for the two-year time frame. There it is an ensemble model. However, this ensemble model consists of the default XGBoost model, but in combination with the default neural network. The two are weighted with 4 and 1 respectively, which makes the output of the XGBoost model four times more important.
        The Optuna models all perform best with the ensemble model, which also includes stacked models. 
        The perform mode, on the other hand, has the ensemble models as the best model types. These also consist of several models but do not include stacked models. This is due to the nature of the perform mode, since it does not train stacked models by default.

        \begin{table}[h]
        \centering
        \begin{tabular}{|c|c|c|c|c|c|c|}
        \hline
        \textbf{Time Frame} & \textbf{1W} & \textbf{1M} & \textbf{6M} & \textbf{1Y}       & \textbf{2Y} & \textbf{5Y} \\ \hline
        \textbf{Explain} &
          \begin{tabular}[c]{@{}c@{}}Default \\ XGBoost\end{tabular} &
          \begin{tabular}[c]{@{}c@{}}Default \\ XGBoost\end{tabular} &
          \begin{tabular}[c]{@{}c@{}}Default \\ XGBoost\end{tabular} &
          \begin{tabular}[c]{@{}c@{}}Default \\ XGBoost\end{tabular} &
          Ensemble &
          \begin{tabular}[c]{@{}c@{}}Default \\ XGBoost\end{tabular} \\ 
        \textbf{Optuna} &
          \textbf{\begin{tabular}[c]{@{}c@{}}Ensemble \\ Stacked\end{tabular}} &
          \textbf{\begin{tabular}[c]{@{}c@{}}Ensemble \\ Stacked\end{tabular}} &
          \textbf{\begin{tabular}[c]{@{}c@{}}Ensemble \\ Stacked\end{tabular}} &
          {\begin{tabular}[c]{@{}c@{}}Ensemble \\ Stacked\end{tabular}} &
          \textbf{\begin{tabular}[c]{@{}c@{}}Ensemble \\ Stacked\end{tabular}} &
          \textbf{\begin{tabular}[c]{@{}c@{}}Ensemble \\ Stacked\end{tabular}} \\ 
        \textbf{Perform}    & Ensemble    & Ensemble    & Ensemble    & \textbf{Ensemble} & Ensemble    & Ensemble    \\ \hline
        \end{tabular}
        \caption{Best Model Type for Classification of Performance}
        \label{tab:BestClassPerf}
        \end{table}
        
        If we look at the run time table \ref{tab:TimeClassPerf}, we see that the high performance of the Optuna models comes at a very high cost. 
        None of the explain models had a longer training time than one and a half minutes. 
        The perform models have an average run time of just under an hour and offer almost the same performance as the Optuna models. 
        The Optuna models themselves have an average run time of almost two and a half hours. The longest Optuna run time was over four hours.
        While the times for the explain and perform modes are fairly consistent, the times for the Optuna vary greatly. This is due to a handy feature of the Optuna framework, as it can abort individual runs if they are not promising and also terminate the entire optimization process if there are no more improvements to be found.
        
        \begin{table}[h]
        \centering
        \begin{tabular}{|c|c|c|c|c|c|c|}
        \hline
        \textbf{Time Frame}    & \textbf{1W} & \textbf{1M} & \textbf{6M} & \textbf{1Y}      & \textbf{2Y} & \textbf{5Y} \\ \hline
        \textbf{Explain}     & 61.01       & 61.06       & 70          & 85.23            & 61.94       & 57.86       \\ 
        \textbf{Optuna} & \textbf{4180.06} & \textbf{6881.92} & \textbf{6014.56} & {10582.86} & \textbf{15660.94} & \textbf{8742.82} \\ 
        \textbf{Perform} & 3411.25     & 3518.4      & 3513.04     & \textbf{3515.91} & 3516.56     & 3537.58     \\ \hline
        \end{tabular}
        \caption{Run Time for Classification of Performance}
        \label{tab:TimeClassPerf}
        \end{table}
        
        \paragraph{The classification of overperformance} behaves in a similar way like the classification of performance. The corresponding tables are \ref{tab:lossClassOver}, \ref{tab:AccClassOver}, \ref{tab:BestClassOver} and \ref{tab:TimeClassOver}. 
        The Optuna models have the best loss, ensemble stacked is the best performing model and the Optuna models take a significant amount of more time to train compared to the other models.
        It is noticeable that although the five-year period again has the best loss and the best accuracy, the loss does not fall and the accuracy does not increase with the advancing time frames. Instead, the loss increases slightly and only falls again from the two-year time frame onwards. 
        In general, the loss and accuracy are significantly worse than in the performance classification. However, this makes sense, since a model must not only predict the performance of the company but this in the context of the global market, respectively with the MSCI World. This certainly makes the task more difficult and is likely to be the main reason for the additional uncertainty. The stock of a company may or may not be affected by changes in the global market. This mechanism, that the performance of a company can be affected to a different extend, then the global economy as a whole by changing market environments, can create this additional uncertainty and is probably the reason for the lower loss and accuracy.

        \begin{table}[h]
        \centering
        \begin{tabular}{|c|c|c|c|c|c|c|}
        \hline
        \textbf{Time Frame} & \textbf{1W}    & \textbf{1M}    & \textbf{6M}    & \textbf{1Y}    & \textbf{2Y}    & \textbf{5Y}   \\ \hline
        \textbf{Explain} & 0.6515 & 0.6651 & 0.673  & {0.673} & 0.6788 & 0.6778 \\ 
        \textbf{Optuna}   & \textbf{0.6422} & \textbf{0.6539} & \textbf{0.6615} & \textbf{0.6621} & \textbf{0.6519} & \textbf{0.62} \\ 
        \textbf{Perform} & 0.6445 & 0.6579 & 0.6655 & 0.6647         & 0.6584 & 0.6409 \\ \hline
        \end{tabular}
        \caption{Log Loss for Classification of Overperformance}
        \label{tab:lossClassOver}
        \end{table}
        
        \begin{table}[h]
        \centering
        \begin{tabular}{|c|c|c|c|c|c|c|}
        \hline
        \textbf{Time Frame} & \textbf{1W}     & \textbf{1M}     & \textbf{6M}     & \textbf{1Y}     & \textbf{2Y}     & \textbf{5Y}     \\ \hline
        \textbf{Explain} & 0.6149 & 0.599  & 0.5814 & 0.6005          & 0.6367 & 0.7102 \\ 
        \textbf{Optuna}   & \textbf{0.6314} & \textbf{0.6141} & \textbf{0.6073} & \textbf{0.6222} & \textbf{0.6707} & \textbf{0.7543} \\ 
        \textbf{Perform} & 0.6287 & 0.6106 & 0.5987 & {0.6183} & 0.6592 & 0.7396 \\ \hline
        \end{tabular}
        \caption{Accuracy for Classification of Overperformance}
        \label{tab:AccClassOver}
        \end{table}
        
        \begin{table}[h]
        \centering
        \begin{tabular}{|c|c|c|c|c|c|c|}
        \hline
        \textbf{Time Frame} &
          \textbf{1W} &
          \textbf{1M} &
          \textbf{6M} &
          \textbf{1Y} &
          \textbf{2Y} &
          \textbf{5Y} \\ \hline
        \textbf{Explain} &
          Ensemble &
          Ensemble &
          Ensemble &
          Ensemble &
          \begin{tabular}[c]{@{}c@{}}Default \\ XGBoost\end{tabular} &
          \begin{tabular}[c]{@{}c@{}}Default \\ XGBoost\end{tabular} \\ 
        \textbf{Optuna} &
          \textbf{\begin{tabular}[c]{@{}c@{}}Ensemble \\ Stacked\end{tabular}} &
          \textbf{\begin{tabular}[c]{@{}c@{}}Ensemble \\ Stacked\end{tabular}} &
          \textbf{\begin{tabular}[c]{@{}c@{}}Ensemble \\ Stacked\end{tabular}} &
          \textbf{\begin{tabular}[c]{@{}c@{}}Ensemble \\ Stacked\end{tabular}} &
          \textbf{\begin{tabular}[c]{@{}c@{}}Ensemble \\ Stacked\end{tabular}} &
          \textbf{\begin{tabular}[c]{@{}c@{}}Ensemble \\ Stacked\end{tabular}} \\ 
        \textbf{Perform} &
          Ensemble &
          Ensemble &
          Ensemble &
          Ensemble &
          Ensemble &
          Ensemble \\ \hline
        \end{tabular}
        \caption{Best Model Type for Classification of Overperformance}
        \label{tab:BestClassOver}
        \end{table}
        
        \begin{table}[h]
        \centering
        \begin{tabular}{|c|c|c|c|c|c|c|}
        \hline
        \textbf{Time Frame} & \textbf{1W}      & \textbf{1M}      & \textbf{6M}      & \textbf{1Y}       & \textbf{2Y}       & \textbf{5Y}      \\ \hline
        \textbf{Explain} & 72.16   & 77.97   & 74.22   & 71.71 & 61.21  & 58.69  \\ 
        \textbf{Optuna}   & \textbf{4181.76} & \textbf{5137.91} & \textbf{8327.81} & \textbf{18797.62} & \textbf{15284.75} & \textbf{9722.12} \\ 
        \textbf{Perform} & 3489.16 & 3511.91 & 3501.26 & 3576  & 3513.8 & 3533.6 \\ \hline
        \end{tabular}
        \caption{Run Time for Classification of Overperformance}
        \label{tab:TimeClassOver}
        \end{table}
        
        \paragraph{For the regression of performance,} we see a similar picture as in the classification tasks, but with more deviation. The tables with loss and accuracy are \ref{tab:RmseRegPerf} and \ref{tab:AccRegPerf}.
        Most Optuna models are still the most accurate, but only in four out of six cases. The one-year time frame is rather eye-catching. Here the explain model has the best loss and the best accuracy. For the one-week time frame, the explain model also has a better loss than the perform and for one month the explain model has a better loss than the Optuna. Generally, the difference between the perform, Optuna and the explain models is smaller than for the classification tasks. 
        If we look at the evaluation loss table \ref{tab:EvalRegOver} of the models, meaning the loss used in training to identify the best model, we see a very different picture. The Optuna models consistently perform the best. This is a strong indication that due to the very long training sessions with Optuna, the hyperparameters have adapted to the evaluation dataset. Classically, overfitting occurs mainly between training and evaluation datasets. However, with hyperparameter tuning, the models can indirectly adapt to the evaluation dataset and overfit. This is because the models that are further trained are the ones that perform best on the evaluation dataset \cite{Bengio.2000}. The explain models have no risk of overfitting the evaluation dataset, as they are not optimized with hyperparameter tuning. Merely, the models trained in explain are then combined with the ensemble algorithm.
        
        \begin{table}[h]
        \centering
        \begin{tabular}{|c|c|c|c|c|c|c|}
        \hline
        \textbf{Time Frame}   & \textbf{1W}     & \textbf{1M}     & \textbf{6M}     & \textbf{1Y}     & \textbf{2Y}     & \textbf{5Y}     \\ \hline
        \textbf{Explain} & 0.9271 & 0.925           & 0.9036 & \textbf{0.9237} & 0.9256 & 0.9483 \\ 
        \textbf{Optuna} & \textbf{0.8965} & {0.9387} & \textbf{0.8669} & {1.0331} & \textbf{0.8882} & \textbf{0.8644} \\ 
        \textbf{Perform} & 0.9489 & \textbf{0.9135} & 0.8841 & 0.9715          & 0.8958 & 0.9004 \\ \hline
        \end{tabular}
        \caption{RMSE for Regression of Performance}
        \label{tab:RmseRegPerf}
        \end{table}
        
        \begin{table}[h]
        \centering
        \begin{tabular}{|c|c|c|c|c|c|c|}
        \hline
        \textbf{Time Frame} & \textbf{1W} & \textbf{1M} & \textbf{6M} & \textbf{1Y} & \textbf{2Y} & \textbf{5Y} \\ \hline
        \textbf{Explain}   & 0.933954    & 0.925219    & 0.894861    & 0.916413    & 0.929378    & 0.932131    \\ 
        \textbf{Optuna} & \textbf{0.892464} & \textbf{0.889246} & \textbf{0.870423} & \textbf{0.894123} & \textbf{0.889967} & \textbf{0.857933} \\ 
        \textbf{Perform}   & 0.9018      & 0.910267    & 0.889528    & 0.905881    & 0.898852    & 0.877212    \\ \hline
        \end{tabular}
        \caption{Evaluation Loss for Regression of Performance}
        \label{tab:EvalRegPerf}
        \end{table}
        
        In general, the regression task seems to be more difficult than the classification task. The accuracy is everywhere somewhat smaller than with the classification. Of course, this makes sense, since the classification models were specifically trained for the binary classification of the companies and the regression models have to deal with the more difficult task of predicting a continuous value.
        
        \begin{table}[h]
        \centering
        \begin{tabular}{|c|c|c|c|c|c|c|}
        \hline
        \textbf{Time Frame} & \textbf{1W}     & \textbf{1M}     & \textbf{6M}     & \textbf{1Y}     & \textbf{2Y}     & \textbf{5Y}     \\ \hline
        \textbf{Explain} & 0.6168 & 0.613          & 0.6211 & \textbf{0.6358} & 0.6221 & 0.6934 \\ 
        \textbf{Optuna}   & \textbf{0.6372} & {0.6192} & \textbf{0.6526} & {0.5618} & \textbf{0.6735} & \textbf{0.7327} \\ 
        \textbf{Perform} & 0.6207 & \textbf{0.625} & 0.6365 & 0.5816          & 0.6518 & 0.7051 \\ \hline
        \end{tabular}
        \caption{Accuracy for Regression of Performance}
        \label{tab:AccRegPerf}
        \end{table}
        
        The best model types, which are shown in table \ref{tab:BestRegPerf}, are again a rather uniform picture. Again, the ensemble models perform best. 
        The one-year ensemble model of the explain mode consists of a default XGBoost with a weighting of 4 and a simple decision tree with a weighting of 1.

        \begin{table}[h]
        \centering
        \begin{tabular}{|c|c|c|c|c|c|c|}
        \hline
        \textbf{Time Frame} & \textbf{1W} & \textbf{1M}       & \textbf{6M} & \textbf{1Y}      & \textbf{2Y} & \textbf{5Y} \\ \hline
        \textbf{Explain} &
          \begin{tabular}[c]{@{}c@{}}Default \\ XGBoost\end{tabular} &
          \begin{tabular}[c]{@{}c@{}}Default \\ XGBoost\end{tabular} &
          \begin{tabular}[c]{@{}c@{}}Default \\ XGBoost\end{tabular} &
          \textbf{Ensemble} &
          Ensemble &
          \begin{tabular}[c]{@{}c@{}}Default \\ XGBoost\end{tabular} \\ 
        \textbf{Optuna} &
          \textbf{\begin{tabular}[c]{@{}c@{}}Ensemble \\ Stacked\end{tabular}} &
          {\begin{tabular}[c]{@{}c@{}}Ensemble \\ Stacked\end{tabular}} &
          \textbf{\begin{tabular}[c]{@{}c@{}}Ensemble \\ Stacked\end{tabular}} &
          \begin{tabular}[c]{@{}c@{}}Ensemble \\ Stacked\end{tabular} &
          \textbf{\begin{tabular}[c]{@{}c@{}}Ensemble \\ Stacked\end{tabular}} &
          \textbf{\begin{tabular}[c]{@{}c@{}}Ensemble \\ Stacked\end{tabular}} \\ 
        \textbf{Perform}    & Ensemble    & \textbf{Ensemble} & Ensemble    & Ensemble & Ensemble    & Ensemble    \\ \hline
        \end{tabular}
        \caption{Best Model Type for Regression of Performance}
        \label{tab:BestRegPerf}
        \end{table}

        \paragraph{The regression of overperformance}is very consistent with the regression of performance. In general, the loss and accuracy are slightly worse compared to the two classification tasks. 
        Again, the explain models perform better than the classification. Optuna is again not the best alternative for the one-month and one-year periods. 
        Looking at the evaluation loss, overfitting can also be seen here, which explains the poor performance. 
        Also the sample of the best model types is the same as in the regression of the performance\footnote{A detailed showcase of all figures regarding the regression of overperformance can be found in the Appendix  \ref{tab:RmseRegOver}, \ref{tab:AccRegOver}, \ref{tab:BestRegOver}, \ref{tab:EvalRegOver}, \ref{tab:TimeRegOver}}. 
        
        \paragraph{As a model example,}we will break down the model for classifying performance for the two-year period, which has been improved with the Optuna training-mode. We have chosen this model because it is one of the more complex ensemble stacked models and thus provides a good example. It takes advantage of all ensemble techniques, feature engineering methods, and almost all model types.
        Table \ref{tab:class2YEnsStac} shows the content of the said ensemble stacked model. The weighting shows the relevance of each model to the ensemble model. Feature Engineering shows if and which methods were used to modify or reduce the input feature space. Stacked indicates whether the model is stacked or not. 
        In total, nine models are included in the ensemble stacked model. The nine are composed of five model types, namely three CatBoost, two Neural Network, two Random Forest, one Extra Trees and one Ensemble model.
        The Ensemble model is left out for now, as it itself consists of different model types and we will discuss it later. 
        In general, the hyperparameters of all models were optimized with Optuna. Half of them use SelectedFeature as feature engineering and only one model uses GoldenFeatures. Both feature engineering methods are explained in \ref{subsec:backgroundFeature}.
        All but the first CatBoost model are stacked. 
        The CatBoost and RandomForest models have the highest weights, making them the most influential models for this classification task. We will confirm that these models perform best in the classification tasks in the upcoming paragraph.
        Even if only to a comparably small extent, another ensemble model also influences the output of the entire ensemble model.
        The composition of this model is listed in table \ref{tab:class2YEns}.
        
        \begin{table}[h]
        \centering
        \begin{tabular}{|c|c|c|c|c|}
        \hline
        \textbf{Model Type} &
          \textbf{Weight} &
          \textbf{\begin{tabular}[c]{@{}c@{}}Optuna \\ Optimized\end{tabular}} &
          \textbf{\begin{tabular}[c]{@{}c@{}}Feature \\ Engineering\end{tabular}} &
          \textbf{Stacked} \\ \hline
        \textbf{CatBoost}      & 1 & Yes & -                & No  \\ 
        \textbf{CatBoost}      & 2 & Yes & GoldenFeatures   & Yes \\ 
        \textbf{CatBoost}      & 7 & Yes & SelectedFeatures & Yes \\ 
        \textbf{NeuralNetwork} & 1 & Yes & SelectedFeatures & Yes \\ 
        \textbf{NeuralNetwork} & 1 & Yes & -                & Yes \\ 
        \textbf{RandomForest}  & 9 & Yes & SelectedFeatures & Yes \\ 
        \textbf{RandomForest}  & 5 & Yes & -                & Yes \\ 
        \textbf{ExtraTrees}    & 2 & Yes & SelectedFeatures & Yes \\ 
        \textbf{Ensemble}      & 3 & -   & -                & -   \\ \hline
        \end{tabular}
        \caption{Example Enumeration of all Models Included in an Ensemble Stacked Model}
        \label{tab:class2YEnsStac}
        \end{table}  
        
        This ensemble model consists of seven models, which include four different model types. Here, as well, a fair half of the models use SelectedFeatures as feature engineering. None of these models is stacked, as this only happens at a later stage of the training process, when the ensemble model was created.
        The weights are more diversified. The most important models are LightGBM and XGBoost models.
        
        \begin{table}[h]
        \centering
        \begin{tabular}{|c|c|c|c|}
        \hline
        \textbf{Model Type} & \textbf{Weight} & \textbf{Optuna Optimized} & \textbf{Feature Engineering} \\ \hline
        \textbf{LightGBM}   & 3 & Yes & -                \\ 
        \textbf{LightGBM}   & 1 & Yes & SelectedFeatures \\ 
        \textbf{XGBoost}    & 2 & Yes & -                \\ 
        \textbf{XGBoost}    & 4 & Yes & SelectedFeatures \\ 
        \textbf{CatBoost}   & 2 & Yes & -                \\ 
        \textbf{CatBoost}   & 1 & Yes & SelectedFeatures \\ 
        \textbf{ExtraTrees} & 1 & Yes & SelectedFeatures \\ \hline
        \end{tabular}
        \caption{Example Enumeration of all Models Included in an Ensemble Model}
        \label{tab:class2YEns}
        \end{table}

        \paragraph{To showcase the overall best model types,}we now highlight the model types that perform best, valued by evaluation loss. Here we exclude ensemble models and focus only on model types, which may consist of several models, but always of the same model type.  
        As shown in table \ref{tab:overallBestModel}, the CatBoost model dominates for the classification tasks. In total, this is the model with the best performance seven times out of the 12 models. The second most common model is RandomForest with a total of only three occurrences, closely followed by ExtraTress with two occurrences.
        
        For the regression tasks, we move away from the gradient boosted model. CatBoost appears only twice. Instead, the two Forests of Randomized Trees have the upper hand. 
        RandomForest and ExtraTrees occur a total of eight times. The default model XGBoost is also found twice, but only for the periods where the Optuna and perform mode have suffered from overfitting.
        
        What is very apparent is that stacked models deliver the best performance across the board, with very few exceptions. Stacked models appear a total of 20 times in the 24 models. The four times where the stacked models did not give the best results were cases of overfitting, which we encountered in some of the regression tasks. 
        What is almost as common are models where the number of input features has been reduced by the SelectedFeatures algorithm. A total of 16 models can benefit from the smaller feature space.
        The GoldenFeautre algorithm, on the other hand, rarely seems to give the models an advantage. In total, models with the GoldenFeature algorithm can be found only three times.

        \begin{table}[h]
        \centering
        \resizebox{\textwidth}{!}{%
        \begin{tabular}{|c|c|c|c|c|c|c|}
        \hline
        \textbf{Time Frame} & \textbf{1W} & \textbf{1M} & \textbf{6M} & \textbf{1Y} & \textbf{2Y} & \textbf{5Y} \\ \hline
        \textbf{Class. Perf.} &
          \begin{tabular}[c]{@{}c@{}}Optuna \\ CatBoost \\ GoldenFeatures \\ SelectedFeatures \\ Stacked\end{tabular} &
          \begin{tabular}[c]{@{}c@{}}Optuna \\ CatBoost \\ SelectedFeatures \\ Stacked\end{tabular} &
          \begin{tabular}[c]{@{}c@{}}Optuna \\ CatBoost \\ SelectedFeatures \\ Stacked\end{tabular} &
          \begin{tabular}[c]{@{}c@{}}Optuna \\ RandomForest \\ SelectedFeatures \\ Stacked\end{tabular} &
          \begin{tabular}[c]{@{}c@{}}Optuna \\ RandomForest \\ SelectedFeatures \\ Stacked\end{tabular} &
          \begin{tabular}[c]{@{}c@{}}Optuna \\ CatBoost \\ Stacked\end{tabular} \\ \hline
        \textbf{Class. Over.} &
          \begin{tabular}[c]{@{}c@{}}Optuna \\ CatBoost \\ SelectedFeatures \\ Stacked\end{tabular} &
          \begin{tabular}[c]{@{}c@{}}Optuna \\ CatBoost \\ SelectedFeatures \\ Stacked\end{tabular} &
          \begin{tabular}[c]{@{}c@{}}Optuna \\ RandomForest \\ Stacked\end{tabular} &
          \begin{tabular}[c]{@{}c@{}}Optuna \\ ExtraTrees \\ SelectedFeatures \\ Stacked\end{tabular} &
          \begin{tabular}[c]{@{}c@{}}Optuna\\ CatBoost \\ SelectedFeatures \\ Stacked\end{tabular} &
          \begin{tabular}[c]{@{}c@{}}Optuna \\ ExtraTrees \\ SelectedFeatures \\ Stacked\end{tabular} \\ \hline
        \textbf{Regr. Perf.} &
          \begin{tabular}[c]{@{}c@{}}Optuna \\ ExtraTrees \\ SelectedFeatures \\ Stacked\end{tabular} &
          \begin{tabular}[c]{@{}c@{}}Perform\\ CatBoost \\ GoldenFeatures \end{tabular} &
          \begin{tabular}[c]{@{}c@{}}Optuna \\ RandomForest \\ SelectedFeatures \\ Stacked\end{tabular} &
          \begin{tabular}[c]{@{}c@{}}Default \\ XGBoost\end{tabular} &
          \begin{tabular}[c]{@{}c@{}}Optuna \\ RandomForest \\ SelectedFeatures \\ Stacked\end{tabular} &
          \begin{tabular}[c]{@{}c@{}}Optuna \\ ExtraTrees \\ SelectedFeatures \\ Stacked\end{tabular} \\ \hline
        \textbf{Regr. Over.} &
          \begin{tabular}[c]{@{}c@{}}Optuna \\ CatBoost \\ GoldenFeatures \\ SelectedFeatures \\ Stacked\end{tabular} &
          \begin{tabular}[c]{@{}c@{}}Perform\\ RandomForest \end{tabular} &
          \begin{tabular}[c]{@{}c@{}}Optuna \\ ExtraTrees \\ Stacked\end{tabular} &
          \begin{tabular}[c]{@{}c@{}}Default \\ XGBoost\end{tabular} &
          \begin{tabular}[c]{@{}c@{}}Optuna \\ ExtraTrees \\ SelectedFeatures \\ Stacked\end{tabular} &
          \begin{tabular}[c]{@{}c@{}}Optuna \\ RandomForest \\ Stacked\end{tabular} \\ \hline
        \end{tabular}%https://www.overleaf.com/project/62d69375b485a8b9b6fe0e86
        }
        \caption{Best non-Ensemble Models Over all Time Frames}
        \label{tab:overallBestModel}
        \end{table}
        
        \subsubsection{Backtesting results}
        Our backtesting results are now discussed in this subsection. We have backtested our models to analyze not only their accuracy but also the theoretical returns generated. The goal is to use as realistic as possible evaluation methods to conclude a reasonable application of these models. However, it must be made clear from the outset that the theoretical figures shown here are in no way directly transferable to past or even future returns. Although the backtesting tests were performed on the test sub-dataset, so we see unbiased performance of the models, in the real market, many more variables have an impact than we have simulated or are able to simulate. 
        The models we are talking about in this subsection are always the ones that had the best test loss for the respective task and time frame. The models are summarized in tables \ref{tab:bestModelsClassPerf}, \ref{tab:bestModelsClassOver}, \ref{tab:bestModelsRegrPerf}, \ref{tab:bestModelsRegrOver}.
        
        \begin{table}[p]
        \centering
        \begin{tabular}{|c|c|c|c|c|c|c|}
        \hline
        \textbf{Time Frame} & \textbf{1W} & \textbf{1M} & \textbf{6M} & \textbf{1Y} & \textbf{2Y} & \textbf{5Y} \\ \hline
        \textbf{Mode}     & Optuna & Optuna & Optuna & Optuna & Optuna & Optuna \\ 
        \textbf{Loss}     & 0.6309 & 0.6246 & 0.6057 & 0.5999 & 0.5851 & 0.5029 \\ 
        \textbf{Accuracy} & 0.6385 & 0.64   & 0.6558 & 0.6559 & 0.6861 & 0.7588 \\ \hline
        \end{tabular}
        \caption{Summary of the Best Models for Classification of Performance}
        \label{tab:bestModelsClassPerf}
        \end{table}
        
        \begin{table}[p]
        \centering
        \begin{tabular}{|c|c|c|c|c|c|c|}
        \hline
        \textbf{Time Frame} & \textbf{1W} & \textbf{1M} & \textbf{6M} & \textbf{1Y} & \textbf{2Y} & \textbf{5Y} \\ \hline
        \textbf{Mode}     & Optuna & Optuna & Optuna & Optuna & Optuna & Optuna \\ 
        \textbf{Loss}     & 0.6422 & 0.6539 & 0.6615 & 0.6621 & 0.6519 & 0.62   \\ 
        \textbf{Accuracy} & 0.6314 & 0.6141 & 0.6073 & 0.6222 & 0.6707 & 0.7543 \\ \hline
        \end{tabular}
        \caption{Summary of the Best Models for Classification of Overperformance}
        \label{tab:bestModelsClassOver}
        \end{table}
        
        \begin{table}[p]
        \centering
        \begin{tabular}{|c|c|c|c|c|c|c|}
        \hline
        \textbf{Time Frame} &  \textbf{1W} & \textbf{1M} & \textbf{6M} & \textbf{1Y} & \textbf{2Y} & \textbf{5Y} \\ \hline
        \textbf{Mode}     & Optuna & Perform & Optuna & Explain & Optuna & Optuna \\ 
        \textbf{Loss}     & 0.8965 & 0.9135  & 0.8669 & 0.9237  & 0.8882 & 0.8644 \\ 
        \textbf{Accuracy} & 0.6372 & 0.625   & 0.6526 & 0.6358  & 0.6735 & 0.7327 \\ \hline
        \end{tabular}
        \caption{Summary of the Best Models for Regression of Performance}
        \label{tab:bestModelsRegrPerf}
        \end{table}
        
        \begin{table}[p]
        \centering
        \begin{tabular}{|c|c|c|c|c|c|c|}
        \hline
        \textbf{Time Frame} &  \textbf{1W} & \textbf{1M} & \textbf{6M} & \textbf{1Y} & \textbf{2Y} & \textbf{5Y} \\ \hline
        \textbf{Mode}     & Optuna & Perform & Optuna & Explain & Optuna & Optuna \\ 
        \textbf{Loss}     & 0.9149 & 0.9387  & 0.9506 & 0.972   & 0.9207 & 0.8804 \\ 
        \textbf{Accuracy} & 0.6269 & 0.6027  & 0.6055 & 0.5896  & 0.6476 & 0.7402 \\ \hline
        \end{tabular}
        \caption{Summary of the Best Models for Regression of Overperformance}
        \label{tab:bestModelsRegrOver}
        \end{table}
        
        \paragraph{Our four approaches}to backtest our models are described in this paragraph. The basic principle behind our approaches is to combine many weak learners in order to find a learner that is superior to the individual weak learners. As described below, you will see how we have further combined the models we have trained.
        
        A naive approach, in which we just buy at each share buyback announcement and sell again at the end of the respective period. 
        
        The second approach evaluates the share buyback announcement in advance for the respective time frame. A buy signal is only issued if all four models of the respective period issue a buy signal. So, the two classification models for performance and overperformance would have to return a one and the two regression models for performance and overperformance would have to return a number greater than or equal to 0. For example, we would buy shares of a company when their share buyback announcement was given a buy signal by the four models for the one-month period. We then sell the shares after exactly one month. We do this independently of all other periods. So, it is very possible that we have bought the same company for the one-week period as well as for the one-month period, as long as the respective four models have issued a buy signal.
        
        The last two approaches are based on combining the predictions for several time frames.
        To do this, we first combined the 24 models for the six time frames by combining the output of the four models for each of the six time frames in the same way as in our second approach. 
        We then combined the models using seven for-loops. Six for-loops for the six time frames and one for-loops for the target time frames, namely for how long the shares should be held. 
        This resulted in 279,936 different combinations. After filtering out all the duplicates with the same performance for the same time target, we were left with 378 different combinations of time frames and time targets. As an example, a combination of the time frames of one week, six months, and one year, with the target frame of one week, could exist. This means that with this combination, a buy signal for a share buyback is only issued if the four models of the one-week, six-months, and one-year time frame all together give a buy signal. 
        If this happens, the share is bought and held for one week.
        
        We have evaluated the large number of combinations according to two different criteria, which results in the last two approaches.
        The first combination approach is the valuation by the Sharpe Ratio\footnote{https://www.investopedia.com/terms/s/sharperatio.asp}.
        It is important to note that we have not used the classical calculation of the Sharpe Ratio and the figures shown later cannot be directly compared with other Sharpe Ratios outside this thesis. The Sharpe Ratio is calculated by dividing the excess return over the risk-free rate by the standard deviation (which stands for the risk) of the portfolio \cite{JasonFernando.}. The exact equation can be found below. Normally $R_p$ is the return, $R_f$ is the risk-free rate, and $\sigma_p$ is the standard deviation of the excess return of the portfolio.

        \begin{equation}
            Sharpe\ Ratio=\frac{\ R_p-R_f}{\sigma_p}
        \end{equation}
        
        We have not chosen the excess return over the risk-free rate, but the excess return over the MSCI World, meaning the delta in performance between the MSCI World and the stock under consideration for a specific time frame. In our case is $R_p$ still the return, but $R_f$ is the performance of the MSCI World and $\sigma_p$ is the standard deviation of the excess return, compared to the MSCI World. 
        The reason for this is mainly that we focus on the overperformance in comparison to the MSCI World in this thesis and the implementation of the backtesting algorithm. 
        This has the advantage that we can compare the risk of our portfolio with the performance of the MSCI World. The disadvantage is that we do not get a neutral assessment compared to the risk-free interest rate and cannot compare ourselves with other strategies outside this work.
        Thus, we have chosen the best combination for each of the six time frames, based on their Sharpe Ratio. 
        
        For the last approach, we did not evaluate the combination of models according to their Sharpe Ratios, but according to the Value at Risk\footnote{https://www.investopedia.com/terms/v/var.asp} (VaR). The VaR indicates how high the loss of the investment can be at a certain probability \cite{WillKenton.}.
        In our case, this means that the value indicates at a probability of 5\%, how high the underperformance can be compared to the MSCI World. So, a VaR of -0.1382 for the one-month time frame means that with this combination, with a probability of 5\%, one could underperform the MSCI World by 13.82\%.
        
        \paragraph{Our naive approach}is visualized as the table \ref{tab:backtestNaive}. The mean shows the average overperformance over the respective period. STD stands for standard deviation. The median is the 50\% percentile of all trades by overperformance, and accuracy indicates how many percent of the trades had a positive overperformance. Sum shows the summed overperformance percent points of all trades. Sharpe Ratio and Value at Risk are as described in the paragraph above.
        
        In general, an overperformance can be seen on average, but it quickly becomes an underperformance at the median. This is also shown by the accuracy, which becomes smaller with longer time frames. 
        Overall, a similar picture is drawn as already described in detail in subsection \ref{subsec:perfAndOver}.
        The Sharpe Ratio becomes smaller and smaller over time and approaches zero. Thus, the additional risk compared to the MSCI World is less and less justified.

        \begin{table}[h]
        \centering
        \begin{tabular}{|c|c|c|c|c|c|c|}
        \hline
        \textbf{Time Frame}         & \textbf{1W} & \textbf{1M} & \textbf{6M} & \textbf{1Y} & \textbf{2Y} & \textbf{5Y} \\ \hline
        \textbf{Mean}         & 0.0137 & 0.01694 & 0.0253  & 0.0476 & 0.0773 & 0.0925  \\ 
        \textbf{STD}          & 0.0759 & 0.12    & 0.292   & 0.4518 & 0.7163 & 1.2335  \\ 
        \textbf{Median}       & 0.0067 & 0.003   & -0.0147 & 0.0261 & 0.0642 & -0.1696 \\ 
        \textbf{Accuracy}  & 0.557  & 0.5153  & 0.4708  & 0.4638 & 0.4407 & 0.3949  \\ 
        \textbf{Sum}          & 141.02 & 173.95  & 259.74  & 488.79 & 794.11 & 950.21  \\ 
        % \textbf{Count}        & 10272  & 10272   & 10272   & 10272  & 10272  & 10272   \\ 
        \textbf{Sharpe Ratio} & 0.181  & 0.1411  & 0.0866  & 0.1053 & 0.1079 & 0.07499 \\ 
        \textbf{Value at Risk} & -0.114      & -0.1811     & -0.4565     & -0.6979     & -1.1047     & -1.9428     \\ \hline
        \end{tabular}
        \caption{Backtesting of the Naive Approach}
        \label{tab:backtestNaive}
        \end{table}
        
        Our second approach already shows much better results than the first approach. 
        All values, except for the standard deviation, improved significantly, as shown in the table \ref{tab:backtestTime}. 
        The mean rises greatly, especially over longer periods of time. The median is positive throughout all time frames and also increases with time. The accuracy is at least above 62\% for all time periods.
        Although the standard deviation has also increased, the mean increases significantly more and thus results in a significantly better Sharpe Ratio. The value at risk also improved slightly.
        
        \begin{table}[h]
        \centering
        \begin{tabular}{|c|c|c|c|c|c|c|}
        \hline
        \textbf{Time Frame}         & \textbf{1W} & \textbf{1M} & \textbf{6M} & \textbf{1Y} & \textbf{2Y} & \textbf{5Y} \\ \hline
        \textbf{Mean}         & 0.0299  & 0.0475 & 0.1258 & 0.2196 & 0.4068  & 1.0037  \\ 
        \textbf{STD}          & 0.07718 & 0.1224 & 0.3223 & 0.5274 & 0.8534  & 1.6018  \\ 
        \textbf{Median}       & 0.019   & 0.032  & 0.0664 & 0.1044 & 0.2105  & 0.5487  \\ 
        \textbf{Accuracy}  & 0.6479  & 0.6358 & 0.6192 & 0.6232 & 0.6628  & 0.7508  \\ 
        \textbf{Sum}          & 183.23  & 229.34 & 390.83 & 629.94 & 1232.89 & 2271.34 \\ 
        % \textbf{Count}        & 6138    & 4833   & 3107   & 2869   & 3031    & 2263    \\ 
        \textbf{Sharpe Ratio} & 0.3868  & 0.3876 & 0.3903 & 0.4163 & 0.4766  & 0.6266  \\ 
        \textbf{Value at Risk} & -0.0975     & -0.1545     & -0.406      & -0.657      & -1.0013     & -1.6393     \\ \hline
        \end{tabular}
        \caption{Backtesting of the Time Wise Optimized Approach}
        \label{tab:backtestTime}
        \end{table}

        For the third approach, we started to combine models, which were trained for different time frames. We applied the estimates of models from multiple time frames to a single time frame, in order to improve the evaluation figures again. 
        The exact results can be seen in table \ref{tab:backtestSR}. 
        To show which models were combined we added the row \textit{build} to the following tables.
        For example, the build {1W 1M 6M 1Y} for the one week time frame indicates that when the four models\footnote{The 24 used models can be found in the tables \ref{tab:bestModelsClassPerf}, \ref{tab:bestModelsClassOver}, \ref{tab:bestModelsRegrPerf}, \ref{tab:bestModelsRegrOver}} of these four time frames, all give a buy signal, the company is bought and then sold again after one week.
        As described above, we were able to identify a total of 378 different strategies by combining the different time frames. In order to select the best from this number of strategies, we focused on the Sharpe Ratio for the third approach and selected the combinations with the best Sharpe Ratios for the six time frames. 
        The strategies selected in this way again show a significant improvement over our first approach, as well as our second. The following is a comparison of the third approach, but only with the second approach.
        The mean, as well as the median, have increased noticeably. The accuracy is also above 70\% for almost all time windows.
        The Sharpe Ratio is also consistently above 0.5. 
        However, the standard deviation has slightly increased again, except for the one-year time frame. 
        Unfortunately, Value at Risk could hardly or not at all improve. 
        Most noticeable is the decrease of the sum. Although the quality of the trades has increased, as can be seen from the higher mean and higher accuracy, significantly fewer trades seem to have been executed. We achieve more often an overperformance with one trade and thus also reduce the risk, but we trade altogether less and miss out on potential returns. This results in a lower sum of overperformance percentage points.

        \begin{table}[h]
        \centering
        \begin{tabular}{|c|c|c|c|c|c|c|}
        \hline
        \textbf{Time Frame}         & \textbf{1W} & \textbf{1M} & \textbf{6M} & \textbf{1Y} & \textbf{2Y} & \textbf{5Y} \\ \hline
        \textbf{Build} &
          \begin{tabular}[c]{@{}c@{}}\{1W 1M \\ 6M 1Y\}\end{tabular} &
          \begin{tabular}[c]{@{}c@{}}\{1W 1M \\ 6M 1Y \\ 2Y\}\end{tabular} &
          \begin{tabular}[c]{@{}c@{}}\{1M 6M \\ 1Y 2Y\}\end{tabular} &
          \begin{tabular}[c]{@{}c@{}}\{1W 1M \\ 1Y 2Y \\ 5Y\}\end{tabular} &
          \begin{tabular}[c]{@{}c@{}}\{1M 6M \\ 1Y 2Y \\ 5Y\}\end{tabular} &
          \begin{tabular}[c]{@{}c@{}}\{6M 2Y \\ 5Y\}\end{tabular} \\ 
        \textbf{Mean}          & 0.0455      & 0.0704      & 0.1778      & 0.2782      & 0.5506      & 1.1766      \\ 
        \textbf{STD}           & 0.0851      & 0.1327      & 0.3358      & 0.5002      & 0.8813      & 1.7465      \\ 
        \textbf{Median}        & 0.0295      & 0.0435      & 0.1063      & 0.1933      & 0.3193      & 0.6615      \\ 
        \textbf{Accuracy}   & 0.7055      & 0.6868      & 0.6913      & 0.7054      & 0.7561      & 0.7718      \\ 
        \textbf{Sum}           & 62.88       & 71.19       & 199.31      & 201.15      & 359.01      & 1216.65     \\ 
        % \textbf{Count}         & 1382        & 1012        & 1121        & 723         & 652         & 1034        \\ 
        \textbf{Sharpe Ratio}  & 0.5349      & 0.5303      & 0.5294      & 0.5561      & 0.6248      & 0.6737      \\ 
        \textbf{Value at Risk} & -0.0949     & -0.1485     & -0.3763     & -0.5472     & -0.9035     & -1.7051     \\ \hline
        \end{tabular}
        \caption{Backtesting of the Sharpe Ratio Optimized Approach}
        \label{tab:backtestSR}
        \end{table}
        
        Our fourth and final approach is also based on the 378 different combinations but selected according to the best value at risk. The values for this can be found in table \ref{tab:backtestVaR}.
        Essentially, the results are very similar to those of the share ratio optimized strategy. However, most of the values are slightly worse. 
        Obviously, the Value at Risk has improved, but only slightly. What has also improved is the standard deviation. 
        Apart from that, the remaining values have deteriorated. 
        For the one-year, two-year and five-year periods the sum has increased, but for the rest, it has almost halved. 
        The mean and median have also worsened, but only by about 0.02pp for most of the time frames.
        Unfortunately, the accuracy has also dropped for most periods.

        \begin{table}[h]
        \centering
        \begin{tabular}{|c|c|c|c|c|c|c|}
        \hline
        \textbf{Time Frame}         & \textbf{1W} & \textbf{1M} & \textbf{6M} & \textbf{1Y} & \textbf{2Y} & \textbf{5Y} \\ \hline
        \textbf{Build} &
          \begin{tabular}[c]{@{}c@{}}\{1W 1M \\ 1Y 5Y\}\end{tabular} &
          \begin{tabular}[c]{@{}c@{}}\{1W 1M \\ 2Y 5Y\}\end{tabular} &
          \begin{tabular}[c]{@{}c@{}}\{1W 6M \\ 1Y 2Y \\ 5Y\}\end{tabular} &
          \begin{tabular}[c]{@{}c@{}}\{1W 1Y \\ 2Y 5Y\}\end{tabular} &
          \begin{tabular}[c]{@{}c@{}}\{1W 1M \\ 2Y 5Y\}\end{tabular} &
          \{2Y 5Y\} \\ 
        \textbf{Mean}          & 0.039       & 0.0583      & 0.1566      & 0.261       & 0.4768      & 1.098       \\ 
        \textbf{STD}           & 0.0759      & 0.1191      & 0.3061      & 0.4862      & 0.8166      & 1.6546      \\ 
        \textbf{Median}        & 0.0244      & 0.039       & 0.104       & 0.1825      & 0.2771      & 0.6526      \\ 
        \textbf{Accuracy}   & 0.6824      & 0.666       & 0.6884      & 0.6906      & 0.7334      & 0.7712      \\ 
        \textbf{Sum}           & 31.33       & 55.28       & 103.49      & 218.46      & 452.5       & 1862.28     \\ 
        % \textbf{Count}         & 809         & 949         & 661         & 837         & 949         & 1696        \\ 
        \textbf{Sharpe Ratio}  & 0.5141      & 0.4892      & 0.5115      & 0.5369      & 0.5839      & 0.6636      \\ 
        \textbf{Value at Risk} & -0.0862     & -0.1382     & -0.3484     & -0.5412     & -0.8705     & -1.632      \\ \hline
        \end{tabular}
        \caption{Backtesting of the Value at Risk Optimized Approach}
        \label{tab:backtestVaR}
        \end{table}

\chapter{Conclusion}
In this chapter, we summarize the findings of this thesis, highlight its limitations, and provide a perspective for future work.

    \section{Discussion}
    
        \subsection{Classification of share buyback announcements}
        The first problem we faced in this thesis was the classification of financial news. Here, we trained a RegEx filter as well as the NLP model DistilBERT on a small, hand-labeled dataset. The results were convincing for both approaches. We achieved an accuracy of 88\% with the RegEx filter and 90\% with DistilBERT. Even though the training process has a lot of potential for improvement, we were able to show that a classification of share buyback announcements related news can be well automated with already simple methods and that more complex NLP models can be well suited for this task as well, even though we used a dataset of only 500 hand-labeled samples. This enabled us to label historical finance news to create a big dataset about share buyback announcements, as well as the automatic classification of future news, as soon as they are released.
        
        \subsection{General impact of share buyback announcements}
        In order to find a general overperformance in connection with share buybacks, we have analyzed the dataset we generated in great detail. In comparison with the MSCI World, we were able to identify an average overperformance across all time frames analyzed. The median, on the other hand, indicated a clear underperformance. This picture is consistent throughout the entire thesis and leads us to assume that there is a minority of companies that clearly outperform the market, thus lifting the average, while the majority of companies underperform the market.
        However, after we no longer compared all companies with the MSCI World but sorted them into six market cap classes and compared the classes individually with the best matching index, we were no longer able to identify any significant average overperformance. The median, on the other hand, remained negative throughout and is decreasing further over time.
        Nevertheless, we were able to find a statistical excess in return, despite the market cap adjustment. To achieve this, we filtered out share buybacks that only intended to repurchase a small number of shares. The resulting overperformance was very similar to the initial picture found when compared to the MSCI World. An overperformance that became stronger over time could be seen. The median remained negative for most market cap classes, but not as strongly negative as before.
        We also found an interesting correlation when comparing share buybacks during bull and bear markets.
        Companies that carry out share buybacks in times of a bear market perform significantly better in comparison to the overall market than companies that announce share buybacks in a bull market. Thus, companies that announce share buybacks in times of crisis seem to recover better from these crises than the rest of the market. 
        
        \subsection{Classification and regression of future return}
        Using the dataset, we generated, and the knowledge gained from the analysis, we trained several machine learning models. 
        From the generated dataset, we created training datasets for 24 different tasks. For these 24 tasks, we trained a total of 72 machine learning models, from which we selected the best 24. 
        We combined these 24 models in a variety of ways to find the best possible set for predicting future returns of companies after announcing a share buyback over six different time frames. 
        The 24 models are composed of four models for each of the six time frames. 
        To test the capabilities of these combined models, we backtested them on a test dataset. The results we have are very promising.
        By combining these four models, we were able to generate an overperformance for each time frame compared to the MSCI World. We also could beat the general overperformance, we saw in the analysis of share buybacks. Thus, we achieve a significantly higher overperformance in the mean and median. The calculated Sharpe Ration could be increased drastically. We were also able to lower the value at risk. For example, we achieved an average excess return of 21.9\% for the one-year period, with an accuracy of 62.3\%.
        To go one step further, we have combined the models that were trained for specific time frames. This additionally led to an improvement of all values. For the one-year period, we were able to increase the average excess return to 27.8\%, with an accuracy of 70.5\%.  We were able to determine accuracies of up to 77\% and a clear overperformance. 
        All together, we were able to show that we can significantly improve the performance of our models, by combining their output, even though they are trained for different tasks and time frames.

        % Ob das gute investments sind muss jeder für sich selber entscheiden Risko bereitschaft 
        
    \section{Limitations}
    With our approach, we tried to create a dataset that is as large as possible. For this, we used two sources, each of which is an aggregation of information, which helps us maximize our coverage. Nevertheless, our dataset still contains limitations. First, despite our best efforts, we cannot claim the dataset to be complete. We do not have a comparative dataset with which to compare completeness. Also, data featuring the number of repurchased shares is limited in helping us verify the completeness of our data. This is because an announcement does not necessarily force the repurchase of shares and does not reflect the number of shares being repurchased. Instead, this information only covers the money invested in repurchases. However, we believe that we have achieved high coverage of all share buyback announcements within the category of English-speaking or globalized countries. MarketScreener aggregates a large sum of financial news websites, with its biggest focus on English-speaking countries \cite{MarketScreener.}. With this data as a basis, we can assume a high coverage of applicable countries, based on the categorization mentioned above. Another issue is the comparability with indices. Most of the comparisons in this thesis have been made using MSCI World as a reference index. While this is representative of larger companies in industrialized countries, it struggles to directly reflect the composition of our dataset. Improvements have been attempted by comparing market cap-specific indices however, these are not optimal, as they adjust for company size, but not for the country of origin or industry. It seems impossible for us to find an index that can perfectly reflect our company composition. Another limitation of our dataset is the time scale. Our dataset covers the period from January 2005 to April 2022. While these 17 years cover various global economic events, it only partially captures major macroeconomic cycles and trends. Thus, our statistical statements, as well as the models we have trained, are based only on the last 17 years. This makes our results vulnerable to changes in longer-term cycles, for instance, the long-term debt cycle of over 50-75 years \cite{NathanMetheny.}.
    
    Another limitation is related to our models for the classification and regression of future returns. While we have tried to generate robust results by training a large sum of models on different tasks and combining them, the models are still based on a singular dataset. This method does not address biases that may arise from the data. Since our test dataset is also sourced from the same dataset as our training dataset, these biases may also be found during testing, and are therefore hardly noticeable during the evaluation. But most importantly, results from backtesting must be viewed with caution. Backtested returns do not reflect real returns in any way. Our backtesting algorithm does not consider a variety of possible influences that real market trading includes.  Metrics like transaction costs, liquidity of the traded asset, or realistic fill prices were not considered. Liquidity especially, can have a large impact on real markets. Most of the stocks in our dataset come from small companies and are therefore often difficult to trade, especially when the transactions are time-critical \cite{Investor.gov.}. 
    Whether the investment strategies we found justify their risk, each must decide for themselves and cannot be evaluated by us. 
    
    Another missing point is the comparability with other state-of-the-art solutions. The field of quantitative trading has minimal transparency in comparison to other machine learning applications, as information and strategy are the foundation of any algorithm and disclosing such can jeopardize the profitability of a project. Thus, we have limited resources for comparison \cite{mcinturf2011keeping}.

    \section{Outlook}
    With this thesis, we analyzed share buybacks in great detail. Nevertheless, there are other aspects that could be further explored, expanded, or addressed in a different way, which could be dealt with in future work.
    For one thing, there is a high degree of potential in the classification of financial news for future work. The NLP techniques we created are based on a rather small dataset. Extending the dataset would not only yield better results, but it would also open the door for hyperparameter tuning, the reasonable use of much larger Deep Learning models, and the expansion of classifications to different news types.
    The trained classification and regression models for the prediction of returns also offer potential for improvement. On the one hand, it is possible to experiment with other or further optimized training methods. Furthermore, the explainability of these models can be improved, referring to Explainable Artificial Intelligence (XAI). Currently, these models are a black box to us. What we refer to as a "single model" can consists of a number of complex models, some which consist of sub-models. This complexity makes it impossible for us to understand the reasoning of the models for a decision. By applying different explainability tools, it would allow us to gain insight into the rationale behind the decisions of our models \cite{EjiroOnose.}. 
    Also, the data given to the models to predict future returns can be extended. Besides experimenting with other feature engineering techniques, one can also include a temporal dimension in the dataset. Currently, we only look at a company at the time it has announced a share buyback. Including the financials of the company from other points in time could improve the accuracy of the models. Also, things like the development of outstanding shares after the announcement of a share buyback could have a strong predictive power.
    The dataset could also be extended with macroeconomic values. In the statistical analysis, we have seen the relevance of the VIX. Additional macroeconomic values may also provide additional value to the forecasts.

    % groößeres dataset für nlp 
    % explainability
    % development of buybacks (not only snapshot of company but also 3M later)
    % mehr macroeconomics 
    
    % ensemble von den zeitabhängigen kombinationen
% TODO: add more chapters here

\appendix{}

\microtypesetup{protrusion=false}
\listoffigures{}
\listoftables{}
\listofalgorithms{}
\microtypesetup{protrusion=true}
\printbibliography{}

\chapter{Appendix}\label{chapter:appendix}

\begin{table}[h]
\centering
\begin{tabular}{|c|c|c|c|c|}
    \hline
\textbf{Size} &
\textbf{Count} &
\textbf{Mean in \$} &
\textbf{Standard Deviation} &
\textbf{Median in \$} \\
    \hline
\textbf{Nano}    & 9,990  & 24.6 M  & 13.8 M  & 24.2 M  \\
\textbf{Mirco}   & 15,771 & 140.8 M & 70.3 M  & 124.3 M \\
\textbf{Small}   & 15,305 & 865.1 M & 465.5 M & 735.9 M \\
\textbf{Mid}     & 8,738  & 4.6 B   & 2.2 B   & 4.0 B   \\
\textbf{Large}   & 5,134  & 34.7 B  & 34.0 B  & 21.5 B  \\
\textbf{Mega}    & 112   & 358.2 B & 305.8 B & 250.7 B  \\
\textbf{Overall} & 55,050 & 5.0 B   & 25.5 B  & 368.7 M \\
    \hline
\end{tabular}
\caption{Distribution of Announcements by Market Capitalization}
\label{tab:marketCapDist}
\end{table}

\begin{table}[h]
\centering
\begin{tabular}{|c|c|c|c|c|c|}
    \hline
\textbf{Country} &
  \textbf{Announcement} &
  \textbf{Company} &
  \textbf{Announcement} &
  \textbf{Announcements}  \\
 &
  \textbf{Count} &
  \textbf{Count} &
  \textbf{\%} &
  \textbf{per Company}  \\
    \hline
\textbf{Japan}          & 8,978 & 2,594 & 16.31\% & 3.46  \\
\textbf{United States}  & 8,114 & 2,359 & 14.74\% & 3.44  \\
\textbf{Hong Kong}      & 5,174 & 985  & 9.40\%  & 5.25  \\
\textbf{United Kingdom} & 3,876 & 580  & 7.04\%  & 6.68  \\
\textbf{South Korea}    & 3,521 & 1,191 & 6.40\%  & 2.96  \\
\textbf{China}          & 3,335 & 1,671 & 6.06\%  & 2.00  \\
\textbf{Canada}         & 3,020 & 598  & 5.49\%  & 5.05  \\
\textbf{Malaysia}       & 2,634 & 433  & 4.78\%  & 6.08  \\
\textbf{Singapore}      & 1,410 & 280  & 2.56\%  & 5.04  \\
\textbf{France}         & 1,314 & 276  & 2.39\%  & 4.76  \\
\textbf{Taiwan}         & 1,226 & 487  & 2.23\%  & 2.52  \\
    \hline
\end{tabular}
\caption{Distribution of Announcements by Country}
\label{tab:country}
\end{table}

\begin{table}[h]
\centering
\begin{tabular}{|c|c|c|c|c|}
    \hline
\textbf{Primary} &
  \textbf{Announcement} &
  \textbf{Company} &
  \textbf{Announcement} &
  \textbf{Announcements} \\
\textbf{Industry} &
  \textbf{Count} &
  \textbf{Count} &
  \textbf{\%} &
  \textbf{per Industry} \\
    \hline
\textbf{Industrial Machinery}               & 1932 & 583 & 3.51\% & 3.31 \\
\makecell{\textbf{Asset Management} \\ \textbf{and Custody Banks}} & 1695 & 394 & 3.08\% & 4.30 \\
\makecell{\textbf{Construction} \\ \textbf{and Engineering}}       & 1693 & 456 & 3.08\% & 3.71 \\
\makecell{\textbf{Packaged Foods} \\ \textbf{and Meats}}           & 1525 & 409 & 2.77\% & 3.73 \\
\textbf{\textbf{Application Software}}               & 1257 & 394 & 2.28\% & 3.19 \\
\makecell{\textbf{Real Estate} \\ \textbf{Development}}            & 1248 & 309 & 2.27\% & 4.04 \\
\makecell{\textbf{Trading Companies} \\ \textbf{and Distributors}} & 1238 & 289 & 2.25\% & 4.28 \\
\textbf{\textbf{Pharmaceuticals}}                    & 1124 & 351 & 2.04\% & 3.20 \\
\makecell{\textbf{IT Consulting} \\\textbf{and Other Services}}   & 1004 & 298 & 1.82\% & 3.37 \\
\makecell{\textbf{Real Estate} \\ \textbf{Operating Companies}}    & 969  & 225 & 1.76\% & 4.31 \\
    \hline
\end{tabular}
    \caption{Distribution of Announcements by Primary Industry}
    \label{tab:primany}
\end{table}

\begin{table}[h]
\centering
\begin{tabular}{|c|c|c|c|c|c|c|}
\hline
\textbf{Timeframe} &
  \textbf{Count} &
  \textbf{Mean} &
  \textbf{\begin{tabular}[c]{@{}c@{}}Standard\\ Deviation\end{tabular}} &
  \textbf{\begin{tabular}[c]{@{}c@{}}25\%\\ Percentile\end{tabular}} &
  \textbf{Median} &
  \textbf{\begin{tabular}[c]{@{}c@{}}75\%\\ Percentile\end{tabular}} \\ \hline
\textbf{1 Week}   & 11,054 & 0.0179 & 0.1250 & -0.0382 & 0.0072  & 0.0604 \\ 
\textbf{1 Month}  & 11,039 & 0.0316 & 0.2087 & -0.0652 & 0.0078  & 0.0963 \\ 
\textbf{6 Months} & 10,271 & 0.1190 & 0.6010 & -0.1508 & 0.0195  & 0.2425 \\ 
\textbf{1 Year}   & 10,242 & 0.2098 & 1.0787 & -0.2223 & 0.0273  & 0.3659 \\ 
\textbf{2 Years}  & 8,971  & 0.2757 & 1.2985 & -0.3624 & 0.0066  & 0.5157 \\ 
\textbf{5 Years}  & 7,510  & 0.4397 & 2.6439 & -0.7466 & -0.1270 & 0.7920 \\ \hline
\end{tabular}
\caption{Table of Overperformance After Announcement While VIX Over 25}
\label{tab:bullbear_over}
\end{table}

\begin{figure}[h]
    \centering
    \includegraphics[scale=0.3]{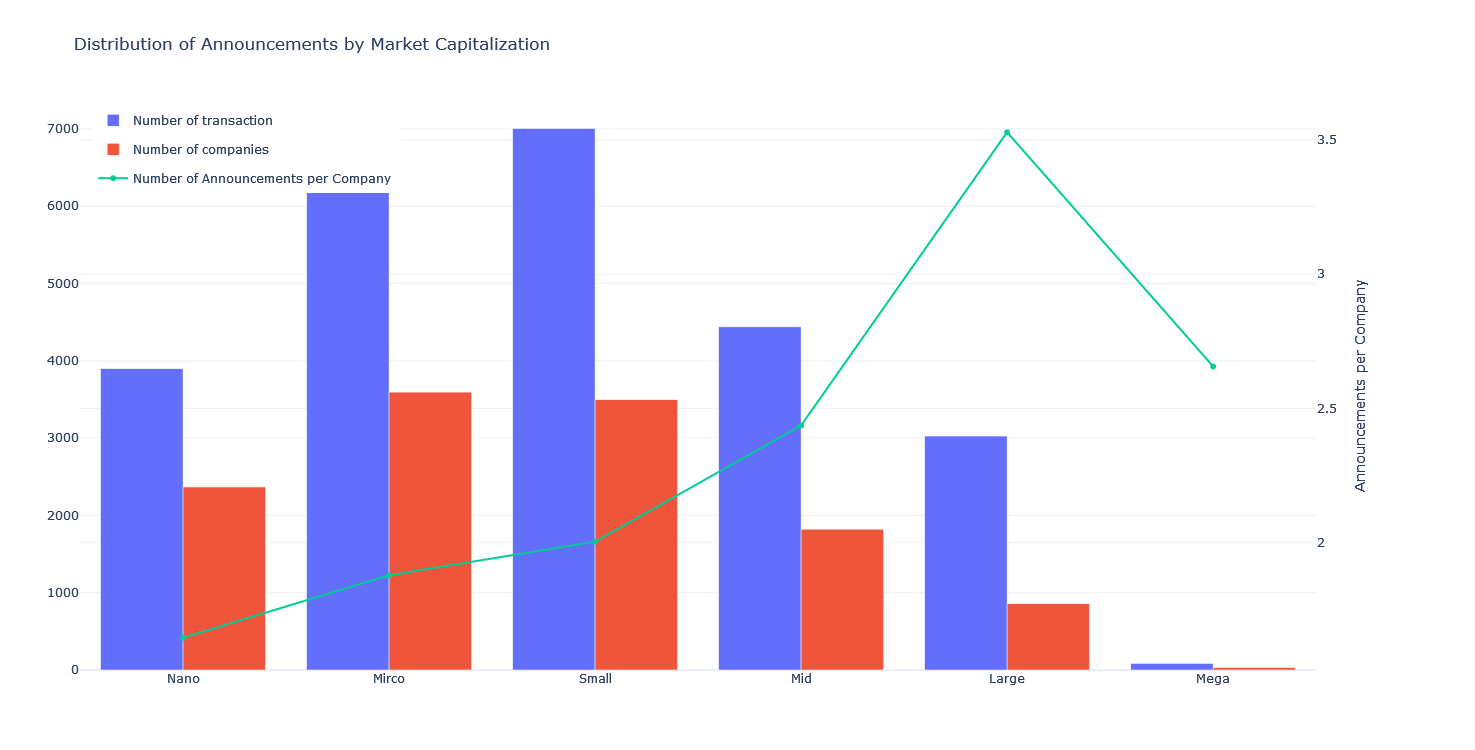}
    \caption{Distribution of Announcements by Market Capitalization with only Announced Transaction Size >1\%}
    \label{fig:dist_marketcap_big_transactions}
\end{figure}

% Note: It may be necessary to compile the document several times to get a multi-page table to line up properly
\begin{landscape}
\begin{longtable}[c]{|c|c|c|c|c|}
\hline
\textbf{Input} &
  \textbf{Unit} &
  \textbf{Time origin} &
  \textbf{Type} &
  \textbf{Explaination} \\ \hline
\endfirsthead
\multicolumn{5}{c}%
{{\bfseries Table \thetable\ continued from previous page}} \\
\hline
\textbf{Input} &
  \textbf{Unit} &
  \textbf{Time origin} &
  \textbf{Type} &
  \textbf{Explaination} \\ \hline
\endhead
\textbf{Country} &
  - &
  - &
  String &
  Country the comapny is based in \\ \hline
\textbf{Primary Industry} &
  - &
  - &
  String &
  Primary industry of  the company \\ \hline
\textbf{Month of announcement} &
  - &
  - &
  int &
  Month of announcement from 1-12 \\ \hline
\textbf{Day of announcement} &
  - &
  - &
  int &
  Day of announcement from 1-31 \\ \hline
\textbf{EBIT} &
  M - USD &
  Latest Quarter &
  float64 &
  - \\ \hline
\textbf{Short Interest} &
  \% &
  Latest &
  float64 &
  \begin{tabular}[c]{@{}c@{}}Number of shorted shares \\ divided by number of \\ shares outstanding \cite{CoryMitchell.Short.}\end{tabular} \\ \hline
\textbf{Float} &
  \% &
  Latest &
  float64 &
  \begin{tabular}[c]{@{}c@{}}Percentage of freely \\ tradable outstanding shares\end{tabular} \\ \hline
\textbf{Revenue} &
  M - USD &
  LTM &
  float64 &
  - \\ \hline
\textbf{Gross Profit} &
  M - USD &
  LTM &
  float64 &
  - \\ \hline
\textbf{Return on Capital} &
  \% &
  LTM &
  float64 &
  \begin{tabular}[c]{@{}c@{}}EBIT * (1 - .375) \\ / Average Total Capital\end{tabular} \\ \hline
\textbf{Price / Earnings} &
  \% &
  Latest Quarter &
  float64 &
  Last sale price devided by EPS \\ \hline
\textbf{VIX} &
  - &
  Latest &
  float64 &
  \begin{tabular}[c]{@{}c@{}}Last close price of VIX \\ at day of annoucement\end{tabular} \\ \hline
\textbf{Total Enterprise Value} &
  M - USD &
  Latest &
  float64 &
  Capital IQs calculation of TEV \\ \hline
\textbf{Total Debt to Total Assets Ratio} &
  \% &
  Latest Quarter &
  float64 &
  - \\ \hline
\textbf{Year Foundned} &
  - &
  - &
  int &
  Year the company was founded \\ \hline
\textbf{Marketcap} &
  M - USD &
  Latest &
  float64 &
  Marketcap of the company \\ \hline
\textbf{Shares Outstanding} &
  M - Shares &
  Latest &
  float64 &
  - \\ \hline
\textbf{Basic EPS - USD} &
  USD &
  LTM &
  float64 &
  - \\ \hline
\textbf{Beta - 1Y} &
  - &
  Latest &
  float64 &
  \begin{tabular}[c]{@{}c@{}}52 Week Volatility in \\ comparison to the market\end{tabular} \\ \hline
\textbf{Beta - 2Y} &
  - &
  Latest &
  float64 &
  \begin{tabular}[c]{@{}c@{}}104 Week Volatility in \\ comparison to the market\end{tabular} \\ \hline
\textbf{Beta - 5Y} &
  - &
  Latest &
  float64 &
  \begin{tabular}[c]{@{}c@{}}60 Month Volatility in \\ comparison to the market\end{tabular} \\ \hline
\textbf{BV/P} &
  USD &
  Latest Quarter &
  float64 &
  Book value per share \\ \hline
\textbf{Last Sale Price} &
  USD &
  Latest &
  float64 &
  \begin{tabular}[c]{@{}c@{}}Last close price of stock\\  at day of announcement\end{tabular} \\ \hline
\textbf{Relative Strenght Index} &
  - &
  Latest &
  float64 &
  RSI over last 14 days based on close\footnote{https://www.investopedia.com/terms/r/rsi.asp}\\ \hline
\textbf{Total Transaction Value} &
  M - USD &
  - &
  float64 &
  Size of announced buyback \\ \hline
\textbf{TEV / EBIT} &
  \% &
  Latest &
  float64 &
  - \\ \hline
\textbf{TEV / Revenue} &
  \% &
  Latest &
  float64 &
  - \\ \hline
\textbf{TEV / Gross Profit} &
  \% &
  Latest &
  float64 &
  - \\ \hline
\textbf{Return on Assets} &
  \% &
  LTM &
  float64 &
  \begin{tabular}[c]{@{}c@{}}EBIT * (1 - .375) \\ / Average Total Assets\end{tabular} \\ \hline
\textbf{Return on Equity} &
  \% &
  LTM &
  float64 &
  \begin{tabular}[c]{@{}c@{}}EBIT * (1 - .375) \\ / Average Total Equity\end{tabular} \\ \hline
\textbf{Return on Invested Capital} &
  \% &
  LTM &
  float64 &
  \begin{tabular}[c]{@{}c@{}}(Net Income - Total Dividends Paid) / \\ Total Capital\end{tabular} \\ \hline
\textbf{Return on Common Capital} &
  \% &
  LTM &
  float64 &
  Net Income Average Common Equity \\ \hline
\textbf{Gross Margin} &
  \% &
  LTM &
  float64 &
  Gross Margin / Total Revenues \\ \hline
\textbf{SG\&A Margin} &
  \% &
  LTM &
  float64 &
  SG\%A Expense / Total Revenues \\ \hline
\textbf{EBIT Margin} &
  \% &
  LTM &
  float64 &
  EBIT / Total Revenues \\ \hline
\textbf{Total Debt/Equity} &
  \% &
  Latest Quarter &
  float64 &
  - \\ \hline
\textbf{Total Debt/Capital} &
  \% &
  Latest Quarter &
  float64 &
  - \\ \hline
\textbf{Total Debt/EBITDA} &
  \% &
  LTM &
  float64 &
  - \\ \hline
\textbf{Altman Z Score} &
  - &
  LTM &
  float64 &
  Risk of bankruptcy \cite{WillKenton.Z.} \\ \hline
\textbf{Total Revenues - 1Y Growth} &
  \% &
  LTM &
  float64 &
  - \\ \hline
\textbf{Gross Profit - 1Y Growth} &
  \% &
  LTM &
  float64 &
  - \\ \hline
\textbf{Total Assets} &
  M - USD &
  Latest Quarter &
  float64 &
  - \\ \hline
\textbf{Total Current Liabilities} &
  M - USD &
  Latest Quarter &
  float64 &
  - \\ \hline
\textbf{3M Price Volatility} &
  \% &
  Latest &
  float64 &
  \begin{tabular}[c]{@{}c@{}}Standard deviation of price \\ over the last 3 months\end{tabular} \\ \hline
\textbf{6M Price Volatility} &
  \% &
  Latest &
  float64 &
  \begin{tabular}[c]{@{}c@{}}Standard deviation of price \\ over the last 6 months\end{tabular} \\ \hline
\textbf{52W Low Price} &
  USD &
  Latest &
  float64 &
  \begin{tabular}[c]{@{}c@{}}Price low of the\\  last 52 weeks\end{tabular} \\ \hline
\textbf{52W High Price} &
  USD &
  Latest &
  float64 &
  \begin{tabular}[c]{@{}c@{}}Price high of the\\  last 52 weeks\end{tabular} \\ \hline
\textbf{ROCE} &
  \% &
  Latest Quarter &
  float64 &
  \begin{tabular}[c]{@{}c@{}}EBIT / \\ (Total Assets \\ - Total Current Liabilities)\end{tabular} \\ \hline
\textbf{Transaction size \%} &
  \% &
  Latest &
  float64 &
  \begin{tabular}[c]{@{}c@{}}Ratio between transaction size \\ and market cap\end{tabular} \\ \hline
\textbf{52W Low Price \%} &
  \% &
  Latest &
  float64 &
  Last Sale Price / 52W Low Price \\ \hline
\textbf{52W High Price \%} &
  \% &
  Latest &
  float64 &
  Last Sale Price / 52W High Price \\ \hline
\textbf{Insider Ownership (\%)} &
  \% &
  Latest Quarter &
  float64 &
  \begin{tabular}[c]{@{}c@{}}Percentage of shares outstanding \\ help by insiders\end{tabular} \\ \hline
\textbf{Insider Ownership} &
  M - Shares &
  Latest Quarter &
  float64 &
  Number of shares held by insiders \\ \hline
\textbf{Institutional Ownership (\%)} &
  \% &
  Latest Quarter &
  float64 &
  \begin{tabular}[c]{@{}c@{}}Percentage of shares outstanding \\ held by institutional shareholders\end{tabular} \\ \hline
\textbf{Institutional Ownership} &
  M - Shares &
  Latest Quarter &
  float64 &
  \begin{tabular}[c]{@{}c@{}}Number of shares held \\ by institutional shareholders\end{tabular} \\ \hline
\textbf{Cash and Equivalents} &
  M - USD &
  Latest Quarter &
  float64 &
  - \\ \hline
\textbf{Short Term Investments} &
  M - USD &
  Latest Quarter &
  float64 &
  - \\ \hline
\textbf{Long Term Debt} &
  M - USD &
  Latest Quarter &
  float64 &
  - \\ \hline
\textbf{LT and ST Debt \%} &
  \% &
  Latest &
  float64 &
  \begin{tabular}[c]{@{}c@{}}(Long Term Debt \\ + Total Current Liabilities) \\ / Market Cap\end{tabular} \\ \hline
\caption{List of all Input Variables}
\label{tab:inputValues}\\
\end{longtable}
\end{landscape}

\begin{table}[h]
\centering
\begin{tabular}{|c|c|c|c|c|c|c|}
\hline
\textbf{Run time} & \textbf{1W}      & \textbf{1M} & \textbf{6M}      & \textbf{1Y} & \textbf{2Y}       & \textbf{5Y}      \\ 
\hline
\textbf{Explain} & 45.17  & 35.83            & 36.19   & \textbf{40.14} & 47.65   & 47.53   \\ 
\textbf{Optuna}   & \textbf{3735.28} & 4235.11     & \textbf{4779.69} & 20839.01    & \textbf{13473.52} & \textbf{7614.79} \\ 
\textbf{Perform} & 3288.5 & \textbf{3330.37} & 3446.01 & 3429.2         & 3509.38 & 3501.52 \\ 
\hline
\end{tabular}
\caption{Run Time for Regression of Overperformance}
\label{tab:TimeRegOver}
\end{table}

\begin{table}[h]
\centering
\begin{tabular}{|c|c|c|c|c|c|c|}
\hline
\textbf{Run time} & \textbf{1W}      & \textbf{1M} & \textbf{6M}       & \textbf{1Y} & \textbf{2Y}       & \textbf{5Y}       \\ \hline
\textbf{Explain} & 36.29   & 30.73            & 51.61  & \textbf{42.7} & 37.44   & 41.83  \\ 
\textbf{Optuna}   & \textbf{3862.36} & 4437.05     & \textbf{26845.88} & 12360.83    & \textbf{12676.78} & \textbf{10465.85} \\ 
\textbf{Perform} & 3406.39 & \textbf{3353.96} & 3465.9 & 3509.52       & 3506.72 & 3467.6 \\ \hline
\end{tabular}
\caption{Run Time for Regression of Performance}
\label{tab:TimeRegPer}
\end{table}

\begin{table}[h]
\centering
\begin{tabular}{|c|c|c|c|c|c|c|}
\hline
\textbf{RMSE}   & \textbf{1W}     & \textbf{1M} & \textbf{6M}     & \textbf{1Y} & \textbf{2Y}     & \textbf{5Y}     \\ \hline
\textbf{Explain} & 0.9322 & 0.9468          & 0.9605 & \textbf{0.972} & 0.9603 & 0.972  \\ 
\textbf{Optuna} & \textbf{0.9149} & 0.9455      & \textbf{0.9506} & 1.0349      & \textbf{0.9207} & \textbf{0.8804} \\ 
\textbf{Perform} & 0.9417 & \textbf{0.9387} & 0.955  & 0.996          & 0.9278 & 0.9234 \\ \hline
\end{tabular}
\caption{RMSE for Regression of Overperformance}
\label{tab:RmseRegOver}
\end{table}

\begin{table}[h]
\centering
\begin{tabular}{|c|c|c|c|c|c|c|}
\hline
\textbf{Accuracy} & \textbf{1W}     & \textbf{1M} & \textbf{6M}     & \textbf{1Y} & \textbf{2Y}     & \textbf{5Y}     \\ \hline
\textbf{Explain} & 0.6142 & 0.5963          & 0.5854 & \textbf{0.5896} & 0.5995 & 0.6815 \\ 
\textbf{Optuna}   & \textbf{0.6269} & 0.5991      & \textbf{0.6055} & 0.5298      & \textbf{0.6476} & \textbf{0.7402} \\ 
\textbf{Perform} & 0.6212 & \textbf{0.6027} & 0.6009 & 0.5413          & 0.6359 & 0.7151 \\ \hline
\end{tabular}
\caption{Accuracy for Regression of Overperformance}
\label{tab:AccRegOver}
\end{table}

\begin{table}[h]
\centering
\begin{tabular}{|c|c|c|c|c|c|c|}
\hline
\textbf{Eval Loss} & \textbf{1W} & \textbf{1M} & \textbf{6M} & \textbf{1Y} & \textbf{2Y} & \textbf{5Y} \\ \hline
\textbf{Explain}   & 0.935208    & 0.950918    & 0.958817    & 0.967021    & 0.960077    & 0.947985    \\ 
\textbf{Optuna} & \textbf{0.912458} & \textbf{0.928326} & \textbf{0.957233} & \textbf{0.950138} & \textbf{0.920071} & \textbf{0.866867} \\ 
\textbf{Perform}   & 0.920179    & 0.938236    & 0.963948    & 0.962316    & 0.928606    & 0.89475     \\ \hline
\end{tabular}
\caption{Evaluation Loss for Regression of Overperformance}
\label{tab:EvalRegOver}
\end{table}

\begin{table}[h]
\centering
\begin{tabular}{|c|c|c|c|c|c|c|}
\hline
\textbf{Best model} &
  \textbf{1W} &
  \textbf{1M} &
  \textbf{6M} &
  \textbf{1Y} &
  \textbf{2Y} &
  \textbf{5Y} \\ \hline
\textbf{Explain} &
  \begin{tabular}[c]{@{}c@{}}Default\\ XGBoost\end{tabular} &
  \begin{tabular}[c]{@{}c@{}}Default \\ XGBoost\end{tabular} &
  Ensemble &
  \textbf{Ensemble} &
  Ensemble &
  \begin{tabular}[c]{@{}c@{}}Default \\ XGBoost\end{tabular} \\ 
\textbf{Optuna} &
  \textbf{\begin{tabular}[c]{@{}c@{}}Ensemble \\ Stacked\end{tabular}} &
  \begin{tabular}[c]{@{}c@{}}Ensemble \\ Stacked\end{tabular} &
  \textbf{\begin{tabular}[c]{@{}c@{}}Ensemble \\ Stacked\end{tabular}} &
  \begin{tabular}[c]{@{}c@{}}Ensemble \\ Stacked\end{tabular} &
  \textbf{\begin{tabular}[c]{@{}c@{}}Ensemble \\ Stacked\end{tabular}} &
  \textbf{\begin{tabular}[c]{@{}c@{}}Ensemble \\ Stacked\end{tabular}} \\ 
\textbf{Perform} &
  Ensemble &
  \textbf{Ensemble} &
  Ensemble &
  Ensemble &
  Ensemble &
  Ensemble \\ \hline
\end{tabular}
\caption{Best Model Type for Regression of Overperformance}
\label{tab:BestRegOver}
\end{table}

\end{document}